\documentclass[12pt,nofootinbib,eqsecnum,longbibliography,prd]{revtex4-2}

\usepackage{amsmath,amssymb}
\usepackage[american]{babel}
\usepackage{microtype}
\usepackage{color}
\usepackage{graphicx}
\usepackage{hyperref}
\hypersetup{colorlinks, citecolor=blue, linkcolor=black, urlcolor=blue}
\usepackage{bm}

\newcommand{\ket}[1]{|#1\rangle}

\newcommand{\braket}[2]{\langle #1|#2\rangle}
\renewcommand{\v}[1]{\mathbf{#1}}

\newcommand{\etav}{\bm{\eta}}
\newcommand{\del}{\partial}

\usepackage{xpatch}
\makeatletter
\def\l@subsubsection#1#2{}
\patchcmd{\@ssect@ltx}
    {\addcontentsline{toc}{#1}{\protect\numberline{}#8}}
    {}
    {}
    {}

\pretocmd{\subsection}{\addtocontents{toc}{\protect\addvspace{-2.5\p@}}}{}{}
\makeatother

\begin{document}

\title{Interactions of Particles with ``Continuous Spin'' Fields}

\author{Philip Schuster}
\email{schuster@slac.stanford.edu}
\affiliation{SLAC National Accelerator Laboratory, 2575 Sand Hill Road, Menlo Park, CA 94025, USA}
\author{Natalia Toro}
\email{ntoro@slac.stanford.edu}
\affiliation{SLAC National Accelerator Laboratory, 2575 Sand Hill Road, Menlo Park, CA 94025, USA}
\author{Kevin Zhou}
\email{knzhou@stanford.edu}
\affiliation{SLAC National Accelerator Laboratory, 2575 Sand Hill Road, Menlo Park, CA 94025, USA}

\date{\today}

\begin{abstract}
Powerful general arguments allow only a few families of long-range interactions, exemplified by gauge field theories of electromagnetism and gravity. However, all of these arguments presuppose that massless fields have zero spin scale (Casimir invariant) and hence exactly boost invariant helicity. This misses the most general behavior compatible with Lorentz symmetry. We present a Lagrangian formalism describing interactions of matter particles with bosonic ``continuous spin'' fields with arbitrary spin scale $\rho$. Remarkably, physical observables are well approximated by familiar theories at frequencies larger than $\rho$, with calculable deviations at low frequencies and long distances. For example, we predict specific $\rho$-dependent modifications to the Lorentz force law and the Larmor formula, which lay the foundation for experimental tests of the photon's spin scale. We also reproduce known soft radiation emission amplitudes for nonzero $\rho$. The particles' effective matter currents are not fully localized to their worldlines when $\rho\neq 0$, which motivates investigation of manifestly local completions of our theory. Our results also motivate the development of continuous spin analogues of gravity and non-Abelian gauge theories. Given the correspondence with familiar gauge theory in the small $\rho$ limit, we conjecture that continuous spin particles may in fact mediate known long-range forces, with testable consequences.
\end{abstract}

\maketitle
\newpage
\tableofcontents
\addtocontents{toc}{\vspace{-1.0\baselineskip}}

\section{Motivation and Overview}
\label{Sec:intro}
Relativity and quantum mechanics imply that long-range forces are mediated by massless particles, which were first classified by Wigner~\cite{Wigner:1939cj}. Powerful restrictions on their interactions have been derived from the covariance of amplitudes for soft radiation emission~\cite{Weinberg:1964ev,Weinberg:1964ew,Weinberg:1965rz}, as well as the consistency of coupling to perturbative general relativity~\cite{Weinberg:1980kq} and gauge theories, e.g.~see Refs.~\cite{Berends:1984rq,Berends:1984wp,Bekaert:2010hp}. These results underlie the common belief that familiar gauge theories and general relativity encompass the full range of possibilities for long-distance physics in nature.   

However, these arguments ignored the most general type of massless particle in Wigner's classification, a ``continuous spin'' particle (CSP) with nonzero spin Casimir $W^2 = - \rho^2$, where we call $\rho$ the spin scale. The bosonic CSP has an infinite tower of integer helicity states. Just as with ordinary massive particles, the helicity states are mixed under Lorentz transformations by an amount controlled by the spin scale. As $\rho \to 0$ one smoothly recovers the familiar massless states with Lorentz invariant helicity. 

Recent results suggest that this limiting behavior applies well beyond kinematics. In Refs.~\cite{Schuster:2013pxj,Schuster:2013vpr}, the arguments of Ref.~\cite{Weinberg:1964ev} were first extended to bosonic CSPs, yielding well-behaved soft factors. Furthermore, Lorentz covariance and unitarity imply the soft factors are scalar-like, vector-like, or tensor-like. In each case, for $\rho \ll \omega$ the soft factors reduce to those for minimally coupled massless scalars, photons, and gravitons respectively, with the other helicities decoupling. This ``helicity correspondence'' raises the intriguing possibility that the massless particles in our universe may in fact be CSPs, with deviations from familiar theories in the deep infrared. 

More recently, Ref.~\cite{Schuster:2014hca} constructed the first gauge field theory for a bosonic CSP, which reduces as $\rho \to 0$ to a sum of free actions for each integer helicity, e.g.~a Maxwell action for $h = \pm 1$ and a Fierz--Pauli action for $h = \pm 2$. (For reviews and discussion, see Refs.~\cite{Rivelles:2014fsa,Rivelles:2016rwo,Bekaert:2017khg,Najafizadeh:2017tin}.) This continuous spin field action was quickly generalized to lower dimensions~\cite{Schuster:2014xja} and the fermionic~\cite{Bekaert:2015qkt} and supersymmetric \cite{Najafizadeh:2019mun,Najafizadeh:2021dsm} cases, illustrating the robustness of its approach of encoding spin as orientation in an auxiliary ``vector superspace.'' Other formalisms, including constrained metric-like, frame-like, and BRST~\cite{Metsaev:2018lth,Buchbinder:2018yoo} formulations, have also been used to construct actions for fermionic~\cite{Alkalaev:2018bqe,Buchbinder:2020nxn} and supersymmetric~\cite{Buchbinder:2019esz,Buchbinder:2019kuh,Buchbinder:2019sie,Buchbinder:2022msd} continuous spin fields, as well as those in higher-dimensional~\cite{Zinoviev:2017rnj,Alkalaev:2017hvj,Burdik:2019tzg} and (A)dS~\cite{Metsaev:2016lhs,Metsaev:2017ytk,Metsaev:2017myp,Khabarov:2017lth,Metsaev:2019opn,Metsaev:2021zdg} spaces. Relations between these formulations and the vector superspace formulation are discussed in Refs.~\cite{Rivelles:2014fsa,Alkalaev:2017hvj,Najafizadeh:2019mun}.

The key outstanding physical question is to understand how continuous spin fields couple to matter. Interactions of continuous spin fields with matter fields were studied in Refs.~\cite{Metsaev:2017cuz,Bekaert:2017xin,Metsaev:2018moa,Rivelles:2018tpt}, but they were gauge invariant only to leading order, like the Berends--Burgers--van Dam currents for higher spin fields~\cite{Berends:1985xx}. Furthermore, the currents that could reduce to minimal couplings as $\rho \to 0$ do not exist for matter fields of equal mass. All of the other currents are nonminimal: they correspond to higher-dimension operators such as charge radii, with vanishing soft factors. While they are a valuable first step, they are less interesting phenomenologically as they do not capture the leading, inverse-square forces we observe. 

In this paper, we aim to resolve both of these problems, putting the study of matter interactions on a firm footing. We depart from previous work by coupling the continuous spin field to a spinless matter particle with worldline $z^\mu(\tau)$, via the action
\begin{align} \label{eq:introaction}
S = - m \int d\tau + \frac12 &\int d^4x \, [d^4\eta] \, \left(\delta'(\eta^2+1)(\del_x\Psi)^2 + \frac{1}{2} \, \delta(\eta^2+1)(\Delta \Psi)^2\right) \nonumber \\
+ &\int d^4x \, [d^4 \eta] \, \delta'(\eta^2 + 1) \, \Psi(\eta, x) \int d\tau \, j(\eta, x - z(\tau), \dot{z}(\tau)).
\end{align}
The first line contains the free gauge field action of Ref.~\cite{Schuster:2014hca}, which is integrated over a bosonic superspace $(x^\mu, \eta^\mu)$, and depends on $\rho$ via the operator $\Delta = \del_x \cdot \del_\eta + \rho$. The final term couples the field to a current sourced by the matter particle. The interaction is exactly gauge invariant when $j$ satisfies the local continuity condition $\Delta j = 0$ up to total $\tau$-derivatives. We classify all solutions to this equation as scalar-like, vector-like, or non-minimal, where the first two families reduce to minimal scalar or vector couplings as $\rho \to 0$. 

For each choice of $j$, we can use the action~\eqref{eq:introaction} to compute various observables. ``Integrating out'' the field $\Psi$ by solving its equations of motion yields an effective action for the matter particles, which contains static and velocity-dependent potentials. Evaluating the action for a given $\Psi$ yields the force on a matter particle in a background field, while evaluating it for a given $z^\mu(\tau)$ yields the field produced by a moving particle. Remarkably, we find that observables involving only null modes of the field, such as radiation emission and forces in a radiation background, are universal. That is, for all scalar-like or vector-like currents the results depend only on $\rho$, and not on the details of the current. 

\begin{figure}[t]
\includegraphics[width=0.49\columnwidth]{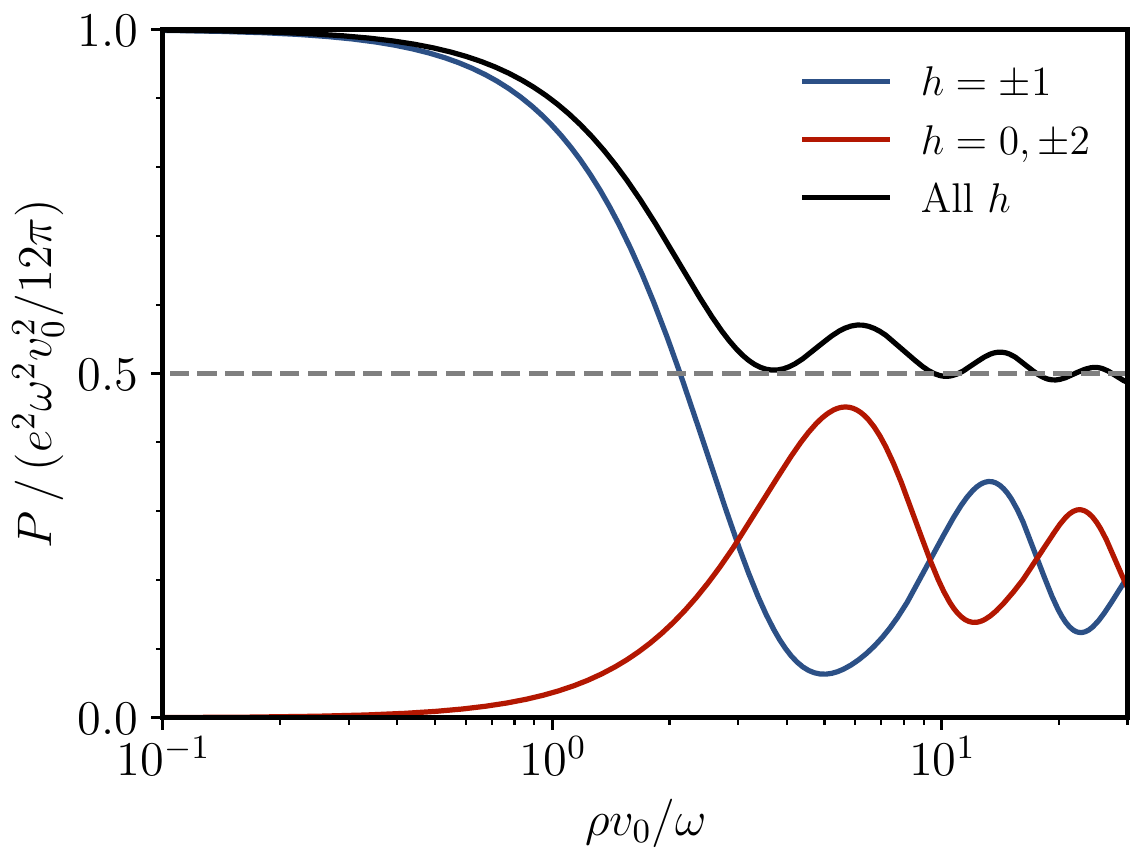}
\caption{Power radiated by an nonrelativistic oscillating charged particle, normalized to the ordinary Larmor result, for a photon with spin scale $\rho$. We show power emitted into the ordinary helicities $h = \pm 1$, the ``nearest neighbor'' helicities $h = 0, \pm 2$, and the total in all integer helicities.}
\label{fig:intro_rad_power}
\end{figure}

Universality allows us to predict specific deviations from familiar results in electromagnetism. For example, if the photon is a CSP, the force per charge on a particle with any vector-like current in an $h = \pm 1$ plane wave background with angular frequency $\omega$ is 
\begin{align}  \label{eq:introshowcase:force}
\frac{\v{F}}{e} = \left( \v{E} + \v{v} \times \v{B}\right) - \frac{\rho^2}{4 \omega^2} \left( (\v{E} \cdot \v{v}_\perp) \v{v}_{\perp} + \v{E} \, v_\perp^2/2\right) + \ldots
\end{align}
where $\v{v}_{\perp} = \v{v} - (\v{v} \cdot \hat{\v{k}}) \hat{\v{k}}$ is the particle's velocity transverse to the direction of propagation~$\hat{\v{k}}$. The force can be written exactly in terms of Bessel functions, and is well-behaved for arbitrary $\rho/\omega$. As another example, if such a particle performs nonrelativistic sinusoidal motion with characteristic velocity $v_0$ and frequency $\omega$, the total power it emits into radiation modes is
\begin{equation} \label{eq:introshowcase:rad}
P = \frac{e^2 \omega^2 v_0^2}{12 \pi} \left(1 - \frac{9}{80} \frac{\rho^2 v_0^2}{\omega^2} + \ldots \right)
\end{equation}
where the leading term matches the standard Larmor formula, and the first correction includes radiation emitted into helicities $h = 0, \pm 2$ and a reduction of radiation emitted into $h = \pm 1$. As shown in Fig.~\ref{fig:intro_rad_power}, we can compute the power for arbitrary $\rho v_0 / \omega$, and it always remains finite, even in the limit $\rho v_0 \gg \omega$ where radiation is produced with many helicities. 

The force between two particles is not universal, depending on the details of the current, but still obeys helicity correspondence. For example, we will show that some simple vector-like currents predict no deviations from Coulomb's law, while others predict corrections at long distances. The underlying reason that many equally ``minimal'' currents exist is that these currents are not fully localized to the particle's worldline. We expect this feature can be removed in a more fundamental description involving additional ``intermediate'' fields.  Exploring such descriptions is a key next step, and might in turn identify certain preferred currents. However, with our present theory we can already recover manifestly causal particle dynamics assuming appropriate boundary conditions for $j$.

Another key next step would be the formulation of continuous spin fields with non-Abelian symmetry, which are necessary to consistently describe tensor-like currents, or to embed a CSP photon within the Standard Model. Such developments would be exciting because of the possible relevance of continuous spin to outstanding problems in fundamental physics. In particular, the enhanced gauge symmetry of a continuous spin field could shed light on the cosmological constant and hierarchy problems, while the presence of weakly coupled ``partner'' polarizations and modifications to long-range force laws may have bearing on dark matter and cosmic acceleration. Currently these applications are only speculative, but the results of this paper already make it possible to probe the spin scale of the photon experimentally.

Remarkably, observable consequences of nonzero $\rho$ have never been considered before this work, besides the pioneering attempt of Ref.~\cite{Iverson:1971vy} to identify neutrinos with fermionic CSPs. The underlying reason is that continuous spin physics has been shrouded in confusion since its inception. It is often assumed that the infinite number of polarization states would render physical observables divergent, but the helicity correspondence implies that almost all of these states decouple. Thus, real experiments will measure finite values for, e.g.~Casimir forces and the heat capacity of the vacuum. All helicities are emitted in Hawking radiation, but the total power remains finite because of the falling greybody factors for high helicity modes. Finally, as we have already stated above, forces remain finite despite being mediated by an infinite number of helicity states, and radiation emission is finite even in the $\rho \to \infty$ limit where all helicities are produced. 

Confusion has also stemmed from field-theoretic ``no-go'' results. Several authors showed in the 1970s that it was impossible to construct Lorentz invariant, local theories of gauge invariant fields that created and annihilated CSPs~\cite{Yngvason:1970fy,Iverson:1971hq,Chakrabarti:1971rz,Abbott:1976bb,Hirata:1977ss}. However, under such restrictive assumptions it would also be impossible to construct electromagnetism, which requires a gauge potential, or general relativity, which in addition has no local gauge invariant operators. There are also plentiful no-go theorems concerning interactions of ``higher spin'' fields with $|h|>2$~\cite{Bekaert:2010hw}. These theorems indeed imply we cannot write currents which reduce to minimal couplings of higher spin fields in the $\rho \to 0$ limit, but they do not pose any obstruction to the existence of scalar-like or vector-like currents. 

The paper is organized as follows. In section~\ref{sec:primer}, we review the kinematics of CSPs and the action for the bosonic continuous spin field. In section~\ref{sec:worldlines}, we couple this field to a current sourced by a matter particle and identify families of scalar-like and vector-like currents. In section~\ref{sec:spacetime} we investigate the localization of simple currents in spacetime, and show how appropriate choices of current yield manifestly causal dynamics. In section~\ref{sec:forces}, we compute static and velocity-dependent forces between a pair of matter particles, as well as the force on a matter particle in a radiation background, which is universal. In section~\ref{sec:radiation}, we find the universal radiation emitted from an accelerating particle in arbitrary motion, and use the special case of an abruptly kicked particle to recover the soft factors found in Refs.~\cite{Schuster:2013pxj,Schuster:2013vpr}. Finally, in section~\ref{sec:conclusion} we discuss future directions and speculative applications. Detailed derivations are collected in appendices, which are referenced throughout the text.

\subsubsection*{Conventions}

We work in natural units, $\hbar = c = \epsilon_0 = 1$. We denote the Minkowski metric by $g_{\mu\nu}$, and use a $(+---)$ metric signature. Fourier transforms obey $f(k) = \int d^4x \, e^{i k \cdot x} f(x)$, so that a spacetime derivative $\del^\mu = \del_x^\mu$ becomes $- i k^\mu$. Vector superspace coordinates are naturally defined with raised indices, so that $\eta^\mu = (\eta^0, \etav)$, and similarly we write $k^\mu = (\omega, \v{k})$ and $x^\mu = (t, \v{r})$. Vector superspace derivatives are always written as $\del^\mu_\eta$, with the $\eta$ explicit. 

When manipulating tensors, symmetrizations and antisymmetrizations of $n$ indices are defined without factors of $1/n!$, complete contractions of symmetric tensors are denoted by a dot, and a prime denotes a contraction with the metric, i.e.~a trace. For example, for a totally symmetric rank $3$ tensor $S^{\mu\nu\rho}$, we have $S^{(\mu\nu\rho)} = 6 S^{\mu\nu\rho}$, $S \cdot S = S^{\mu\nu\rho} S_{\mu\nu\rho}$, and $S'^\mu = g_{\nu\rho} S^{\mu\nu\rho}$. Two primes denotes a double trace, i.e.~a contraction with two copies of the metric. 

For a null four-vector $k^\mu$, we will often introduce a basis of null complex frame vectors $(k^\mu, q^\mu, \epsilon_+^\mu, \epsilon_-^\mu)$, where $\epsilon_+^* = \epsilon_-$, and the only nonzero inner products are $\epsilon_+ \cdot \epsilon_- = -2$ and $q \cdot k = 1$. (This normalization differs from Refs.~\cite{Schuster:2013pxj,Schuster:2013vpr} but matches Ref.~\cite{Schuster:2014hca}.) This implies
\begin{equation} \label{eq:frame_metric}
g_{\mu\nu} = (k_\mu q_\nu + q_\mu k_\nu) - \frac12 (\epsilon_{+,\mu} \epsilon_{-,\nu} + \epsilon_{-,\mu} \epsilon_{+,\nu}).
\end{equation}
We define the Levi--Civita symbol to obey $\epsilon^{0123} = 1$, and fix the handedness of the basis by demanding $\epsilon^{\mu\nu\rho\sigma} = k^{[\mu} q^\nu \epsilon_+^\rho \epsilon_-^{\sigma]} / 2 i$. As a concrete example, if $k^\mu = (\omega, 0, 0, \omega)$, we can choose
\begin{equation} \label{eq:frame_choice}
q^\mu = (1/2 \omega, 0, 0, -1/2 \omega), \qquad \epsilon_\pm^\mu = (0, 1, \pm i, 0).
\end{equation}

\addtocontents{toc}{\vspace{-0.2\baselineskip}}
\section{The Free Theory}
\label{sec:primer}
In this section, we set up the theory of a free bosonic continuous spin field. We begin in subsection~\ref{sec:primer_particles} by reviewing the continuous spin states that such a field must create and annihilate. In subsection~\ref{sec:primer_fields} we write down the action and immediately specialize to $\rho = 0$, where familiar gauge theories are recovered as special cases. Finally, in subsection~\ref{sec:primer_csp} we consider nonzero $\rho$ and discuss the resulting qualitative changes.

\subsection{Continuous Spin States in Context}
\label{sec:primer_particles}

In a relativistic quantum theory, states transform under translations $P^\mu$ and rotations and boosts $J^{\mu\nu}$, which generate the connected Poincare group. The state space of an elementary particle is an irreducible unitary representation of this group. For any such particle, we can always diagonalize translations to yield states obeying $P^\mu |k, \sigma \rangle = k^\mu |k, \sigma \rangle$, where $k^\mu$ is the particle's four-momentum and $\sigma$ denotes its internal ``polarization'' state. The nontrivial physical content of each representation is given by the action of ``little group'' transformations~\cite{Wigner:1939cj}, which keep the four-momentum constant but may change the internal state. In general, the little group is generated by the elements of the Pauli--Lubanski vector $W^\mu = \frac{1}{2} \epsilon^{\mu\nu\rho\sigma} J_{\nu\rho} P_\sigma$, which has three independent components because $k \cdot W \, |k, \sigma \rangle = 0$. Representations are classified by the values of the Casimir invariants $P^2 = m^2$ and $W^2 = - \rho^2$, where $m$ is the mass and we call $\rho$ the spin scale.

For massive particles, with $P^2 > 0$, we can boost the particle's momentum to $k^\mu = (m, 0, 0, 0)$. In this case, the nonzero components of the Pauli--Lubanski vector are the rotation generators $\v{W} = m \v{J}$. The little group is therefore the rotation group $SO(3)$, whose projective representations are characterized by an integer or half-integer spin $S \geq 0$, corresponding to spin scale $\rho = m \sqrt{S(S+1)}$. These are the usual massive spin-$s$ particles, often encountered in the quantum mechanics of atoms and nuclei. 

This familiar setting allows us to give the spin scale an intuitive physical meaning. 
Label the states of a moving particle as $|k, h \rangle$ where the helicity $h = -s, \ldots, s$ is the projection of the spin in the particle's direction of motion, $\mathbf{J} \cdot \hat{\mathbf{k}}$. Then an infinitesimal Lorentz boost by $\v{v}_\perp$ in a transverse direction mixes these states into each other, $U(\v{v}_\perp) \, |k, h \rangle = \sum_{h'} c_{hh'} |k', h' \rangle$ where the mixing coefficients for $h \neq h'$ scale as positive powers of $\rho \, v_\perp/k^0$. In other words, the spin scale characterizes the mixing of helicity states under Lorentz transformations. 

We can carry these lessons to the case of massless particles, $P^2 = 0$, where the little group generators for a general momentum $k^\mu$ are
\begin{equation}
R = q \cdot W, \quad T_\pm = \epsilon_\pm \cdot W
\end{equation}
and $q$ and $\epsilon_\pm$ are the frame vectors defined in the conventions. These operators obey the commutation relations $[R, T_\pm] = \pm T_\pm$, with $T_+^\dagger = T_-$, which is the algebra of $\mathrm{ISO}(2) = E(2)$, the isometry group of a plane. (See Ref.~\cite{Schuster:2013pxj} for further discussion.) Concretely, if $k^\mu = (\omega, 0, 0, \omega)$ and we choose the other frame vectors as in~\eqref{eq:frame_choice}, then $R = \mathbf{J} \cdot \hat{\mathbf{k}}$ is a rotation about the axis of the three-momentum and $T_\pm$ are combinations of transverse boosts and rotations. By diagonalizing the helicity operator $R$, we can work with states obeying $R \,|k, h \rangle = h \, |k, h \rangle$, just as in the massive case. 

Since the helicity operator is not inherently Lorentz invariant, we expect $T_\pm$ to mix helicity states by an amount controlled by the spin scale. Indeed, the commutation relations yield
\begin{align} \label{eq:Tpm_action}
T_\pm \, |k,h \rangle &= \rho \, |k,h \pm 1 \rangle.
\end{align}
This equation implies that the ladder of helicity states does not end! That is, a single massless particle generically has \textit{infinitely} many polarizations, which is tied to the non-compactness of the little group $E(2)$. To avoid this situation, textbook treatments specialize to $\rho = 0$, sometimes implicitly, recovering the familiar irreducible representations which have only a single internal state $|k, h \rangle$ with Lorentz invariant helicity. 

By contrast, the ``continuous spin'' representations are characterized by a nonzero spin scale $\rho$, which can take on a continuum of real values. Their historical name often leads to the misconception that we are allowing the \textit{helicity} to take on continuous values, but $h$ must always be integer or half-integer to yield a projective representation of the Poincare group. A bosonic continuous spin representation contains states of all integer $h$, a fermionic continuous spin representation contains all half-integer $h$, and supersymmetry unifies the two into one multiplet~\cite{Najafizadeh:2019mun}. One can Fourier transform the helicity basis to reach the ``angle'' basis $\ket{k, \theta} = \sum_h e^{i h \theta} \, \ket{k, h}$, where the little group transformations take a simpler form~\cite{Schuster:2013pxj}.  Although the angle basis provides a geometric picture for the little group action, we will work primarily in the helicity basis, as it offers the most natural description of physical reactions.

For completeness, we note that the continuous spin representations can also be constructed by starting with a massive particle representation, and taking the simultaneous limits $m \to 0$ and $S \to \infty$ with $mS$ fixed~\cite{Khan:2004nj}. However, that description will not be useful in this work. 

\subsubsection*{From States to Fields}

In the nonrelativistic limit, it is straightforward to describe the interactions of atoms and nuclei with arbitrarily high spin. However, while these particles evidently have no trouble interacting in our Lorentz invariant universe, we have a more difficult time describing such processes using the standard tool of relativistic quantum field theory. 

The basic problem is that the building blocks of quantum field theory are local fields transforming in representations of the Lorentz group, but these fields must create and annihilate particles transforming in little group representations. Outside of the simplest cases, there will be a mismatch in the number of degrees of freedom. For example, a massive spin $1$ particle has $3$ polarizations, but is usually described by a vector field $A^\mu(x)$ with $4$ components. To remove the extra degree of freedom, one must use a Lagrangian whose equations of motion yield a single constraint, such as $\del \cdot A=0$. More generally, a bosonic spin $s$ particle with $2s+1$ degrees of freedom can be described by a symmetric rank $s$ tensor field with $(s+3)(s+2)(s+1)/6$ degrees of freedom, provided one chooses a special Lagrangian which yields an intricate set of constraints. This approach was pioneered by Singh and Hagen for arbitrary spin~\cite{Singh:1974qz,Singh:1974rc}, but to this day it remains challenging to use. 

The situation is even more subtle for massless particles. The CPT theorem implies that the helicities of massless particles always come in opposite-sign pairs, where $h = \pm 1$ for the photon and $h = \pm 2$ for the graviton. To describe the two photon polarizations with a four-component vector field one must use a Lagrangian with a built-in redundancy, or ``gauge symmetry,'' so that the two unwanted degrees of freedom decouple or are non-dynamical. Describing the graviton with a symmetric tensor field $h_{\mu\nu}(x)$ requires removing eight degrees of freedom, and hence a more complex gauge symmetry. More generally, gauge field theories for free particles of any helicity were developed by Fang and Fronsdal~\cite{Fronsdal:1978rb,Fang:1978wz}, and involve high-rank symmetric tensors. As with the massive case, manipulating these tensors at high ranks is cumbersome. In addition, the gauge symmetries place increasingly severe constraints on the fields' interactions at higher ranks, as we will discuss in subsection~\ref{ssec:worldlines}.

\subsection{Embedding Fields in Vector Superspace}
\label{sec:primer_fields}

\subsubsection*{Vector Superspace Actions}

We now turn to the problem of constructing a free field which creates and annihilates a single bosonic continuous spin particle, which has all integer helicities. Given the preceding discussion, one can expect such a field theory must involve symmetric tensor fields of arbitrarily high rank, and an accompanying set of gauge symmetries to account for extra degrees of freedom. However, manipulating all of these fields would be technically painful. 

Relatively recently, Ref.~\cite{Schuster:2014hca} introduced the first action for a free bosonic continuous spin field, which is the basis for the treatment of interactions in this work. (Predecessors to this result include a covariant equation of motion~\cite{Bekaert:2005in} and an alternative action which propagated a continuum of CSPs~\cite{Schuster:2013pta}.) The key is to avoid the complications of high-rank tensors by introducing fields that depend on an auxiliary four-vector coordinate $\eta^{\mu}$ in addition to the spacetime coordinate $x^\mu$. Lorentz transformations act on this ``vector superspace'' by $x \to \Lambda x$ and $\eta \to \Lambda \eta$, so that the Poincare generators are 
\begin{equation} \label{eq:lorentz_generators}
P^\mu = i \, \del_x^\mu, \quad J^{\mu\nu} = i \, (x^{[\mu} \del_x^{\nu]} + \eta^{[\mu} \del_\eta^{\nu]}).
\end{equation}
A field $\Psi(\eta, x)$ analytic in $\eta$ can be decomposed into component symmetric tensor fields of arbitrarily high rank. For example, a direct Taylor expansion would give $\Psi(\eta,x)=\psi^{(0)}(x)+\eta^{\mu}\psi^{(1)}_{\mu}(x)+\eta^{\mu}\eta^{\nu}\psi^{(2)}_{\mu\nu}(x) + \ldots$, though we will shortly see that a slightly different expansion is more useful. 

The action for the field is~\cite{Schuster:2014hca}
\begin{equation} \label{eq:CSP_action}
S[\Psi]= \frac12 \int d^4x \, [d^4\eta] \, \left(\delta' (\eta^2+1)(\del_x\Psi)^2 + \frac{1}{2} \, \delta(\eta^2+1)(\Delta \Psi)^2\right)
\end{equation}
where $\Delta = \del_\eta \cdot \del_x + \rho$. To explain the notation, the delta function and its derivative imply the action only depends on $\Psi$ and its first $\eta$-derivative on the unit hyperboloid $\eta^2 + 1 = 0$, and thus remains unchanged under the gauge symmetry $\delta \Psi = (\eta^2 + 1)^2 \chi(\eta, x)$. The field $\Psi$ is defined for all $\eta$, but field values far off the unit hyperboloid are pure gauge, with no physical meaning. The regulated integration measure $[d^4 \eta]$ permits standard manipulations such as integration by parts. The $\eta$ integrals in this subsection can be evaluated using
\begin{align}
\int [d^4\eta] \, \delta(\eta^2 + 1) \, F(\eta) &= \left( 1 - \frac18 \del_\eta^2 + \frac{1}{192} \del_\eta^4 - \ldots \right) F(\eta) \bigg|_{\eta = 0} \, ,  \label{eq:genfn_leading_first} \\
\int [d^4\eta] \, \delta'(\eta^2 + 1) \, F(\eta) &= \left( 1 - \frac14 \del_\eta^2 + \hspace{1.025mm}\frac{1}{64}\hspace{1.025mm} \del_\eta^4 - \ldots \right) F(\eta) \bigg|_{\eta = 0} \, .
\label{eq:genfn_leading}
\end{align}
These generating functions are derived in appendix~\ref{app:genFn}, and all other $\eta$-space integration identities needed in this work are derived in appendix~\ref{app:usefulIdentities}. The deeper geometric motivation for the action is discussed in appendix~\ref{app:anacont} and Ref.~\cite{Schuster:2014hca}. 

In the rest of this subsection, we will specialize to $\rho = 0$ and show how~\eqref{eq:CSP_action} contains the action, equation of motion, and gauge symmetry for familiar massless particles of arbitrary integer helicity. For instance, the derivatives on the right-hand sides of~\eqref{eq:genfn_leading_first} and~\eqref{eq:genfn_leading} will produce tensor contractions between the component fields. The purpose of this exercise is to gradually build facility with $\eta$-space computations, and to demonstrate the power of the formalism, which can replace an infinite set of tensor manipulations with a single integral. 

\subsubsection*{Decomposing the Action}

We can write the action in terms of uncoupled component tensor fields by defining 
\begin{align} \label{eq:field_tensors}
\Psi(\eta, x) &= \sum_{n \geq 0} P_{(n)}(\eta) \cdot \phi^{(n)}(x) \\
&= \phi(x) + \sqrt{2} \, \eta^\mu A_\mu(x) + (2 \eta^\mu \eta^\nu - g^{\mu\nu} (\eta^2 + 1)) \, h_{\mu\nu}(x) + \ldots.
\end{align}
where the $\phi^{(n)}$ are symmetric rank $n$ tensors, and the $P_{(n)}$ are the polynomials
\begin{equation}
P_{(n)}^{\mu_1\dots \mu_n} = 2^{n/2} \left( \eta^{\mu_1} \cdots \eta^{\mu_n} - \frac{1}{4 (n-2)!} \, g^{(\mu_1 \mu_2} \eta^{\mu_3} \cdots \eta^{\mu_n)} (\eta^2 + 1) \right). \label{eq:Pn_def}
\end{equation}
It is straightforward to see how the action reduces in simple cases. First, if only $\phi^{(0)} = \phi$ is nonzero, then only the first term of \eqref{eq:CSP_action} is nonzero, giving
\begin{equation}
S[\Psi] = \frac12 \int d^4x \, [d^4\eta] \, \delta'(\eta^2 + 1) (\del_x \phi)^2 = \frac12 \int d^4x \, (\del_x \phi)^2
\end{equation}
which is the action for a massless scalar field. If only $\phi^{(1)}_\mu = A_\mu$ is nonzero, the terms are
\begin{align}
\frac12 \int [d^4\eta] \, \delta' (\eta^2+1) \, (\del_x\Psi)^2 &= -\frac{1}{2} \, \del_{\rho}A_{\mu} \, \del^{\rho}A^{\mu} \\
\frac12 \int [d^4\eta] \, \frac{1}{2}\delta(\eta^2+1)(\del_{\eta}\cdot\del_x\Psi)^2 &= \frac{1}{2} \,(\del_{\mu}A^{\mu})^2 
\end{align}
which sum to the canonically normalized Maxwell action,
\begin{equation}
S[\Psi] = - \frac14 \int d^4x \, F_{\mu\nu} F^{\mu\nu}.
\end{equation}
If both $\phi^{(0)}$ and $\phi^{(1)}$ are nonzero, there is no cross-coupling between them because their product is linear in $\eta$, and only even powers of $\eta$ contribute to the integrals. 

Similarly, if only $\phi^{(2)}_{\mu\nu} = h_{\mu\nu}$ is nonzero, one recovers the Fierz--Pauli action, which describes the metric perturbation in linearized gravity. As explained in Ref.~\cite{Schuster:2014hca}, one can recover a Fronsdal action for each of the $\phi^{(n)}$, and furthermore there are no cross-couplings between fields of different ranks. This can be efficiently derived using orthogonality theorems for the polynomials $P_{(n)}$ shown in appendix~\ref{app:tensors}, though it need not be read to follow the main text.

\subsubsection*{Decomposing the Equation of Motion}

By formally varying \eqref{eq:CSP_action} with respect to $\Psi$, we obtain the equation of motion
\begin{equation} \label{eq:free_eom}
\delta' (\eta^2+1) \, \del_x^2 \Psi -\frac{1}{2} \, \Delta \, (\delta(\eta^2+1) \, \Delta \Psi)=0.
\end{equation}
In general, we can integrate this against $P_{(n)}$ to extract the equation of motion for $\phi^{(n)}$.

For example, if we trivially multiply by $P_{(0)} = 1$ and integrate over $\eta$, the second term in~\eqref{eq:free_eom} does not contribute because it is a total $\eta$ derivative. The first term yields
\begin{align}
0 &= \int [d^4\eta] \, \delta'(\eta^2 + 1) \, \del_x^2 \Psi \\
&= \int [d^4\eta] \, \delta'(\eta^2 + 1) \, \left(\del_x^2 \phi + \sqrt{2} \, \eta^\mu \del_x^2 A_\mu + (2 \eta^\mu \eta^\nu - g^{\mu\nu} (\eta^2 + 1)) \, \del_x^2 h_{\mu\nu} + \ldots \right) \\
&= \del_x^2 \phi
\end{align}
as expected. Above, the contribution of $A_\mu$ vanished since it is linear in $\eta$, and the contribution of $h_{\mu\nu}$ cancels after evaluating the integral. More generally, the decoupling of all other $\phi^{(n)}$ immediately follows from the orthogonality theorems in appendix~\ref{app:tensors}.

To give one more example, integrating the equation of motion against $P_{(1)}^{\mu}=\sqrt{2} \, \eta^{\mu}$ gives
\begin{align}
0 &= \sqrt{2} \int [d^4\eta] \, \delta'(\eta^2 + 1) \, \eta^\mu \, \del_x^2 \Psi - \frac{1}{2} \eta^\mu \, \Delta \, (\delta(\eta^2+1) \, \Delta \Psi) \\
&= \sqrt{2} \int [d^4\eta] \, \delta'(\eta^2 + 1) \, \eta^\mu \, \del_x^2 \Psi + \frac{1}{2} \del_x^\mu \, (\delta(\eta^2+1) \, \Delta \Psi) \\
&= -\del_x^2 A^\mu + \del_x^\mu \del \cdot A
\end{align}
again as expected, where $\phi$ and $h_{\mu\nu}$ decouple by parity considerations, and again the $\phi^{(n)}$ in general decouple by orthogonality. Of course, with more work we recover the linearized Einstein equations at rank $2$, and the Fronsdal equations for general rank $n$. 

\subsubsection*{Gauge Symmetry}

The general gauge symmetry of the action~\eqref{eq:CSP_action} is 
\begin{equation} \label{eq:eps_gauge}
\delta_\epsilon \Psi = D\epsilon(\eta,x) = (\eta\cdot\del_x - \tfrac{1}{2} (\eta^2+1) \Delta) \epsilon(\eta,x).
\end{equation}
where $\epsilon(\eta,x)$ is an arbitrary function analytic in $\eta$. The operator $D$ satisfies two identities which will be used frequently below,
\begin{align}
\Delta (D \epsilon) &= \del_x^2 \epsilon - \frac12 (\eta^2 + 1) \Delta^2 \epsilon, \label{eq:Delta_D_identity} \\
\delta'(\eta^2+1) D \epsilon &= \frac12 \Delta(\delta(\eta^2+1) \, \epsilon). \label{eq:deltapr_D_identity}
\end{align}
These identities, together with integration by parts, can be used to show the  infinitesimal gauge variations of the two terms in the action cancel, as required. 

Once again, the single expression~\eqref{eq:eps_gauge} simultaneously packages the gauge symmetries of all of the component fields, as can be seen by expanding
\begin{equation}
\epsilon(\eta, x) = \sum_{n \geq 0} P_{(n)}(\eta) \cdot \epsilon^{(n)}(x).
\end{equation}
For example, the contribution of $\epsilon^{(0)}$ is
\begin{equation}
\delta_{\epsilon^{(0)}} \Psi = \eta^\mu \del_\mu \epsilon^{(0)}
\end{equation}
which is, up to an overall constant, simply the familiar gauge transformation $\delta A_\mu = \del_\mu \epsilon$ of electromagnetism. Similarly, the contribution of $\epsilon^{(1)}$ is 
\begin{equation} \label{eq:hmunu_gauge}
\delta_{\epsilon^{(1)}} \Psi = \frac{1}{\sqrt{2}} \left(2 \eta^\mu \eta^\nu - g^{\mu\nu} (\eta^2 + 1) \right) (\del_{\mu} \epsilon^{(1)}_\nu)
\end{equation}
which is essentially the gauge transformation $\delta h_{\mu\nu} = \del_{(\mu} \xi_{\nu)}$ of linearized gravity. The pattern continues, with $\epsilon^{(n-1)}$ generating the gauge symmetry for $\phi^{(n)}$. 

\subsubsection*{Gauge Fixing and Plane Wave Solutions}

In electromagnetism, the Lorenz gauge $\del_\mu A^\mu = 0$ simplifies the free equation of motion to $\del^2 A^\mu = 0$. Similarly, in linearized gravity the Lorenz/harmonic gauge $\del_\mu \bar{h}^{\mu\nu} = 0$ simplifies the equation of motion to $\del^2 h^{\mu\nu} = 0$, where $\bar{h}^{\mu\nu} = h_{\mu\nu} - \frac12 h' g_{\mu\nu}$ is the trace-reversed metric perturbation. Both of these gauges are special cases of the ``harmonic'' gauge
\begin{equation} \label{eq:harmonic_def}
\delta(\eta^2 + 1) \, \Delta \Psi = 0
\end{equation}
which simplifies the equation of motion~\eqref{eq:free_eom} to $\delta'(\eta^2 + 1) \, \del_x^2 \Psi = 0$. Concretely, this implies that in harmonic gauge $\del_x^2 \Psi$ must be proportional to $(\eta^2 + 1)^2$. 

Such an ambiguity does not appear in electromagnetism and linearized gravity, but does in higher spin theories, where it arises from the freedom in double traces. We can remove it by using some of the residual gauge symmetry to go to ``strong harmonic'' gauge, where 
\begin{equation} \label{eq:strong_harmonic_def}
\del_x^2 \Psi = 0.
\end{equation}
While these gauge conditions will be crucial throughout the paper, we defer further discussion to the next subsection, where we show that these gauges can be achieved for arbitrary $\rho$. 

The equations above are solved by the null plane waves $\Psi = e^{- i k \cdot x} \, \psi_{h,k}(\eta)$, where $k^\mu$ is a null four-momentum, $h$ is an integer, and 
\begin{equation} \label{eq:rho0_helicity_wavefunctions}
\psi_{h,k}(\eta) \bigg|_{\rho = 0} = \begin{cases} (i \eta \cdot \epsilon_+)^h & h \geq 0 \\ (-i \eta \cdot \epsilon_-)^{-h} & h \leq 0 \end{cases}
\end{equation}
where $\epsilon^{\mu}_{\pm}$ are null frame vectors associated with $k^\mu$, defined in the conventions. As shown in Ref.~\cite{Schuster:2014hca}, the little group generators act as $R \, \psi_{h,k} = h \, \psi_{h,k}$, which implies $h$ is indeed the helicity, and $T_\pm \, \psi_{h,k} = 0$ up to pure gauge terms, which implies the helicity is Lorentz invariant. More concretely, if we further specify to $\epsilon_\pm^0 = 0$, the modes given above are exactly those of radiation gauge in electromagnetism for $h = \pm 1$, and of transverse traceless gauge in linearized gravity for $h = \pm 2$. As we will show in the next subsection, \eqref{eq:rho0_helicity_wavefunctions} gives a complete basis of physical solutions. This implies that the action~\eqref{eq:CSP_action} for $\rho = 0$ indeed contains, for each null momentum, exactly one mode of each integer helicity. 

\subsection{Continuous Spin Fields}
\label{sec:primer_csp}

Generalizing the previous results to arbitrary spin scale is straightforward, as $\rho$ only appears through the operator $\Delta = \del_\eta \cdot \del_x + \rho$. Remarkably, all of our $\eta$-space results, including the equation of motion~\eqref{eq:free_eom}, gauge symmetry~\eqref{eq:eps_gauge}, identities~\eqref{eq:Delta_D_identity} and~\eqref{eq:deltapr_D_identity}, and (strong) harmonic gauge condition~\eqref{eq:harmonic_def} and~\eqref{eq:strong_harmonic_def} are unchanged. The null plane wave solutions are now
\begin{equation} \label{eq:helicity_wavefunctions}
\psi_{h,k}(\eta) = e^{- i \rho \eta \cdot q} \times \begin{cases} (i \eta \cdot \epsilon_+)^h & h \geq 0 \\ (-i \eta \cdot \epsilon_-)^{-h} & h \leq 0 \end{cases}
\end{equation}
where $q^\mu$ is another null frame vector. The integer $h$ still represents the helicity, but now we have $T_\pm \, \psi_{h,k} = \rho \, \psi_{(h \pm 1), k}$ up to pure gauge terms, so that the parameter $\rho$ indeed corresponds to the spin scale. This was shown explicitly in Ref.~\cite{Schuster:2014hca}, though that work used a different phase convention for $T_\pm$, and the phase factor $e^{- i \rho \eta \cdot q}$ was sometimes reversed or dropped.

Again, the solutions~\eqref{eq:helicity_wavefunctions} are a complete basis of physical solutions, in the sense that the general mode expansion for a free bosonic continuous spin field is 
\begin{equation} \label{eq:modeexpansion}
\Psi(\eta,x) = \int \frac{d^3\v{k}}{(2\pi)^3 \, 2|\v{k}|}\sum_h\left( a_h(\v{k})\psi_{h,k}(\eta) e^{- i k \cdot x} + \text{c.c.}\right) \bigg|_{k^0 = |\v{k}|} + D \epsilon
\end{equation}
which is everywhere analytic in $\eta$. Here, $D \epsilon$ is a pure gauge term, and $a_h(\v{k})$ is the amplitude of the mode of helicity $h$ and momentum $\v{k}$. These coefficients contain all of the gauge invariant information in the field. By the orthonormality relation~\eqref{eq:ortho_final} for helicity modes, they can be extracted by projection,
\begin{equation} \label{eq:mode_projection}
a_h(\v{k}) = 2 |\v{k}| \int [d^4 \eta] \, \delta'(\eta^2 + 1) \, \psi^*_{h,k}(\eta) \Psi(\eta, k) \bigg|_{k^0 = |\v{k}|}.
\end{equation}
To see that the pure gauge term does not contribute to the right-hand side, apply~\eqref{eq:deltapr_D_identity} and note that the solutions~\eqref{eq:helicity_wavefunctions} satisfy the harmonic gauge condition. 

These are the key results we will need going forward; in the remainder of this subsection we prove the assertions made above, and discuss further features of the action.

\subsubsection*{Tensor Mixing}

An important qualitative difference for nonzero $\rho$ is the ubiquitous mixing of tensor components. For example, the action now contains mixing terms and apparent mass terms,
\begin{multline} \label{eq:mixing_terms}
S[\Psi] = S[\Psi] \bigg|_{\rho = 0} + \frac{\rho}{\sqrt{2}} \left(\phi \, (\del^\mu A_\mu) + A^\mu (\del_\mu h' - \del^\nu h_{\mu\nu}) + \ldots \right) \\ + \frac{\rho^2}{4} \left(\phi^2 - \frac12 A_\mu A^\mu - \phi \, h' + \frac16 (h')^2 - \frac13 h_{\mu\nu} h^{\mu\nu} + \ldots \right)
\end{multline}
which makes it hard to see that the physical solutions still represent massless particles. The mixing terms vanish in harmonic gauge, but then the harmonic gauge condition itself mixes tensor ranks, as do the solutions~\eqref{eq:helicity_wavefunctions}, which each contain tensors of arbitrarily high rank. Thus, while the tensor expansion is a useful conceptual tool at $\rho = 0$, it is less intuitive at any nonzero $\rho$, and may be misleading if not used with care.

\subsubsection*{Achieving (Strong) Harmonic Gauge}

To think about gauge fixing, it is useful to separate the null and non-null modes of the field. For the null modes, one has $\del_x^2 \Psi = 0$ by definition, and the equation of motion implies the harmonic gauge condition is automatically satisfied for all fields which are bounded at infinity. The (strong) harmonic gauge conditions are only nontrivial statements about non-null modes, on which $\del_x^2$ is invertible.

To see that harmonic gauge can be reached for non-null modes, note that~\eqref{eq:Delta_D_identity} implies
\begin{equation}
\delta_\epsilon (\delta(\eta^2 + 1) \, \Delta \Psi) = \delta(\eta^2 + 1) \, \del_x^2 \epsilon.
\end{equation}
Since $\epsilon$ is arbitrary and $\del_x^2$ is invertible on these modes, this freedom can be used to set $\delta(\eta^2 + 1) \, \Delta \Psi$ to zero. The residual gauge freedom is in the form of $\epsilon$ with $\del_x^2 \epsilon = 0$, which only affects null modes, and $\epsilon$ proportional to $\eta^2 + 1$. We can use the latter freedom to reach strong harmonic gauge, since for $\epsilon = (\eta^2 + 1) \beta(\eta, x)$ we have
\begin{equation}
\delta_\epsilon (\del_x^2 \Psi) = - \frac12 (\eta^2 + 1)^2 \del_x^2 \, \Delta \beta.
\end{equation}
Since $\Delta \beta$ is arbitrary and $\del_x^2$ is invertible, we can indeed set $\del_x^2 \Psi$ to zero for non-null modes, which completely eliminates them. 

Incidentally, gauge transformations of the form $\epsilon = (\eta^2 + 1) \beta$ correspond precisely to the gauge symmetries $\delta_\chi \Psi = (\eta^2 + 1)^2 \chi$ first noted below~\eqref{eq:CSP_action} for $\chi = - \Delta \beta / 2$. The only subtlety of this correspondence is that $\chi$ gauge transformations which fall appropriately at infinity may correspond to $\beta$ which do \textit{not} fall at infinity. For this reason, Ref.~\cite{Schuster:2014hca} and the higher spin literature often introduce $\epsilon$ and $\chi$ gauge transformations as two separate families, despite their redundancy.  

\subsubsection*{Completeness of Helicity Modes}

Working in momentum space, the field $\Psi(\eta, k)$ in strong harmonic gauge is only nonzero for null $k^\mu$. Fixing one such $k^\mu$ for consideration, there is a residual gauge transformation $\epsilon(\eta, k)$, and the operator~\eqref{eq:eps_gauge} that generates gauge transformations is 
\begin{equation} \label{eq:d_def}
D = - i k \cdot \eta - \frac12 (\eta^2 + 1) (- i k \cdot \del_\eta + \rho)).
\end{equation}
It is convenient to pull out phases, defining $\epsilon = -2 i e^{-i \rho \eta \cdot q} \, \tilde{\epsilon}$ and $\Psi = e^{- i \rho \eta \cdot q} \, \tilde{\Psi}$, which simplifies the gauge transformation and the harmonic gauge condition to
\begin{align}
\delta_\epsilon \tilde{\Psi} &= ( (\eta^2 + 1) (k \cdot \del_\eta) - 2 \eta \cdot k) \, \tilde{\epsilon}, \label{eq:simplified_gt} \\
k \cdot \del_\eta \, \tilde{\Psi} &= (\eta^2 + 1) \, \alpha
\end{align}
for some arbitrary $\alpha(\eta, k)$. Now,~\eqref{eq:simplified_gt} implies
\begin{equation}
\delta_\epsilon(k \cdot \del_\eta \, \tilde{\Psi}) = (\eta^2 + 1) (k \cdot \del_\eta)^2 \tilde{\epsilon}.
\end{equation}
To understand this, it is useful to think of $\tilde{\epsilon}$ and $\alpha$ as Taylor series in $\eta \cdot k$, $\eta \cdot q$, and $\eta \cdot \epsilon_\pm$. In these variables we have $k \cdot \del_\eta = \del_{\eta \cdot q}$, so there is enough freedom in $\tilde{\epsilon}$ to cancel off any $\alpha$. The residual gauge freedom is in $\tilde{\epsilon} = f + (\eta \cdot q) g$ for $f$ and $g$ independent of $\eta \cdot q$, for which
\begin{equation} \label{eq:strong_residual_gauge}
\delta_\epsilon \tilde{\Psi} = - 2 (\eta \cdot k) f + (1 - \eta \cdot \epsilon_+ \, \eta \cdot \epsilon_-) g.
\end{equation}
In other words, we can use all of the remaining gauge freedom to remove any terms in $\tilde{\Psi}$ with powers of $\eta \cdot k$, or powers of both $\eta \cdot \epsilon_+$ and $\eta \cdot \epsilon_-$. After removing such terms we are left with $\tilde{\Psi}$ equal to a sum of monomials in $\eta \cdot \epsilon_+$ or $\eta \cdot \epsilon_-$. This shows that the helicity modes~\eqref{eq:helicity_wavefunctions} are a complete basis of physical solutions.

\subsubsection*{A New Spacetime Symmetry}

As a final remark, the vector superspace used to construct continuous spin fields is a bosonic analogue of the fermionic superspace used in supersymmetric theories. As first noted in Ref.~\cite{Rivelles:2016rwo}, the action~\eqref{eq:CSP_action} is invariant under the $\eta$-dependent spacetime translation $\delta x^\mu = \omega^{\mu\nu} \eta_\nu$ for antisymmetric $\omega^{\mu\nu}$, corresponding to the tensorial conserved charge 
\begin{equation}
N^{\mu\nu} = i(\eta^{\mu}\del_x^{\nu}-\eta^{\nu}\del_x^{\mu}).
\end{equation}
This symmetry mixes modes of different helicity, even for $\rho=0$, and it continues to hold when the continuous spin field couples to particles. It would then appear to be a novel extension of spacetime symmetry, which could evade the Coleman--Mandula theorem by virtue of the infinite number of polarization states. While we have not yet found any use for this symmetry, it may be a guide to constructing more complete continuous spin theories.

\addtocontents{toc}{\vspace{-0.2\baselineskip}}
\section{Coupling Matter Particles to Fields}
\label{sec:worldlines}
We can couple spinless matter particles to fields by introducing currents built from the particle's worldline. This is the most useful way to describe low-energy classical experiments, and for familiar theories it can be readily quantized, yielding a worldline formalism equivalent to perturbative quantum field theory. In subsection~\ref{ssec:worldlines} we identify minimal couplings to ordinary fields, and show that almost all non-minimal couplings produce only contact interactions, which do not affect a particle's coupling to radiation. We generalize to continuous spin fields in subsection~\ref{ssec:CSP_currents}, where we enumerate an enormous family of potential currents. Again, we find that the vast majority of these currents do not couple to radiation, allowing us to define families of scalar-like and vector-like currents that couple universally to radiation, and reduce to the familiar minimal scalar and vector couplings as $\rho \to 0$. 

\subsection{Review: Coupling to Ordinary Fields}
\label{ssec:worldlines}

\subsubsection*{Minimal Couplings to Scalar and Vector Fields}

A free matter particle of mass $m$ and worldline $z^\mu(\tau)$ has action
\begin{equation}
S_0 = - m \int \sqrt{\frac{dz^\mu}{d\tau} \frac{dz_\mu}{d\tau}} \, d\tau.
\end{equation}
In this section we take $\tau$ to be the proper time, but all results below can be straightforwardly rewritten for general worldline parametrization by replacing $d\tau$ with $\sqrt{\dot{z}^2} \, d\tau$ and $\dot{z}^\mu = dz^\mu/d\tau$ with $\dot{z}^\mu/\sqrt{\dot{z}^2}$. Now, we can couple the particle to a massless real scalar field $\phi(x)$ by
\begin{equation} \label{eq:scalar_current_coupling}
S_{\mathrm{int}} = \int d^4x \, \phi(x) J(x)
\end{equation}
which yields an equation of motion $\del_x^2 \phi = J$ for the field. The current $J(x)$ can be built from $z^\mu(\tau)$ and its derivatives; if we were considering particles with spin, it could also depend on the spin tensor $S^{\mu\nu}(\tau)$. The minimal coupling corresponds to the simplest possible current,
\begin{equation} \label{eq:scalar_current}
J(x) = g \int d \tau \, \delta^{(4)}(x - z(\tau))
\end{equation}
which is localized to the worldline, and corresponds to a Yukawa interaction in a field-theoretic description of the matter. 

Next, consider the coupling of a particle of charge $e$ to a massless vector field $A_\mu(x)$,
\begin{equation} \label{eq:vector_current_coupling}
S_{\mathrm{int}} = - \!\int d^4x \, A_\mu(x) J^\mu(x).
\end{equation}
Here the simplest possible current $J^\mu$, obeying worldline locality, translation invariance, and current conservation $\del_\mu J^\mu = 0$ (which is required for $S_{\mathrm{int}}$ to be gauge invariant) is
\begin{equation} \label{eq:vector_current}
J^\mu(x) = e \int d \tau \, \dot{z}^\mu(\tau) \, \delta^{(4)}(x - z(\tau)),
\end{equation}
which corresponds to a minimal coupling of scalar matter to a vector field. The field's equation of motion is $\del_x^2 A^{\mu} = J^\mu$ in Lorenz gauge. The current is conserved because its divergence is a total $\tau$ derivative,
\begin{equation} \label{eq:ibp_current_conservation}
\del_\mu J^\mu = e \int d\tau \, (\dot{z}^\mu \del_\mu) \, \delta^{(4)}(x-z(\tau)) = - e \int d\tau \, \frac{d}{d\tau}\delta^{(4)}(x-z(\tau)) = 0
\end{equation}
where the integral vanishes because worldlines have no boundaries. This physically corresponds to assuming particles and antiparticles are always produced or destroyed in pairs.

\subsubsection*{Non-minimal Couplings and Contact Terms}

We are most interested in currents for continuous spin fields which reduce to the currents defined above as $\rho \to 0$. However, our formalism also includes currents which reduce to non-minimal couplings, so to build intuition we first review them for scalar and vector fields. We continue to assume the currents are Lorentz covariant and translationally invariant, so that the only dependence on position is through the combination $x^\mu - z^\mu(\tau)$. We further assume the current can be written as a $\tau$ integral of a function of $z^\mu(\tau)$ and $\dot{z}^\mu(\tau)$, with no explicit dependence on higher $\tau$ derivatives or on $\tau$-nonlocal products. 

For a scalar field, the most general current obeying these assumptions can be constructed from~\eqref{eq:scalar_current} by acting on the delta function with powers of $\del_x^2$ or $\dot{z} \cdot \del_x$. This general current is most compactly expressed in momentum space, 
\begin{equation} \label{eq:nonmin-scalar}
J(k) = \int d\tau \, e^{ik\cdot z(\tau)} f(k^2, k\cdot\dot z(\tau)),
\end{equation}
where the minimal coupling~\eqref{eq:scalar_current} corresponds to $f(k^2, k \cdot \dot{z}) = g$. 

All terms with at least one power of $k^2$ do not produce any fields away from the particles themselves, and thus correspond to contact interactions. To build intuition, we will derive this basic result for $f = g \, k^2$ in several ways. In this case, the Lagrangian for the field is 
\begin{equation}
\mathcal{L} = \frac12 (\del_\mu \phi)^2 - J_0(x) \, \del^2 \phi(x)
\end{equation}
where $J_0$ is the minimal current~\eqref{eq:scalar_current}. In terms of the shifted field $\tilde{\phi} = \phi + J_0$, it becomes
\begin{equation}
\mathcal{L} = \frac12 (\del_\mu \tilde{\phi})^2 - \frac{1}{2} (\del_\mu J_0)^2.
\end{equation}
Since $J_0$ is localized to a particle's worldline, the interaction term is only nonzero when two particles coincide, so it mediates a contact interaction. The shifted field is free, which implies the field $\phi$ is not affected except for exactly on the particle itself. In particular, this coupling does not affect the radiation produced by an accelerating particle. Conversely, it does not affect the particle's motion in the presence of a free background field, including any radiation field. In general, all of these conclusions hold whenever the function $f$ contains a factor of the differential operator $\del^2$ appearing in the field's free equation of motion. 

We can also give some of these terms a more direct physical interpretation. For a particle at rest, $\dot{z}^\mu = (1, \v{0})$, the $\tau$ integration sets $k^0 = 0$, giving $J(\v{k}) = f(-|\v{k}|^2, 0)$. Thus, for static particles terms with powers of $k \cdot \dot{z}$ have no effect, while those proportional to powers of $k^2$ describe the particle's static, spherically symmetric spatial profile. 

The $k^2$ term itself corresponds in field theory to an operator analogous to a charge radius. However, we note that, despite the name, the presence of such an operator does not actually imply a particle has a nonzero radius, any more than a dipole moment implies a particle has a specific length. The corresponding current is still localized, in the sense that it only depends on fields in a neighborhood of the worldline. Particles of finite size can be described by~\eqref{eq:nonmin-scalar}, but only by summing over an infinite series of terms. We highlight this point because the currents introduced in subsection~\ref{ssec:CSP_currents} will, by contrast, be intrinsically delocalized.

Terms with only powers of $k \cdot \dot{z}$ are the only ones that can yield non-contact interactions. To understand them more physically, note that 
\begin{equation} \label{eq:kz_terms}
\int d\tau \, (k \cdot \dot{z})^n \, e^{i k \cdot z} = - i \int d\tau \, (k \cdot \dot{z})^{n-1} \, \frac{d}{d\tau} (e^{i k \cdot z}) = i \int d\tau \, e^{ik \cdot z} \frac{d}{d\tau} (k \cdot \dot{z})^{n-1}.
\end{equation}
This implies that a term with a single power of $k \cdot \dot{z}$ has no effect, and a term with two powers of $k \cdot \dot{z}$ implicitly depends on the acceleration $\ddot{z}$. In general, a power of $k \cdot \dot{z}$ can be exchanged for a $\tau$ derivative by the above manipulation, so term with more powers of $k \cdot \dot{z}$ depends on higher time derivatives of $\dot{z}$. These terms describe the particle's dynamic response to nontrivial motion. In other words, they characterize the particle's non-rigidity, e.g.~distinguishing a rigid charged sphere from a spherical balloon full of charged jelly. To further build intuition, note that the $n = 2$ case corresponds to a coupling in position space 
\begin{equation}
S_{\text{int}} \supset \int d^4x \, d\tau \, (\del_\mu \del_\nu \phi(x)) \, \dot z^\mu \dot z^\nu \, \delta^{(4)}(x-z(\tau)), 
\end{equation}
which has the enhanced shift symmetry $\delta \phi = c + b_\mu x^\mu$  of a galileon field~\cite{Nicolis:2008in}. 

The terms highlighted in~\eqref{eq:kz_terms} are the only ones that can affect the radiation produced by an accelerating particle, though their effect is suppressed at low accelerations. This conclusion might seem surprising because in general, it is certainly possible for a particle's static structure to impact the radiation it produces, e.g.~through dipole and higher moments. However, those terms are not permitted for a spinless particle by rotational symmetry.

In the vector case, gauge invariance imposes the constraint $k_\mu J^\mu(k) = 0$. After suitable subtraction of total $\tau$ derivatives, the most general current can be written as
\begin{equation} \label{eq:nonmin-vector}
J^\mu(k) = \int d\tau \, e^{ik\cdot z} \left( e \, \dot{z}^\mu + f(k^2, k\cdot\dot z) (\dot z^\mu k^2 - k^\mu k\cdot \dot z) \right),
\end{equation}
where we have isolated the minimal coupling~\eqref{eq:vector_current} in the first term. Again, terms in $f$ proportional to powers of $k\cdot\dot z$ describe the source's non-rigidity, and terms with $k^2$ describe the source's static spatial profile, though now the constant term represents the charge radius. 

In contrast to the scalar case, \textit{all} of the non-minimal vector couplings correspond to contact interactions, because they are proportional to $(k^2 \eta^{\mu\nu} - k^\mu k^\nu) \dot{z}_\nu$, and the field's free equation of motion is $(k^2 \eta^{\mu\nu} - k^\mu k^\nu) A_\nu = 0$. None of the non-minimal couplings produce any fields away from the particle itself, and they therefore cannot affect the particle's response to free fields, or the radiation it produces. This is essentially a statement of Gauss's law: the vector radiation from a spherically symmetric particle is determined solely by its charge. 

\subsubsection*{Tensor and Higher Spin Fields}

Matter particles can also couple to higher rank fields, but such fields introduce additional complications. First, at rank $2$, the coupling of a particle to the canonically normalized metric perturbation $h_{\mu\nu}(x)$ in linearized gravity is
\begin{equation} \label{eq:tensor_coupling}
S_{\text{int}} = \frac{\kappa}{2} \int d^4x \, h_{\mu\nu}(x) T^{\mu\nu}(x)
\end{equation}
where $\kappa = \sqrt{32 \pi G}$. The stress-energy tensor of the particle is
\begin{equation} \label{eq:Tmunu_leading}
T^{\mu\nu}(x) = m \int d\tau \, \dot{z}^\mu(\tau) \dot{z}^\nu(\tau) \, \delta^{(4)}(x - z(\tau))
\end{equation}
and its divergence is 
\begin{equation} \label{eq:T_divergence}
\del_\mu T^{\mu\nu} = m \int d \tau \, \ddot{z}^\nu(\tau) \, \delta^{(4)}(x - z(\tau)).
\end{equation}
This presents two well-known, closely related problems: consistency of the equations of motion, and gauge invariance of the action off-shell. 

First, the equation of motion for $h_{\mu\nu}$ in linearized gravity implies $\del_\mu T^{\mu\nu} = 0$, which is only true for non-accelerating particles. Thus, if we regard $z^\mu(\tau)$ as a given background then we cannot consider nontrivial trajectories, while if we regard $z^\mu(\tau)$ as dynamical then the equations of motion are inconsistent~\cite{misner1973gravitation}. The resolution is simply the equivalence principle: in~\eqref{eq:tensor_coupling} we must use the total stress-energy tensor, including that of the fields which cause the particle to accelerate. Full consistency requires an infinite series in $\kappa$ of terms involving $h_{\mu\nu}$, which leads to the nonlinear structure of general relativity~\cite{ortin2004gravity}. However, it is still possible to compute results to leading order in the coupling $\kappa$ in the linearized theory. 

Second, if we regard $z^\mu(\tau)$ as dynamical, gauge invariance of $S_{\text{int}}$ under $\delta h_{\mu\nu} = \del_{(\mu} \epsilon_{\nu)}$ seems to require $\del_\mu T^{\mu\nu} = 0$ \textit{identically}, which is not true for general off-shell $z^\mu(\tau)$. The resolution is to recall that the gauge transformation arises from an infinitesimal diffeomorphism symmetry, which also acts on the particle's position by $\delta z^\mu = \kappa \epsilon^\mu$. Including this contribution implies that $S_0$ also varies under a gauge transformation, rendering the full action gauge invariant to leading order in $\kappa$~\cite{ortin2004gravity}. Achieving full gauge invariance requires generalizing $\delta h_{\mu\nu}$ to an infinite series in $\kappa$, which recovers the nonlinear gauge structure of general relativity~\cite{maggiore2007gravitational}.

These subtleties make it more difficult to generalize~\eqref{eq:Tmunu_leading} to continuous spin fields, even at leading order in $\kappa$. The consistency problem implies we must include additional terms in the stress-energy tensor, which complicates calculations. The gauge invariance problem is more serious, as it is not clear how the more general continuous spin gauge symmetry~\eqref{eq:eps_gauge} is supposed to act on $z^\mu$. Furthermore, it is unclear to what degree one can trust results calculated with an action that is only partially gauge invariant.

For these reasons, we defer the study of ``tensor-like'' continuous spin currents that reduce to~\eqref{eq:Tmunu_leading} as $\rho \to 0$ to future work. In this paper, we exclude tensor-like currents by demanding that both $z^\mu$ and the action be completely gauge invariant. 

As an aside, Ref.~\cite{deWit:1979sib} considered a minimal coupling to higher spin fields,
\begin{equation} \label{eq:highRankCurrent_leading}
J^{\mu_1 \cdots \mu_n}(x) = m g_n \int d \tau \, \dot{z}^{\mu_1}(\tau) \cdots \dot{z}^{\mu_n}(\tau) \, \delta^{(4)}(x - z(\tau)).
\end{equation}
As for rank $2$, the current is conserved when $\ddot{z}^\mu = 0$, and gauge invariance can be restored to leading order in $g_n$ by gauge transforming the worldline, by $\delta z_\mu \propto g_s \, \epsilon^{(n-1)}_{\mu \nu_1 \ldots \nu_{n-2}} \dot{z}^{\nu_1} \cdots \dot{z}^{\nu_{n-2}}$. (For a perspective on these transformations, see Refs.~\cite{Segal:2000ke,Segal:2001di}.) However, we will not consider this further as for $n > 2$ there is no known theory consistent or gauge-invariant to all orders. 

%%%%%%%%%%%%%%
%%%%%%%%%%%%%%
\subsection{Coupling to Continuous Spin Fields}
\label{ssec:CSP_currents}

The free continuous spin field can be coupled to a current by 
\begin{equation} \label{eq:interaction}
S_{\mathrm{int}} = \int d^4x \, [d^4\eta] \, \delta' (\eta^2+1) \, J(\eta, x) \Psi(\eta, x).
\end{equation}
This is a canonical choice, as taking an integral over $\delta(\eta^2 + 1)$ would just yield a special case of~\eqref{eq:interaction} by the identity~\eqref{eq:distro_id}, while integrating over $\delta''(\eta^2 + 1)$ or any other distribution would couple generic currents to field values off the first neighborhood of the hyperboloid, which are pure gauge. To gain intuition for~\eqref{eq:interaction}, note that when $\rho = 0$, defining
\begin{align} \label{eq:current_tensors}
J(\eta, x) &= \sum_{n \geq 0} \bar{P}_{(n)}(\eta) \cdot J^{(n)}(x) \\
&= J(x) + \sqrt{2} \, \eta^\mu J_\mu(x) + (2 \eta^\mu \eta^\nu + g^{\mu\nu}) \, \frac{\kappa \, T_{\mu\nu}(x)}{2} + \ldots.
\end{align}
simultaneously yields the minimal scalar coupling~\eqref{eq:scalar_current_coupling}, vector coupling~\eqref{eq:vector_current_coupling}, and tensor coupling~\eqref{eq:tensor_coupling}. More generally, defining the ``dual'' polynomials by
\begin{equation} \label{eq:Pn_dual_def}
\bar{P}_{(n)}^{\mu_1\dots \mu_n} = 2^{n/2} \left( \eta^{\mu_1} \cdots \eta^{\mu_n} + \frac{1}{4 (n-1)!} \, g^{(\mu_1 \mu_2} \eta^{\mu_3} \cdots \eta^{\mu_n)} \right)
\end{equation}
leads to a sum of canonically coupled currents at each rank, 
\begin{equation} \label{eq:canonical_sint}
S_{\mathrm{int}} = \int d^4x \, \sum_{n \geq 0} (-1)^n \phi^{(n)} \cdot J^{(n)}
\end{equation}
as shown in appendix~\ref{app:tensors}. The current $J$ appears on the right-hand side of the equation of motion~\eqref{eq:free_eom}, so the (strong) harmonic gauge conditions~\eqref{eq:harmonic_def} and~\eqref{eq:strong_harmonic_def} become
\begin{equation} \label{eq:strong_harmonic_def_2}
\delta(\eta^2 + 1) \, \Delta \Psi = 0, \qquad \del_x^2 \Psi = J.
\end{equation}

\subsubsection*{The Continuity Condition}

Demanding gauge invariance of $S_{\mathrm{int}}$ yields a key constraint on the current,
\begin{equation} \label{eq:continuity_deltaForm}  
\delta(\eta^2 + 1) \, \Delta J = 0
\end{equation}
which we term the continuity condition. In this work, we assume it holds off-shell, though note that on-shell, it ensures the two conditions in~\eqref{eq:strong_harmonic_def_2} are consistent. As shown in appendix~\ref{app:tensors}, we can reexpress the continuity condition in terms of tensor components as
\begin{equation} \label{eq:continuity_tensors}
\left\langle \del \cdot J_{(n)} + \frac{\rho}{n \sqrt{2}} (J_{(n-1)} + \frac12 J'_{(n+1)}) \right\rangle = 0
\end{equation}
for all $n \geq 1$, where the brackets denote a trace subtraction.

To build intuition, we note that the trace subtraction is trivial at lower ranks, so that for $\rho = 0$ we recover the ordinary conservation conditions $\del \cdot J_{(1)} = 0$ and $\del \cdot J_{(2)} = 0$, and for higher ranks we recover a ``weak'' conservation condition $\langle \del \cdot J_{(n)} \rangle = 0$ familiar in higher spin theory. However, for nonzero $\rho$ the continuity condition mixes tensor ranks, which implies that any current must contain an infinite tower of nonzero $J_{(n)}$. 

Concretely, consider the $n = 1$ and $n = 2$ cases of the continuity condition, 
\begin{equation}
\del \cdot J_{(1)} = - \frac{\rho}{\sqrt{2}} \left(J_{(0)} + \frac12 J'_{(2)}\right), \qquad \del \cdot J_{(2)} = - \frac{\rho}{2 \sqrt{2}} \left(J_{(1)} + \frac12 J'_{(3)}\right).
\end{equation}
If we wanted to construct a ``scalar-like'' current whose $\rho \to 0$ limit had only $J_{(0)}$ of the form~\eqref{eq:scalar_current}, then the first equation implies $J_{(1)}$ must have a \textit{nonconserved}, order $\rho$ contribution, which then implies $J_{(2)}$ must have a nonconserved, order $\rho^2$ contribution, and so on to arbitrary $J_{(n)}$. Alternatively, to construct a ``vector-like'' current whose $\rho \to 0$ limit has only the conserved $J_{(1)}$ of the form~\eqref{eq:vector_current}, the second equation requires $J_{(2)}$ of order $\rho$, and so on. 

Constructing such a tower of tensors, with appropriate trace subtractions, requires complex combinatorics. Furthermore, the results are physically opaque due to the omnipresent mixing between ranks. For instance, it might seem impossible for $J_{(1)}$ to be nonconserved, given that it couples to $A_\mu$ in the $\rho \to 0$ limit, but this is permitted because of the complex mixing~\eqref{eq:mixing_terms} of $A_\mu$ with other tensor components. For these reasons, we find it much more straightforward to construct currents directly in $\eta$-space, without using a tensor expansion. 

\subsubsection*{Constructing Currents}

We will construct currents from the worldline using the same assumptions we used to find the general scalar current~\eqref{eq:nonmin-scalar} and vector current~\eqref{eq:nonmin-vector}. Specifically, we assume the current in position space has the form
\begin{equation}
J(\eta, x) = \int d\tau \, j(\eta, x - z(\tau), \dot{z}(\tau))
\end{equation}
which corresponds in momentum space to 
\begin{equation} \label{eq:CSP_current}
J(\eta, k) = \int d\tau \, e^{i k \cdot z} f(\eta, k, \dot{z})
\end{equation}
where $f$ is the Fourier transform of $j$. The minimal scalar current~\eqref{eq:scalar_current} corresponds to $f(\eta, k, \dot{z}) = g$, and the minimal vector current~\eqref{eq:vector_current} corresponds to $f(\eta, k, \dot{z}) = \sqrt{2} \, e \, \eta \cdot \dot{z}$.

We assume $z^\mu(\tau)$ is gauge invariant; as discussed above, this excludes currents which reduce to minimal tensor or higher spin couplings as $\rho \to 0$. Then gauge invariance of the full action requires the continuity condition~\eqref{eq:continuity_deltaForm} to hold for arbitrary $z^\mu(\tau)$, which implies 
\begin{equation} \label{eq:continuity_fourier}
(k \cdot \del_\eta + i \rho) f(\eta, k, \dot{z}) = (\eta^2 + 1) \, \alpha(\eta, k, \dot{z})
\end{equation}
where $\alpha$ is an arbitrary function analytic in $\eta$. Strictly speaking, we are also free to add terms to $f$ proportional to $k \cdot \dot{z}$, with no other $\dot{z}$ dependence, as these are total $\tau$ derivative which not contribute to the current; the above equation assumes such terms have been dropped. 

We assume $f$ is analytic in $\eta$, so that it can only contain positive powers of $\eta$, and for nonzero $\rho$ the continuity condition implies that $f$ must have arbitrarily high powers of $\eta$. These must be accompanied by arbitrarily high \textit{negative} powers of $k$, which implies the currents cannot be localized to the worldline. Furthermore, $f$ contains an enormous amount of freedom, even if one fixes the $\rho \to 0$ limit. We will shortly discuss general currents, which are parametrized in appendix~\ref{app:classification}, but let us first build intuition through simple examples. 

%%%%%%%%%%%%%%%%%%
\subsubsection*{Simple Scalar-Like and Vector-Like Currents}

We begin by considering ``scalar-like'' currents with $\alpha = 0$, which all reduce to~\eqref{eq:scalar_current} in the limit $\rho \to 0$. Note that for any vector $V^\mu(k, \dot{z})$, the continuity condition is satisfied by
\begin{equation} \label{eq:exponential_family}
f(\eta, k, \dot{z}) = g \, e^{- i \rho \eta \cdot V / k \cdot V}
\end{equation} 
This family contains two illustrative special cases: the ``temporal'' current 
\begin{equation} \label{eq:J_T_intro}
J_T(\eta, k) = g \int d \tau \, e^{i k \cdot z} \, e^{- i \rho \eta \cdot \dot{z} / k \cdot \dot{z}}
\end{equation}
which corresponds to $V^\mu = \dot{z}^\mu$, and the ``spatial'' current
\begin{equation} \label{eq:J_S_intro}
J_S(\eta, k) = g \int d\tau \, e^{i k \cdot z} \, \exp\left(- i \rho \, \frac{\eta \cdot k - \eta \cdot \dot{z} \, \dot{z} \cdot k}{k^2 - (k \cdot \dot{z})^2} \right)
\end{equation}
which corresponds to $V^\mu = k^\mu - (k \cdot \dot{z}) \dot{z}^\mu$, the linear combination orthogonal to $\dot{z}^\mu$. We can also define vectors which are inhomogeneous in $k^\mu$, such as 
\begin{equation} \label{eq:V_mu_def}
V^\mu_\pm = k^\mu \pm \beta \rho \dot{z}^\mu
\end{equation}
for a dimensionless parameter $\beta$. For real $\beta$, we can construct the ``inhomogeneous'' current
\begin{equation} \label{eq:J_I_def}
J_I(\eta, k) = \frac{g}{2} \int d\tau \, e^{i k \cdot z} \left( e^{- i \rho \eta \cdot V_+ / k \cdot V_+} + e^{- i \rho \eta \cdot V_- / k \cdot V_-} \right).
\end{equation}
Above, two terms are required to ensure the position-space current is real, $J(-k) = J(k)^*$. Note that a pure imaginary $\beta$ would also satisfy the continuity condition, though in that case each term alone would give a real position-space current. 

Though many others exist, the temporal current $J_T$ is the simplest for situations involving radiation, the spatial current $J_S$ has the simplest static limit, and the inhomogeneous current $J_I$ is a relatively simple extension of the two. Note that all three have essential singularities at $k \cdot V = 0$. Such isolated essential singularities are a generic feature of solutions to~\eqref{eq:continuity_fourier}. However, while finite-order poles in $k$ are pathological, we will see these singularities are benign, leading to well-behaved physical results and only a weak nonlocality.

These scalar-like currents can be straightforwardly generalized to ``vector-like'' currents, which reduce to~\eqref{eq:vector_current} as $\rho \to 0$. For example, the vector-like temporal current is
\begin{align} 
J^V_T(\eta, k) &= \sqrt{2} \, e \int d\tau \ e^{i k \cdot z} \, (\eta \cdot \dot{z}) \sum_{m=0}^\infty \frac{1}{(m+1)!} \left( \frac{- i \rho \, \eta \cdot \dot{z}}{k \cdot \dot{z}} \right)^m \label{eq:J_V} \\
&= \frac{\sqrt{2} \, e}{\rho} \int d\tau \ e^{i k \cdot z} \, (ik\cdot\dot{z}) \, (e^{- i \rho \eta \cdot \dot{z} / k \cdot \dot{z}} -1). \label{eq:J_V_exp}
\end{align}
We can drop the $-1$ term because it is a total $\tau$ derivative, leaving an integrand which directly satisfies~\eqref{eq:continuity_fourier}. This unfortunately makes the current appear to diverge in the $\rho \to 0$ limit, but it is convenient since the result is related to $J_T$ by the simple substitution $g \to (\sqrt{2} \, e/\rho) (i k \cdot \dot{z})$. Similarly, we can write a vector-like inhomogeneous current 
\begin{equation} \label{eq:J_I_V}
J_I^V(\eta, k) = \frac{e}{\sqrt{2} \, \beta \rho^2} \int d\tau \, e^{i k \cdot z}  \left( (ik\cdot V_+) \, e^{- i \rho \eta \cdot V_+ / k \cdot V_+} - (ik\cdot V_-) \, e^{- i \rho \eta \cdot V_- / k \cdot V_-} \right)
\end{equation}
where analogous total $\tau$ derivative terms have been dropped. Finally, writing the vector-like spatial current requires constructing a prefactor which is annihilated by $k \cdot \del_\eta$ but also reduces to $\eta \cdot \dot{z}$ in the $\rho \to 0$ limit, again up to total $\tau$ derivatives. The result is 
\begin{multline} \label{eq:J_S^V}
J_S^V(\eta, k) = \sqrt{2} \, e \int d\tau \, e^{i k \cdot z} \left(\frac{(\eta \cdot \dot{z}) k^2 - (\eta \cdot k)(k \cdot \dot{z})}{k^2 - (k \cdot \dot{z})^2} - \frac{i k \cdot \dot{z}}{\rho} \right) \\ \times \exp\left(- i \rho \, \frac{\eta \cdot k - \eta \cdot \dot{z} \, \dot{z} \cdot k}{k^2 - (k \cdot \dot{z})^2} \right).
\end{multline}
Despite appearances, these currents are all vector-like. 

%%%%%%%%%%%%%%%%%%
\subsubsection*{General Currents}

Now that we have seen some examples, let us parametrize the general solution for $f$. The result for $\alpha = 0$ is given by~\eqref{eq:f_0_full} in appendix~\ref{app:classification}, which is equivalent to 
\begin{equation}
f = e^{- i \rho \eta \cdot \dot{z} / k \cdot \dot{z}} \left( g_0(u, k \cdot \dot{z}, k^2) + \left(\eta^2 + 1 - \frac{2 \eta \cdot k \, \eta \cdot \dot{z}}{k \cdot \dot{z}} + \frac{(\eta \cdot \dot{z})^2 \, k^2}{(k \cdot \dot{z})^2} \right) g_1(u, k \cdot \dot{z}, k^2) \right) 
\end{equation}
where $u = \eta \cdot (k - \dot{z} \, k^2 / (k \cdot \dot{z}))$. For $\rho = 0$, this expression efficiently packages the minimal scalar and vector currents, and all non-minimal currents. The general scalar current~\eqref{eq:nonmin-scalar} corresponds to $g_0$ independent of $u$, while the general vector current~\eqref{eq:nonmin-vector} corresponds to $g_0$ with a prefactor of $(k \cdot \dot{z}) u$. The function $g_1$ is less familiar, but if we take $g_1 \propto (k \cdot \dot{z})^2$, which is the simplest case which avoids negative powers of $k$, the result includes a nonminimal tensor current. Similarly, taking the simplest $\alpha \propto (k \cdot \dot{z})^3$ and applying~\eqref{eq:f_a_full}, the result includes a nonminimal current for a rank $3$ higher spin field.

The situation is less clear for nonzero $\rho$, where negative powers of $k$ appear. For instance, the temporal current~\eqref{eq:J_T_intro} has $g_0 = g$, while the spatial current~\eqref{eq:J_S_intro} has $g_0 = g \, e^{i \rho u / ((k \cdot \dot{z})^2 - k^2)}$. Given how complicated $g_0$ can be for even the simplest examples, it is not obvious how to usefully define a ``scalar-like'' current, nor how to extract predictions given the enormous freedom in the general solution. 

However, a remarkable simplification occurs when we consider coupling to null modes of the field, with $k^2 = 0$. In this case, the difference of the spatial and temporal currents is 
\begin{equation}
J_S - J_T = g \int d\tau \, e^{i k \cdot z} e^{-i \rho \eta \cdot \dot{z} / k \cdot \dot{z}} \left( \exp\left(i \rho \frac{\eta \cdot k}{(k \cdot \dot{z})^2} \right) - 1 \right)
\end{equation}
where the factor in parentheses contains only terms proportional to $(\eta \cdot k)^n$ for $n > 0$. Such terms do not contribute to the action in the presence of radiation modes~\eqref{eq:helicity_wavefunctions}, because the $\eta$ integral in this case is proportional to 
\begin{equation}
\int [d^4 \eta] \, \delta'(\eta^2 + 1) \, e^{i \rho \eta \cdot (q - \dot{z} / k \cdot \dot{z})} (\eta \cdot \epsilon_\pm)^{|h|} \, (\eta \cdot k)^n = 0
\end{equation}
which vanishes because $k$ is orthogonal to itself, $\epsilon_\pm$, and $q - \dot{z} / k \cdot \dot{z}$. Thus, the spatial and temporal currents couple to radiation in exactly the same way. 

This phenomenon turns out to be very general. As shown in appendix~\ref{app:classification}, any $f$, including those corresponding to arbitrary $\alpha$, can be written in the form
\begin{equation} \label{eq:general_universality_decomposition}
f(\eta,k,\dot z) = e^{- i \rho \eta \cdot \dot{z} / k \cdot \dot{z}} \, \hat{g}(k\cdot\dot z) + (k^2 + D\,\Delta) X(\eta, k,\dot z)
\end{equation}
where $\hat{g}(k \cdot \dot{z}) = g_0(0, k \cdot \dot{z}, 0)$, $D$ is the operator~\eqref{eq:d_def} generating gauge transformations, and $X$ is regular as $k^2 \to 0$. The action~\eqref{eq:interaction} only depends on the current through the combination $\delta'(\eta^2+1) J$, and under this delta function the second term simplifies as
\begin{equation}
\delta'(\eta^2+1)(k^2 + D\,\Delta)X = \left( \delta'(\eta^2+1) \, k^2 + \frac12 \, \Delta \, \delta(\eta^2 + 1) \, \Delta\right) X
\end{equation}
where we used~\eqref{eq:deltapr_D_identity}. The object acting on $X$ is just, up to a sign, the differential operator in the free equation of motion~\eqref{eq:free_eom}. Thus, by the same logic as in subsection~\ref{ssec:worldlines}, these terms do not affect the radiation a particle emits, or its response to a free field background. These key physical observables are determined solely by the single-variable function $\hat{g}$. 

We can therefore give sharp definitions of classes of currents based on $\hat{g}$:

\begin{itemize}
\item Scalar-like currents have constant $\hat{g}(k \cdot \dot{z}) = g$.
\item Vector-like currents have linear $\hat{g}(k \cdot \dot{z}) = (\sqrt{2} \, e / \rho) (i k \cdot \dot{z})$. 
\item Non-minimal scalar-like currents have $\hat{g} \propto (k \cdot \dot{z})^n$ for $n \geq 2$. They generalize the couplings highlighted by~\eqref{eq:kz_terms}, which characterize the matter particle's non-rigidity.
\item Currents with negative powers of $k \cdot \dot{z}$ are permitted in our analysis but less phenomenologically interesting, because their support is nonlocal even when $\rho = 0$. 
\end{itemize}

This classification is the main result of this subsection, and for the rest of this work we will focus on scalar-like and vector-like currents. A continuous spin field coupled by a scalar-like current is of less phenomenological interest because it reduces as $\rho \to 0$ to a massless scalar field, which has not been observed. However, this case is mathematically simpler and will serve as useful preparation for the case of vector-like currents, which can describe infrared modifications of electromagnetism. 

We have defined scalar-like and vector-like currents so that they couple to radiation universally. Note, however, that the long-range force between particles is \textit{not} universal. For scalar and vector fields, we found that currents proportional to the equation of motion operator produce contact interactions, which only take effect when particles coincide in space. By contrast, a continuous spin current is generally not localized to a particle's worldline, so the currents of particles can overlap even when the particles themselves are well-separated. 

To conclude, we note that while the decompositions above use a ``temporal'' prefactor $e^{- i \rho \eta \cdot \dot{z} / k \cdot \dot{z}}$, this was an arbitrary choice made for later convenience, and does not single out the temporal current as more fundamental. In addition, it is natural to conjecture that in a more complete description, where all sources of stress-energy could to the continuous spin field with equal strength, one might have tensor-like currents with $\hat{g} \propto (k \cdot \dot{z})^2/\rho^2$, where the terms that diverge as $\rho\to 0$ reduce to total derivatives on-shell. We defer further exploration of this conjecture to future work. Finally, in appendix~\ref{sec:fields} we compare our currents to those previously found by working in terms of matter fields.

\addtocontents{toc}{\vspace{-0.2\baselineskip}}
\section{Currents and Interactions in Spacetime}
\label{sec:spacetime}
The nonlocality of the currents found in section~\ref{sec:worldlines} motivates investigating the locality and causality properties of our theory. In subsection~\ref{sec:simple_spacetime}, we compute the profiles of some simple currents in spacetime, and show that they can be confined to the forward or backward light cone if their essential singularities appear only at real frequencies. In subsection~\ref{sec:eom_causality}, we give specific requirements on currents and their boundary conditions for the matter particles to have manifestly causal dynamics. This section can be skipped without loss of continuity. 

\subsection{Simple Currents in Spacetime}
\label{sec:simple_spacetime}

\subsubsection*{Currents of Worldline Elements}

We first compute the inverse Fourier transforms $j(\eta, x, \dot{z})$ for our simple currents, which gives the spacetime profile of the current from a differential element of the worldline. For simplicity we can work in the frame of such a worldline element, where $\dot{z}^\mu = (1, \v{0})$. The results below are illustrated at left in Fig.~\ref{fig:J_loc}, and derived in appendix~\ref{app:es_ints}. 

First, for the scalar-like spatial current, we have
\begin{align} \label{eq:f_S_expr}
j_S(\etav, \v{r}) &= g \int \frac{d\omega \, d^3\v{k}}{(2\pi)^4} \, e^{- i k \cdot x} e^{- i \rho \etav \cdot \hat{\v{k}} / |\v{k}|} \\
&= g \, \delta(t) \, \left(\delta^{(3)}(\v{r}) - \frac{|\rho \etav| (1 - \cos \theta)}{8 \pi r^2} J_2\left(\sqrt{2 |\rho \etav| r (1 - \cos \theta)} \right) \right) \label{eq:f_S}
\end{align}
where $\theta$ is the angle between $\etav$ and $\v{r}$. Above, the $\omega$ integral was trivial, while the $\v{k}$ integral requires handling a benign essential singularity at $\v{k} = 0$. One curious feature of this result is that it is not everywhere analytic in $\eta$. Analytic $f$'s can correspond to nonanalytic $j$'s, but there is still no problem with defining $\eta$-space integration, as explained in appendix~\ref{app:anacont}.

The result~\eqref{eq:f_S} shows that the spatial current of a worldline element is localized in time, but has a part spread over a distance $r \sim 1/\rho$ in space. The delocalized part smoothly goes to zero at large $r$, and in the $\rho \to 0$ limit it becomes \textit{larger} in spatial extent, but also decouples. By contrast, for the scalar-like temporal current the current is localized in space, 
\begin{equation} \label{eq:f_T_initial}
j_T(\eta^0, \v{r}, t) = g \int \frac{d\omega \, d^3\v{k}}{(2\pi)^4} \, e^{- i k \cdot x} e^{- i \rho \eta^0 / \omega} = g \, \delta^{(3)}(\v{r}) \, \int \frac{d\omega}{2 \pi} \, e^{- i \omega t} e^{- i \rho \eta^0 / \omega}.
\end{equation}
The frequency integral now passes through an essential singularity at $\omega = 0$, and thus requires a contour prescription to define. Passing above or below it corresponds to imposing retarded or advanced boundary conditions in position space, respectively. For the former, we find
\begin{equation} \label{eq:f_T}
j_T(\eta^0, \v{r}, t) = g \, \delta^{(3)}(\v{r}) \left(\delta(t) - \theta(t) \, \sqrt{\frac{\rho \eta^0}{t}} J_1\left(2 \sqrt{\rho \eta^0 t}\right) \right). 
\end{equation}
Concretely, this implies the current of a worldline element at position $\v{r}$ with velocity $\v{v}$ is supported on the forward ray $\v{r} + \v{v} T$ for $T > 0$. It varies on the timescale $t \sim 1/\rho$, and is confined to within the forward light cone. 

\begin{figure}[t]
\includegraphics[width=0.3\columnwidth]{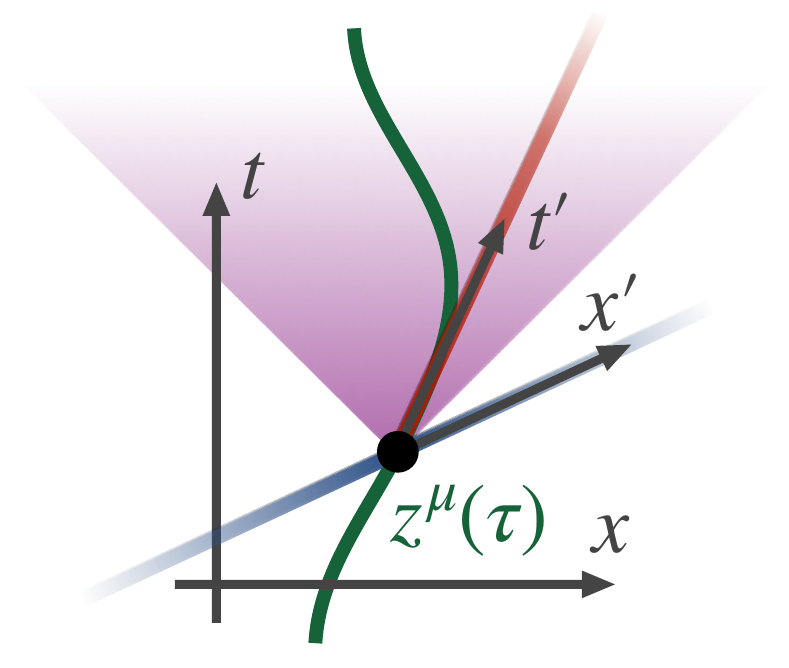} \qquad
\includegraphics[width=0.2\columnwidth]{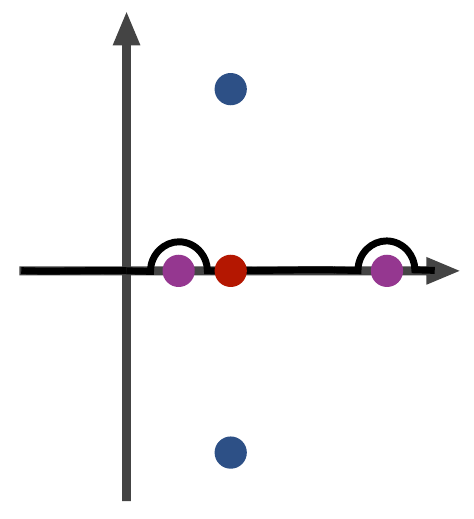}
\caption{{\bf Left:} Localization properties of some simple scalar-like currents. In the rest frame of a worldline element, with coordinates $(\v{x}', t')$, the spatial current (blue) is localized on the spacelike slice $t' = 0$, the retarded temporal current (red) is localized on the timelike ray $\v{x}' = 0$, $t' > 0$, and the retarded inhomogeneous current (purple) is contained on or within the forward light cone. {\bf Right:} Essential singularities in the complex $\omega$ plane for the spatial current, temporal current, and the integral in~\eqref{eq:f_I} for the inhomogeneous current, for a velocity with $\v{k} \cdot \v{v} > 0$. We also show an integration contour corresponding to retarded boundary conditions for the inhomogeneous current.}
\label{fig:J_loc}
\end{figure}

Finally, for the scalar-like inhomogeneous current we have 
\begin{equation}
j_I(\eta, \v{r}, t) = g \, \text{Re} \int \frac{d\omega \, d^3\v{k}}{(2\pi)^4} \, e^{- i k \cdot x} \, \exp \left(- i \rho \frac{\eta^0 (\omega + \rho \beta) - \etav \cdot \v{k}}{\omega^2 + \rho \beta \omega - |\v{k}|^2} \right). \label{eq:f_I}
\end{equation}
For real $\beta$, the integral has two essential singularities at real $\omega$, so a contour prescription is again required to define the current in spacetime. Integrating above both singularities corresponds to retarded boundary conditions, and gives a result confined to $t \geq 0$. While we have not computed the result, we expect it to have a spacetime spread $\sim 1/\rho$ on dimensional grounds, and Lorentz invariance implies it must be confined on or within the forward light cone. Advanced boundary conditions correspond to integrating below both singularities. 

The properties above are qualitatively unchanged for the vector-like currents. For example, for the vector-like spatial current, $j_S^V$ is identical to~\eqref{eq:f_S} but with $g$ replaced with $\sqrt{2} \, e \eta^0$. For the vector-like temporal current~\eqref{eq:J_V_exp}, we have
\begin{equation} \label{eq:f_T^V}
j_T^V(\eta^0, \v{r}, t) = \sqrt{2} \, e \eta^0 \, \delta^{(3)}(\v{r}) \left(\delta(t) - \theta(t) \, \frac{\rho}{t} J_2\left(2 \sqrt{\rho \eta^0 t}\right) \right)
\end{equation}
which has the same localization properties as~\eqref{eq:f_T}.

One cannot ascribe too much significance to the expressions above because $\eta$-space expressions are unphysical. We must always integrate over $\eta$ to produce physical observables, and as we will see in later sections, the results are often the same for currents which appear radically different in $\eta$-space. However, we can infer in general that boundary conditions are needed to define any current with essential singularities at real $\omega$, and retarded or advanced boundary conditions can be imposed for any current whose essential singularities are all at real $\omega$. These features suggest that such currents arise more fundamentally through integrating out locally coupled, auxiliary degrees of freedom. 

\subsubsection*{Currents and Fields of Static Particles}

We can also build intuition by evaluating the total current for a static particle at rest at the origin, and the accompanying continuous spin field. For ordinary fields, these currents are time-independent and localized at the origin, but the continuity condition forbids such solutions for nonzero $\rho$, leading to currents that are either delocalized or time-dependent. 

For the scalar-like spatial current, trivially integrating~\eqref{eq:f_S} gives
\begin{equation} \label{eq:J_S_pos}
J_S(\etav, \v{r}) = g \left( \delta^{(3)}(\v{r}) - \frac{|\rho \etav| (1 - \cos \theta)}{8 \pi r^2} J_2\left(\sqrt{2 |\rho \etav| r (1 - \cos \theta)} \right) \right)
\end{equation}
which is delocalized. The corresponding field, in strong harmonic gauge, is
\begin{equation} \label{eq:Psi_S_pos}
\Psi_S(\etav, \v{r}) = \frac{g}{4 \pi r} \, J_0\left(\sqrt{2 |\rho \etav| r (1 - \cos \theta)} \right).
\end{equation}
For the scalar-like temporal current, formally integrating~\eqref{eq:f_T} yields zero for positive $\eta^0$, but a divergent result for negative $\eta^0$. To make sense of the current, we need to introduce an appropriate infrared regulator. One simple prescription is to turn off the coupling $g$ at early times, replacing it with $g \, \theta(\tau + T)$ for a large positive $T$. This gives a current
\begin{equation} \label{eq:J_T_result}
J_T(\eta^0, t, \v{r}) = g \, \delta^{(3)}(\v{r}) \, J_0\left(2 \sqrt{\rho \eta^0 (t + T)} \right) \theta(t + T)
\end{equation}
and the corresponding field, in strong harmonic gauge, is an outgoing spherical wave
\begin{equation} \label{eq:Psi_T_result}
\Psi_T(\eta^0, t, \v{r}) = \frac{g}{4 \pi r} J_0\left( 2 \sqrt{\rho \eta^0 (t + T - r)} \right) \theta(t + T - r).
\end{equation}
We recover the familiar scalar solution in the limit $\rho \to 0$ and $T \to \infty$, provided that $\rho T \ll 1$. We could also regulate the solution by allowing the charge to oscillate with a low frequency, or by giving the particle nontrivial motion; expressions for the current for arbitrary particle motion are given in appendix~\ref{app:es_ints} but are unenlightening.

For the scalar-like inhomogeneous current, there are a variety of contour prescriptions for the frequency integration, but they all give the same current for a static particle. This is reflected by the fact that we can perform the $\tau$ integration first, which produces a delta function that eliminates the frequency integration, yielding
\begin{equation}
J_I(\eta, \v{r}) = g \int \frac{d^3\v{k}}{(2\pi)^3} \, e^{i \v{k} \cdot \v{r}} e^{- i \rho \etav \cdot \hat{\v{k}} / |\v{k}|} \cos(\eta^0 \beta \rho^2 / |\v{k}|^2)
\end{equation}
which reduces to the spatial current for $\beta \to 0$. Such a manipulation would not have been possible for the temporal current, because for a static particle its essential singularity is at $\omega = 0$ itself. Finally, for the vector-like currents the results are qualitatively similar, though there is also an overall prefactor of $\eta^0$. 

\subsection{Equations of Motion and Causality}
\label{sec:eom_causality}

We have just seen that the current $j$ of an element of a particle's worldline is not local, which seems to violate naive causality. Specifically, generalizing the action \eqref{eq:introaction} to multiple matter particles with worldlines $z_i^\mu(\tau_i)$ yields the equations of motion 
\begin{align}
\del_x^2 \Psi(\eta, x) &= \sum_i\int d\tau_i \, j(\eta, x - z_i(\tau_i), \dot{z}_i(\tau_i)), \label{eq:Psi_eom}\\
m_i \ddot{z}^{\mu}_i(\tau_i) &= - \int d^4x \, [d^4 \eta] \, \delta'(\eta^2 + 1) \, \Psi(\eta, x) \left( \partial_x^\mu + \frac{d}{d\tau_i} \frac{\del}{\del \dot z_i^\mu} \right) j(\eta, x - z_i(\tau_i), \dot{z}_i(\tau_i)).\label{eq:z_eom}
\end{align}
This yields two types of potentially acausal effects. First, if $j$ extends outside the forward lightcone of $z_i$, then~\eqref{eq:Psi_eom} implies the field is sourced acasually: there are reference frames where the field at $x$ depend on the motion of particles at times later than $x^0$. Likewise, if $j$ extends outside the backward lightcone of $z_i$, then~\eqref{eq:z_eom} implies the particle responds acausally: there are frames where its acceleration depends on field values at times later than $z_i^0$. These effects are both avoided if $j$ is local, but this is impossible for nonzero $\rho$. 

\subsubsection*{Manifestly Causal Equations of Motion}

There is, however, a straightforward modification of the equations of motion that renders our theory manifestly causal. We have seen that currents whose essential singularities are at real $\omega$, such as the temporal and inhomogeneous currents, admit retarded or advanced boundary conditions, where $j$ is confined within the forward or backward light cone respectively. The theory is manifestly causal if the current has retarded boundary conditions $j_R$ when sourcing the field, and advanced boundary conditions $j_A$ when responding to it, i.e.~if
\begin{align}
\del_x^2 \Psi(\eta, x) &= \sum_i\int d\tau \, j_R(\eta, x - z_i(\tau_i), \dot{z}_i(\tau_i)), \label{eq:Psi_eom_causal}\\
m_i \ddot{z}^{\mu}_i(\tau_i) &= - \!\int d^4x \, [d^4 \eta] \, \delta'(\eta^2 + 1) \, \Psi(\eta, x) \left( \partial_x^\mu + \frac{d}{d\tau_i} \frac{\del}{\del \dot z_i^\mu} \right) j_A(\eta, x - z_i(\tau_i), \dot{z}_i(\tau_i)). \label{eq:z_eom_causal}
\end{align}
Note, however, that these equations of motion still have the unfamiliar property that their exact solution appears to require knowledge of fields and particle trajectories in the infinite past, not just on a Cauchy surface in spacetime.

Such equations of motion do not directly emerge from our action~\eqref{eq:introaction}, which allows only a single choice of current. But it is reasonable to speculate that this action is merely an effective one, which arises from integrating out ``intermediate'' auxiliary or dynamical fields that interact locally with both the particle and the continuous spin field. In this case, solving the intermediate fields' equations of motion would require a choice of boundary conditions. Inserting such solutions into the full equations of motion would induce precisely the kind of asymmetry in~\eqref{eq:Psi_eom_causal} and~\eqref{eq:z_eom_causal}. Motivated by the form of our currents, we have constructed toy models with similar structure, but we have yet to realize a full Lagrangian model that renders our currents entirely local. 

\subsubsection*{Non-Manifest Causality} 

We should also take care in dismissing the original equations of motion~\eqref{eq:Psi_eom} and~\eqref{eq:z_eom} as irreparably acausal. In particular, the gauge field $\Psi(\eta,x)$ has a huge amount of gauge freedom, and indeed no local gauge-invariant observables can be built from the field. Thus, the apparently acausal sourcing of $\Psi(\eta,x)$ could be a gauge artifact. To avoid this ambiguity, we should define causality in terms of the interactions of matter particles: the theory is causal if, when a particle is instantaneously kicked at some point, the kick cannot affect the trajectories of other particles until they enter the future light cone of that point. 

One can perform a concrete computation to determine if the theory is causal in this sense, for each choice of current $j$. We have not yet rigorously studied this question, but we note the answer is not obvious a priori, as apparently acausal effects can vanish after the $x$ and $\eta$ integrations for certain $j$. A similar cancellation arises in the quantization of the continuous spin field, where the gauge fixed field operators end up obeying causal commutation relations. 

Another possibility is that full consistency requires additional effects absent from~\eqref{eq:introaction}, such as ``contact'' interactions between extended currents that do not involve a continuous spin field, but which exactly cancel acausal effects. Indeed, this is precisely what occurs in electromagnetism in Coulomb gauge, which contains an instantaneous interparticle potential but yet is causal as a whole.

Further investigation of all of these possibilities represent important directions for future work. For now, one can simply define our theory by the manifestly causal equations of motion. In practical terms, the physical results derived below are not affected either way, as they either consider non-accelerating particles -- where retarded and advanced boundary conditions give equal results -- or the coupling of particles to radiation, where the results are completely independent of the details of the current.

\addtocontents{toc}{\vspace{-0.2\baselineskip}}
\section{Forces on Matter Particles}
\label{sec:forces}
We can readily calculate the dynamics of matter particles from the action~\eqref{eq:introaction}. In subsection~\ref{sec:static_force}, we consider two particles interacting via the continuous spin field, and compute both the static interaction potential and the leading velocity-dependent corrections for some example currents. In subsection~\ref{sec:radiation_force} we compute forces from radiation background fields. These are universal for all scalar-like and vector-like currents, depending only on the spin scale $\rho$, and we give results for backgrounds with $|h| \leq 2$. 

\subsection{Static and Velocity-Dependent Interaction Potentials}
\label{sec:static_force}

Consider particles with trajectories $z_i^\mu(\tau)$, coupled to a continuous spin field by currents $J_i$. Each current produces a field $\Psi_i$ which interacts with each other particle $j$, contributing
\begin{align} 
S_{ij} &= \int d^4x \, [d^4\eta] \, \delta'(\eta^2 + 1) \, \Psi_i(\eta, x) J_j(\eta, x) \\
&= - \int \frac{d^4k}{(2\pi)^4} \, [d^4\eta] \, \delta'(\eta^2 + 1) \, \frac{J_i(\eta, k) J_j(\eta, -k)}{(\omega + i \epsilon)^2 - |\v{k}|^2} \\
&= - \int \frac{d^4k}{(2\pi)^4} \, [d^4\eta] \, \delta'(\eta^2 + 1) \, d\tau_i \, d\tau_j \, \frac{e^{- i k \cdot (z_i(\tau_i) - z_j(\tau_j))} f(\eta, k, \dot{z}_i) f^*(\eta, k, \dot{z}_j)}{(\omega + i \epsilon)^2 - |\v{k}|^2} \label{eq:S_int_pair}
\end{align}
to the action, where we have assumed retarded boundary conditions. Evaluating this generic expression in terms of the particle trajectories allows us to compute both static and velocity-dependent potentials. For simplicity we consider two dynamical particles, where the total interaction action is $S_{\text{int}} = (S_{12} + S_{21})/2$ to avoid double counting. Two useful integrals are
\begin{align}
\int \frac{d^3 \v{k}}{(2\pi)^3} \frac{e^{i \v{k} \cdot \v{r}}}{|\v{k}|^{2+n}} &= \frac{1}{(2n)!} \frac{(-r^2)^n}{4 \pi r}, \label{eq:pow_ft} \\
\int \frac{d^3 \v{k}}{(2\pi)^3} \frac{(\v{a} \cdot \hat{\v{k}})(\v{b} \cdot \hat{\v{k}}) \, e^{i \v{k} \cdot \v{r}}}{|\v{k}|^2} &= \frac{\v{a} \cdot \v{b} - (\v{a} \cdot \hat{\v{r}})(\v{b} \cdot \hat{\v{r}})}{8 \pi r} \label{eq:dir_ft}.
\end{align}
The $n = 0$ case of~\eqref{eq:pow_ft} is standard, and can be used to derive~\eqref{eq:dir_ft} using symmetry. For general integer $n > 0$,~\eqref{eq:pow_ft} can be derived inductively by taking the Laplacian of both sides. Technically, in that case there are also distributional corrections at $r = 0$, but this does not affect the computation of long-range forces.

\subsubsection*{Static Potentials}

We warm up by considering static particles, $z_i^\mu(\tau) = (t_i, \v{r}_i)$, interacting by an ordinary scalar field, $f = g$. Then the interaction simplifies to 
\begin{equation}
S_{ij} = - g^2 \int \frac{d^4k}{(2\pi)^4} \, [d^4\eta] \, \delta'(\eta^2 + 1) \, dt_i \, dt_j \, \frac{e^{- i \omega (t_i - t_j)} e^{i \v{k} \cdot (\v{r}_i - \v{r}_j)}}{(\omega + i \epsilon)^2 - |\v{k}|^2}.
\end{equation}
Performing one time integration and stripping off the other to yield a Lagrangian gives
\begin{equation} \label{eq:scalar_static_force}
L_{\text{int}} = g^2 \int \frac{d^3 \v{k}}{(2\pi)^3} \frac{e^{i \v{k} \cdot (\v{r}_1 - \v{r}_2)}}{|\v{k}|^2} = \frac{g^2}{4 \pi r}
\end{equation}
for a pair of particles with separation $\v{r} = \v{r}_1 - \v{r}_2$. This is a familiar $1/r^2$ attractive force. 

Given our previous results~\eqref{eq:Psi_S_pos} and~\eqref{eq:Psi_T_result} for the continuous spin fields of static particles, one might suspect that for nonzero $\rho$ the static potential must be radically different, either deviating from $1/r$ or exhibiting time dependence. However, in the static limit, the $\rho$-dependent phases in the temporal and spatial scalar-like currents simply cancel out, leading to the exact same result. This is a first hint that continuous spin physics can be simpler than the intermediate $\eta$-space expressions suggest. 

In general, while the definition of a scalar-like current ensures the potential contains a piece of the form~\eqref{eq:scalar_static_force}, there can be deviations at short or long distances. In appendix~\ref{app:classification}, we wrote the general solution to the continuity condition for static particles. Using just the $\bar{g}_0$ term in the homogeneous solution~\eqref{eq:general_static_homogeneous}, one can introduce terms in the integrand of~\eqref{eq:scalar_static_force} scaling as $|\v{k}|^n$, which yield contributions to the potential proportional to $1/r^{n+1}$. For positive $n$, this reflects how ordinary nonminimal couplings can modify forces at short distances. The novelty of continuous spin physics is that the continuity condition motivates solutions with an infinite series of negative $n$, which modify forces at long distances. 

To give one simple example, we consider the inhomogeneous current~\eqref{eq:J_I_def}, which satisfies the continuity condition for both real and imaginary $\beta$. For this current there are cross terms whose phases do not cancel, leading to
\begin{align}
L_{\text{int}} &= \frac{g^2}{4} \int \frac{d^3 \v{k}}{(2\pi)^3} \frac{e^{i \v{k} \cdot \v{r}}}{|\v{k}|^2} \int [d^4\eta] \, \delta'(\eta^2 + 1) \, \left(2 + e^{2 i \eta^0 \beta \rho^2 / |\v{k}|^2} + e^{- 2 i \eta^0 \beta \rho^2 / |\v{k}|^2} \right) \\
&= \frac{g^2}{2} \int \frac{d^3 \v{k}}{(2\pi)^3} \frac{e^{i \v{k} \cdot \v{r}}}{|\v{k}|^2} \left(1 + J_0(2 i \beta \rho^2 / |\v{k}|^2) \right)\\
& = \frac{g^2}{4\pi^2} \frac{1}{r} \int_0^{\infty} dx \, \frac{\sin x}{x} \left( 1 + J_0\left(\frac{2 i \beta \rho^2 r^2}{x^2}\right) \right) \\
&= \frac{g^2}{8 \pi r} \left( 1 + \int_{-\infty}^\infty dx\, \frac{e^{ix}}{\pi i \, x} \, J_0 \left(\frac{2 i \beta \rho^2 r^2}{x^2} \right) \right) \label{eq:L_int_beta}
\end{align}
where we applied~\eqref{eq:integral_proof}, then integrated over $\hat{\v{k}}$ and defined $x = |\v{k}| r$.

While it would be tempting to evaluate the integral by Taylor expanding the Bessel function and applying~\eqref{eq:pow_ft} to each term, this would not be legitimate because the integral passes directly through an essential singularity at $x = 0$. However, for imaginary $\beta$ the Bessel function is rapidly oscillating but bounded near $x = 0$, so the integral can be evaluated analytically. A continuation with the desired symmetry properties can be written in terms of Meijer $G$-functions,
\begin{equation} \label{eq:general_expr}
L_{\text{int}} = \frac{g^2}{8\pi r} \left( 1 + \frac{1}{\sqrt{2\pi}} \, G_{6,0}^{0,3}\left( \frac{2^{16}}{\beta^4 (\rho r)^8} ,2\big|^{\frac14 \,\frac34 \,1 \,\frac12 \,1\, 1}\right) \right).
\end{equation}
At real $\beta$, the function~\eqref{eq:general_expr} can be Taylor expanded as
\begin{equation}
L_{\text{int}} = \frac{g^2}{4 \pi r} \left(1 - \frac{4 \sqrt{2 \beta} \, \rho r}{\Gamma(1/4)^2} + \frac{\beta^{3/2} \, (\rho r)^3}{12 \sqrt{2} \, \Gamma(7/4)^2} + \ldots \right)
\end{equation}
The force is always attractive, and at large $r$, the $G$-function falls to zero, leaving an asymptotic potential of half the standard strength. 

All of these qualitative conclusions hold unchanged for vector-like currents. For an ordinary vector field, one picks up a factor of $(\eta^0)^2$ which causes a sign flip, so that like charges repel. Again, the temporal and spatial currents do not produce any deviations from Coulomb's law, and the inhomogeneous current yields long-distance corrections,
\begin{align}
L_{\text{int}} &= -\frac{e^2}{4\pi r} \int_{-\infty}^\infty dx\, \frac{e^{ix}}{\pi i \, x} \, \frac{x^4}{\beta^2 (\rho r)^4} \, \left( J_0 \left(\frac{2 i \beta \rho^2 r^2}{x^2} \right) - 1 \right)\\
&=  \frac{e^2}{4\pi r} \left( \frac{1}{\sqrt{2\pi}} \, G_{6,0}^{0,3}\left( \frac{2^{16}}{\beta^4 (\rho r)^8} ,2\big|^{\frac14 \,\frac34 \,1 \,\frac12 \,2\, 2}\right) \right).
\end{align}
This result can again be Taylor expanded for real $\beta$, giving
\begin{equation}
L_{\text{int}} = - \frac{e^2}{4 \pi r} \left(1 - \frac{\sqrt{\beta} \, \rho r}{\sqrt{2} \, \Gamma(9/4)^2} + \ldots \right)
\end{equation}
while at large $r$ it oscillates and alternates in sign.

\subsubsection*{Velocity-Dependent Potentials}

We computed static interaction potentials by evaluating the action for static particles, but there are also important velocity-dependent effects. We can include them by evaluating the action for moving but nonaccelerating particles, yielding a velocity-dependent potential. These potentials in turn neglect acceleration-dependent effects, such as radiation reaction, but yield a good description of the dynamics when the acceleration is small. 

To aid the reader, we first review the velocity-dependent potential for an ordinary scalar field. For nonaccelerating particles, the worldline is $x_i^\mu(\tau) = (t_i, \v{r}_i^0 + \v{v}_i t_i)$ where $t_i = \gamma_i \tau$, so 
\begin{align}
S_{ij} &= - g^2 \int \frac{d^4k}{(2\pi)^4} \, \frac{dt_i \, dt_j}{\gamma_i \gamma_j} \frac{e^{- i \omega (t_i - t_j)} e^{i \v{k} \cdot (\v{r}_i^0 - \v{r}_j^0)} e^{i \v{k} \cdot (\v{v}_i t_i - \v{v}_j t_j)}}{(\omega + i \epsilon)^2 - |\v{k}|^2} \\
&= g^2 \int \frac{d^3\v{k}}{(2\pi)^3} \, \frac{dt_j}{\gamma_i \gamma_j} \frac{e^{i \v{k} \cdot \v{r}_{ij}(t_j)}}{|\v{k}|^2 - (\v{k} \cdot \v{v}_i)^2}. 
\end{align}
where doing the $t_i$ integral set $\omega = \v{k} \cdot \v{v}_i$, and the separation is $\v{r}_{ij}(t) = \v{r}_i^0 + \v{v}_i t - (\v{r}_j^0 + \v{v}_j t)$. Stripping off the $t_j$ integration yields the Lagrangian for particle $j$ in the field of particle $i$, so that the interaction Lagrangian of two dynamical particles with separation $\v{r}(t)$ is
\begin{align}
L_{\text{int}} &= \frac{g^2}{2\gamma_1 \gamma_2} \int \frac{d^3\v{k}}{(2\pi)^3} \frac{e^{i \v{k} \cdot \v{r}(t)}}{|\v{k}|^2} \left( \frac{1}{1- (\hat{\v{k}} \cdot \v{v}_1)^2} + \frac{1}{1- (\hat{\v{k}} \cdot \v{v}_2)^2} \right) \\
&= \frac{g^2}{4 \pi r} \left(1 - \frac{v_1^2}{4} - \frac{v_2^2}{4} - \frac{(\v{v}_1 \cdot \hat{\v{r}})^2}{4} - \frac{(\v{v}_2 \cdot \hat{\v{r}})^2}{4} + O(v^4) \right)
\end{align}
where we used~\eqref{eq:dir_ft}. A similar calculation for a vector field yields the potential
\begin{equation}
L_{\text{int}} = - \frac{e^2}{4 \pi r} \left(1 - \v{v}_1 \cdot \v{v}_2 + \frac{v_1^2}{4} + \frac{v_2^2}{4} - \frac{(\v{v}_1 \cdot \hat{\v{r}})^2}{4} - \frac{(\v{v}_2 \cdot \hat{\v{r}})^2}{4} + O(v^4) \right)
\end{equation}
which includes magnetic interactions and retardation effects. While it may look unfamiliar, it is equivalent to the textbook Darwin Lagrangian~\cite{jackson1999classical}
\begin{equation}
L_{\text{int}} = - \frac{e^2}{4 \pi r} \left( 1 - \frac{\v{v}_1 \cdot \v{v}_2 + (\v{v}_1 \cdot \hat{\v{r}}) (\v{v}_2 \cdot \hat{\v{r}})}{2} + O(v^4) \right)
\end{equation}
since, up to accelerations, it differs by the total time derivative of $(e^2/16 \pi)(\v{v}_1 \cdot \hat{\v{r}} - \v{v}_2 \cdot \hat{\v{r}})$.

For nonzero $\rho$ our formalism readily yields expressions for velocity-dependent potentials in momentum space, but evaluating the Fourier transform is complicated by the essential singularity terms. We will therefore consider only the scalar spatial current, where the integrals are simplest. The answer will be a series in the independent dimensionless variables $v^2$ and $\rho v r$, so to avoid clutter we will neglect terms suppressed by a power of $v^2$. 

Defining $\v{v}_{ij} = \v{v}_j - \v{v}_i$, and letting $V_i$ be the vector defined below~\eqref{eq:J_S_intro}, we have
\begin{align}
L_{ij} &\approx g^2 \int \frac{d^3\v{k}}{(2\pi)^3} \frac{e^{i \v{k} \cdot \v{r}_{ij}(t)}}{|\v{k}|^2} \int [d^4 \eta] \, \delta'(\eta^2 + 1) \, e^{- i \rho \eta \cdot V_i / k \cdot V_i} e^{i \rho \eta \cdot V_j / k \cdot V_j} \bigg|_{\omega = \v{k} \cdot \v{v}_i} \\
&\approx g^2 \int [d^4 \eta] \, \delta'(\eta^2 + 1) \int \frac{d^3 \v{k}}{(2\pi)^3} \frac{e^{i \v{k} \cdot \v{r}_{ij}(t)}}{|\v{k}|^2} \, e^{-i \rho \eta^0 \v{k} \cdot \v{v}_{ij} / |\v{k}|^2}. 
\end{align}
where we dropped terms of order $v^2$ inside the exponent. Next, the $\v{k}$ integral can be evaluated using the same method as used to evaluate~\eqref{eq:Psi_S_setup}. Defining $x = \rho v_{ij} r_{ij} (1 - \cos \theta)$, where $\theta$ is the angle between $\v{r}_{ij}$ and $\v{v}_{ij}$, we have
\begin{equation}
L_{ij} \approx \frac{g^2}{4 \pi r_{ij}} \int [d^4 \eta] \, \delta'(\eta^2 + 1) \, J_0\left( \sqrt{2 \eta^0 x} \right) = \frac{g^2}{4 \pi r_{ij}} \, I_0(\sqrt{x}) J_0(\sqrt{x})
\end{equation}
where we performed the $\eta$ integral by directly using the generating function~\eqref{eq:primed_generating_function}. Thus, at leading nontrivial order, the velocity-dependent potential for a pair of matter particles is 
\begin{equation}
L_{\text{int}} = \frac{g^2}{4 \pi r} \left(1 - \frac{\rho^2 r^2}{32} \left( |\v{v}_i - \v{v}_j| - (\v{v}_i - \v{v}_j) \cdot \hat{\v{r}} \right)^2 + O(v^2, (\rho v r)^4) \right).
\end{equation}
For particles moving along a line, the new term has no effect, while for transverse motion it becomes important at long distances, $\rho v r \sim 1$, just like the corrections to static forces. 

\subsection{Forces From Background Radiation Fields}
\label{sec:radiation_force}

\subsubsection*{The Interaction Lagrangian}

Next, we consider the dynamics of a particle in a free background field $\Psi$, by evaluating the interaction action~\eqref{eq:interaction} in terms of the mode expansion coefficients defined in~\eqref{eq:modeexpansion}, 
\begin{equation} \label{eq:S_int_background}
S_{\text{int}} = \text{Re} \int \frac{d^3 \v{k}}{(2\pi)^3 \, |\v{k}|} \, [d^4 \eta] \, \delta'(\eta^2 + 1) \, \sum_h a_h(\v{k}) \psi_{h,k}(\eta) J(\eta, k)^* \bigg|_{k^0 = |\v{k}|}
\end{equation} 
where taking the real part adds on the negative frequency part of the field. As explained below~\eqref{eq:general_universality_decomposition}, this action is independent of $X(\eta, k, \dot{z})$, and thus depends only on $\hat{g}(k \cdot \dot{z})$. 

For concreteness, we note this can be shown more directly. The $k^2 X$ term in~\eqref{eq:general_universality_decomposition} vanishes since $k$ is null, and the remaining term is of the form $D \xi$. However, the identity~\eqref{eq:deltapr_D_identity} implies
\begin{equation}
\int [d^4 \eta] \, \delta'(\eta^2 + 1) \, (D \xi) \, \Psi = \frac12 \int [d^4 \eta] \, \Delta(\delta(\eta^2 + 1) \, \xi) \, \Psi
\end{equation}
and integrating by parts yields an integrand proportional to $\delta(\eta^2 + 1) \, (\Delta \Psi)$, which vanishes in harmonic gauge. It thus vanishes in any gauge, since the action is gauge invariant.

In any case, we can keep just the $\hat{g}$ term in the current, yielding
\begin{align} 
S_{\text{int}} &= \text{Re} \sum_h \int d \tau \, \frac{d^3 \v{k}}{(2\pi)^3 \, |\v{k}|} \, [d^4 \eta] \, \delta'(\eta^2 + 1) \, \hat{g}^* \, a_h(\v{k}) e^{- i k \cdot z} \, \psi_{h,k}(\eta) \, e^{i \rho \eta \cdot \dot{z} / k \cdot \dot{z}} \bigg|_{k^0 = |\v{k}|} \\
&= \text{Re} \sum_h \int d \tau \, \frac{d^3 \v{k}}{(2\pi)^3 \, |\v{k}|} \, \hat{g}^* \, a_h(\v{k}) e^{- i k \cdot z} \, \left( \frac{\epsilon_+ \cdot V}{|\epsilon_+ \cdot V|} \right)^h J_h(\sqrt{-V^2}) \bigg|_{k^0 = |\v{k}|} 
\end{align} 
where we performed the $\eta$ integral using~\eqref{eq:helicity_identity} with $V = \rho((\dot{z} / k \cdot \dot{z}) - q)$. We can make this more physically transparent by specializing to $\epsilon_\pm^0 = 0$ and simplifying using the properties of the null frame vectors. The result is the Lagrangian
\begin{equation} \label{eq:general_int_V}
L_{\text{int}} = \text{Re} \sum_h \sqrt{1-v^2} \int \frac{d^3 \v{k}}{(2\pi)^3 \, |\v{k}|} \hat{g}^* a_h(\v{k}) e^{- i (|\v{k}| t - \v{k} \cdot \v{r})} \left( \frac{- \bm{\epsilon}_+ \cdot \v{v}_\perp}{v_\perp} \right)^h J_h\left( \frac{\rho v_\perp}{|\v{k}| - \v{k} \cdot \v{v}} \right) 
\end{equation}
whose integral over coordinate time $t$ is the interaction action. Above, $\v{r}(t)$ and $\v{v}(t)$ are the position and velocity of the particle, and $\v{v}_\perp$ is the velocity in the plane transverse to $\hat{\v{k}}$. The action is smooth in $\v{v}$, despite its dependence on the magnitude of $\v{v}_\perp$, because of the phase in front of the Bessel function. 

The expression~\eqref{eq:general_int_V} is our general result, but for simplicity we will specialize to monochromatic plane waves traveling in the $\hat{\v{k}} = \hat{\v{z}}$ direction, by taking
\begin{equation}
a_h(\v{k}) = \omega_0 \, \bar{a}_h \, (2\pi)^3 \delta^{(3)}(\v{k} - \omega \hat{\v{z}}).
\end{equation}
The resulting Lagrangians are 
\begin{equation}
L_{\text{int}}^S = g \sqrt{1-v^2} \, \text{Re} \sum_h \bar{a}_h e^{- i \omega_0 (t - z) } \left( \frac{-v_x - i v_y}{v_\perp} \right)^h J_h\left( \frac{\rho v_\perp}{\omega_0 (1 - v_z)} \right) 
\end{equation}
for any scalar-like current, and 
\begin{equation}
L_{\text{int}}^V = \frac{\sqrt{2} \, e}{\rho} \, \text{Re} \sum_h \bar{a}_h e^{- i \omega_0 (t - z) } (- i \omega_0 (1 - v_z)) \left( \frac{-v_x - i v_y}{v_\perp} \right)^h J_h\left( \frac{\rho v_\perp}{\omega_0 (1 - v_z)} \right) 
\end{equation}
for any vector-like current, where above $z$ now stands for $\v{r} \cdot \hat{\v{z}}$ and $v_\perp = \sqrt{v_x^2 + v_y^2}$. 

We will work with three specific backgrounds below. First, we will consider a helicity zero background with $\bar{a}_0 = \phi_0$, which reduces to a scalar field $\phi = \phi_0 \cos(\omega_0 (t - z))$ as $\rho \to 0$. Next, a background with nonzero $\bar{a}_1$ reduces as $\rho \to 0$ to a circularly polarized electromagnetic wave. For simplicity, we will consider the combination of $h = \pm 1$ backgrounds $\bar{a}_{-1} = - \bar{a}_1 = A_0/\sqrt{2}$, which reduces as $\rho \to 0$ to the linearly polarized wave
\begin{align}
\v{A} &= - A_0 \sin(\omega_0 (t - z)) \, \hat{\v{x}}, \\
\v{E} &= \omega_0 A_0 \cos(\omega_0 (t - z)) \hat{\v{x}}, \label{eq:E_def} \\
\v{B} &= \omega_0 A_0 \cos(\omega_0 (t - z)) \hat{\v{y}}. \label{eq:B_def}
\end{align}
By varying the ratio $\bar{a}_1/\bar{a}_{-1}$, we could also produce backgrounds with an arbitrary elliptical polarization. Finally, we consider a helicity $\pm 2$ background with $\bar{a}_2 = \bar{a}_{-2} = h$, which reduces as $\rho \to 0$ to a ``plus'' polarized gravitational wave, $h_{xx} = - h_{yy} = h \cos (\omega_0 (t-z))$.

\subsubsection*{Extracting the Force}

We can straightforwardly compute forces with the Lagrangians above, but there are multiple natural definitions of force. To build intuition, let us first consider particles minimally coupled to scalar, vector, or tensor fields with $\rho = 0$. The equations of motion are
\begin{align}
(m - g \phi) \ddot{z}^\mu &= - g (g^{\mu\nu} - \dot{z}^\mu \dot{z}^\nu) \del_\nu \phi, \label{eq:scalar_particle_eom} \\
m \ddot{z}^\mu &= e \dot{z}_\nu F^{\mu\nu}, \label{eq:vector_particle_eom} \\
\ddot{z}^\mu &= \kappa \, \dot{z}^\nu \dot{z}^\rho (\del_\nu h_{\rho}^{\ \mu} + \del_\rho h_{\nu}^{\ \mu} - \del^\mu h_{\nu \rho})/2 \label{eq:tensor_particle_eom}
\end{align}
where a dot denotes a derivative with respect to proper time $\tau$, and all fields are evaluated at $z^\mu(\tau)$. These are straightforwardly derived from the interaction actions~\eqref{eq:scalar_current_coupling},~\eqref{eq:vector_current_coupling}, and~\eqref{eq:tensor_coupling}, which correspond to Lagrangians
\begin{equation}
L_{\text{int}}(\v{r}(t), \v{v}(t)) = \begin{cases} g \, \phi \, \sqrt{1 - v^2} & \text{scalar} \\ e \, \v{A} \cdot \v{v} & \text{vector}, \text{radiation gauge} \\
(\kappa m/2) \, h^{ij} v_i v_j / \sqrt{1-v^2} & \text{tensor}, \text{transverse\ traceless\ gauge} \end{cases}.
\end{equation}
All three fields produce qualitatively different effects. While the vector field yields the familiar Lorentz force, a scalar field can affect the particle's inertia, which can even cause singular accelerations when $\phi = m/g$, at which point our entire description of matter as particles breaks down. This effect comes from the $v^2$ term in the scalar Lagrangian. By contrast, there is no $v^2$ term for a tensor field in transverse traceless gauge, but in this case the coordinate acceleration of a free particle initially at rest is zero, indicating that the coordinate system itself stretches with the gravitational wave. This is an unnatural way to describe laboratory experiments with rigid detectors, for which we should instead use the proper detector frame, where the tensor field produces an ordinary Newtonian force~\cite{maggiore2007gravitational}. 

For continuous spin backgrounds, we choose to define the force by $\v{F} = d\v{p} / dt$ where $\v{p} = m \v{v} / \sqrt{1-v^2}$ is the unperturbed three-momentum, as this is the most natural quantity for a vector background. The Euler--Lagrange equation yields 
\begin{equation}
\v{F} = \frac{\del L_{\text{int}}}{\del \v{r}} - \frac{d}{dt} \frac{\del L_{\text{int}}}{\del \v{v}} = \frac{\del L_{\text{int}}}{\del \v{r}} - \left(\frac{\del}{\del t} + \v{v} \cdot \frac{\del}{\del \v{r}} + \v{a} \cdot \frac{\del}{\del \v{v}} \right) \frac{\del L_{\text{int}}}{\del \v{v}}.
\end{equation}
This is only an implicit equation, since $\v{a}$ in the final term is itself determined by the force. In addition, this final term contains the inertia-modifying effects of a scalar field, which can produce singular accelerations. We avoid both of these issues by assuming the particle experiences only a single weak background field. In this case, both $\v{a}$ and $L_{\text{int}}$ start at first order in the field, so the last term is second order and can be dropped. (Of course, we are also implicitly ignoring backreaction effects, such as radiation reaction forces.) 

Working in Euclidean notation, the resulting forces for scalar, vector, and tensor fields are
\begin{align}
\v{F} &\approx g (\nabla \phi + \v{v} \dot{\phi} - (v^2/2) \nabla \phi + \v{v} (\v{v} \cdot \nabla) \phi + O(v^3)), \label{eq:nr_scalar_force} \\
\v{F} &= e (\v{E} + \v{v} \times \v{B}), \label{eq:nr_vector_force} \\
F_i &= \kappa m (- \dot{h}_{ij} v_j + v_j v_k (\del_i h_{jk} - \del_j h_{ik} - \del_k h_{ij})/ 2 + O(v^3)) \label{eq:nr_tensor_force}
\end{align}
respectively, where a dot now represents a time derivative. As discussed above, this force is not the most natural quantity for a scalar or tensor background, but we can still use it as a starting point to compute corrections at nonzero $\rho$. We denote the order $\rho^n$ contribution to the force on a particle with a scalar-like or vector-like current by $\v{F}^{S,n}$ and $\v{F}^{V,n}$, respectively. 

\subsubsection*{Scalar-Like Currents}

We now warm up by considering a particle with a scalar-like current. First, in the helicity $0$ background defined above, the Lagrangian is 
\begin{equation}
L_{\text{int}}^S = g \phi_0 \cos(\omega_0(t - z)) \, \sqrt{1 - v^2} J_0\left( \frac{\rho v_\perp}{\omega_0 (1 - v_z)} \right).
\end{equation}
The leading term of the Bessel produces the standard force~\eqref{eq:nr_scalar_force}, which in this case is
\begin{equation}
\v{F}^{S,0}_{h=0} = g \omega_0 \phi_0 \sin(\omega_0 (t - z)) (\hat{\v{z}} - \v{v} + O(v^2)).
\end{equation}
More generally, $\hat{\v{z}}$ would be the direction of propagation $\hat{\v{k}}$ of the plane wave, by rotational symmetry. Next, the quadratic term in the Bessel gives the leading correction,
\begin{equation} \label{eq:leading_correction_scalar}
\v{F}^{S,2}_{h=0} = - \frac{\rho^2}{2 \omega_0^2} \, g \omega_0 \phi_0 \sin(\omega_0 (t - z)) (\v{v}_\perp + O(v^2)).
\end{equation}
For the same particle in the helicity $\pm 1$ background, the leading term in the Lagrangian is 
\begin{equation}
L_{\text{int}}^S = \frac{g \rho}{\sqrt{2} \, \omega_0} A_0 \cos(\omega_0 (t - z)) \frac{\sqrt{1-v^2}}{1 - v_z} \, v_x + O(\rho^2)
\end{equation}
which corresponds to a force 
\begin{align}
\v{F}^{S,1}_{|h|=1} &= \frac{g \rho}{\sqrt{2}} \, A_0 \sin(\omega_0 (t - z)) (\hat{\v{x}} + 2 v_x \hat{\v{z}} + O(v^2)) \\
&= - \frac{g \rho}{\sqrt{2}} \left(\v{A} + 2 (\v{v} \cdot \v{A}) \hat{\v{k}} \right).
\end{align}
The two terms here are like an electric and magnetic force respectively, but their relative normalization differs from the usual Lorentz force, and the ``magnetic'' force instead points along the direction of propagation of the wave. Furthermore, both forces are $\pi/2$ out of phase with the usual definitions~\eqref{eq:E_def} and~\eqref{eq:B_def} of the electric and magnetic fields. 

Thus, $\rho$-dependent corrections to forces generally have novel direction and velocity dependence. For a higher helicity background $h > 1$, the Lagrangian would start at order $\bar{a}_h (\rho v/\omega)^{h}$, with nontrivial tensor structure in $v$, and yields a leading force of order $(\rho \bar{a}_h) (\rho v/\omega)^{h-1}$. 

\subsubsection*{Vector-Like Currents}

We now turn to the more phenomenologically interesting case of particles with vector-like currents. We first consider the helicity $0$ background, where
\begin{equation}
L_{\text{int}}^V = - \frac{\sqrt{2} \, e \, \omega_0 }{\rho} \, \phi_0 \sin (\omega_0 (t - z)) (1 - v_z) J_0\left( \frac{\rho v_\perp}{\omega_0 (1 - v_z)} \right).
\end{equation}
In this case there should be no force when $\rho = 0$, but the Lagrangian instead appears to diverge as $\rho \to 0$, due to the constant term in the Bessel. The resolution is familiar from subsection~\ref{ssec:CSP_currents}: the apparently divergent term is a total time derivative, and thus has no physical effect. Discarding this term, the leading correction is from the quadratic term, 
\begin{equation}
\v{F}^{V,1}_{h=0} = - \frac{e \rho}{\sqrt{2}} \, \phi_0 \cos(\omega_0 (t - z)) (\v{v}_\perp + O(v^2)).
\end{equation}
This has the same velocity dependence as~\eqref{eq:leading_correction_scalar}, which is not a coincidence. In the nonrelativistic limit, the leading correction to the force is always purely transverse, due to the dependence of the Bessel functions on $v_\perp$, and for a $h = 0$ background that force must be directly proportional to $\v{v}_\perp$ by rotational symmetry. 

Next, for the helicity $\pm 1$ background, we have
\begin{equation}
L_{\text{int}}^V = e \v{A} \cdot \v{v} + \frac{e \rho^2}{8 \omega_0^2} A_0 \sin(\omega_0(t-z)) (v_x v_\perp^2 + O(v^4)) + O(\rho^4)
\end{equation}
where the first term recovers the Lorentz force law~\eqref{eq:nr_vector_force}, and the second term yields
\begin{align}
\v{F}^{V,2}_{|h|=1} &= - \frac{e \rho^2}{4 \omega_0} A_0 \cos(\omega_0(t-z)) \left(\frac{3 v_x^2 + v_y^2}{2} \, \hat{\v{x}} + v_x v_y \, \hat{\v{y}} + O(v^3) \right) \\
&= - \frac{e \rho^2}{4 \omega_0^2} \left((\v{E} \cdot \v{v}_\perp) \v{v}_{\perp} + \v{E} \, v_\perp^2/2 + O(v^3) \right) \label{eq:Lorentz_change}
\end{align}
as quoted in~\eqref{eq:introshowcase:force}, where terms involving $\v{B}$ appear at order $v^3$. While this correction is suppressed by $(\rho v/\omega_0)^2$ relative to the familiar electric force, it has a distinctive direction and velocity dependence which may be detectable. For instance, it includes a term parallel to $\v{v}_\perp$. Note that~\eqref{eq:Lorentz_change} only applies to radiation backgrounds; it does not apply to situations with static electromagnetic fields, where the force correction is not universal.

Finally, for the helicity $\pm 2$ background, we have
\begin{equation}
L_{\text{int}}^V = \frac{e \rho}{2 \sqrt{2} \, \omega} \, h \sin(\omega_0(t-z)) \, (v_y^2 - v_x^2 + O(v^3))
\end{equation}
which corresponds to a force 
\begin{equation}
\v{F}^{V,1}_{|h| = 2} = \frac{e \rho}{\sqrt{2}} \, h \cos(\omega_0 (t-z)) (v_x \hat{\v{x}} - v_y \hat{\v{y}} + O(v^2)).
\end{equation}
This has the same velocity dependence as the leading force~\eqref{eq:nr_tensor_force} from a gravitational wave, but shifted $\pi/2$ out of phase. More generally, in a higher helicity background $h > 2$, the Lagrangian starts at order $v \bar{a}_h (\rho v/\omega)^{h-1}$ and yields a leading force of order $(\rho v \bar{a}_h) (\rho v/\omega)^{h-2}$. 

\subsubsection*{Discussion}

In the above discussion we have restricted to plane wave backgrounds. Generalizing to arbitrary radiation backgrounds, such as wavepackets, is computationally straightforward and may offer a useful perspective on the localization properties of the interactions. We have also neglected inertia-modifying effects, which, e.g.~would arise for a particle with a vector-like current in an $h = 0$ background. Searching for such effects would provide information about the amplitude of low-frequency background radiation in all of the helicity modes. 

It is also interesting to consider if there is a natural continuous spin counterpart to a uniform static electric or magnetic field. Some care must be taken in defining this for several reasons. First, the enormous gauge redundancy~\eqref{eq:eps_gauge} and the absence of local gauge-invariant observables obscures what a ``uniform field'' should mean. Second, the linearly varying potentials of uniform fields in electromagnetism, such as $A^\mu(\v{x},t)=(E z, \v{0})$, have no clear continuous spin counterpart that satisfies the free equation of motion~\eqref{eq:free_eom}. Third, as we have seen in subsection~\ref{ssec:worldlines}, the currents of source particles are generically spatially delocalized and thus overlap the probe particle, so requiring that ``uniform'' fields satisfy a free equation of motion is not even justified. (The temporal current is not spatially delocalized, but it is not time-independent and does not give rise to a static field.)

As a result, free background fields are not a useful concept for describing static forces. One should instead compute these non-universal interparticle forces directly, as we have already done in subsection~\ref{sec:static_force}.  There, we saw that the static force law for spatial and temporal currents was exactly $1/r$, so for these currents all \emph{electrostatic} forces are unchanged. We have also seen that the inhomogeneous current has a modified static force law, and that continuous spin fields generically have modified velocity-dependent force laws.  

\addtocontents{toc}{\vspace{-0.2\baselineskip}}
\section{Radiation Emission From Matter Particles}
\label{sec:radiation}
In this section we investigate the continuous spin radiation emitted by an accelerating particle. First, we review radiation in ordinary scalar and vector fields in subsection~\ref{sec:radiation_scalar_vector}. We cover this standard material carefully, in a way that directly generalizes to the nonlocal currents of continuous spin fields in subsection~\ref{sec:radiation_CSP}. In subsection~\ref{sec:radiation_computation} we apply these results to particles in nonrelativistic motion, yielding concrete deviations from the Larmor formula, and show the radiated power is well-behaved for arbitrary accelerations. Along the way, we make contact with previous work by recovering the soft factors for CSP emission, by considering the radiation emitted by an instantaneously kicked particle.

\subsection{Review: Scalar and Vector Radiation}
\label{sec:radiation_scalar_vector}

\subsubsection*{The Radiation Field and Soft Factors}

We first consider the radiation produced in a massless scalar field $\phi$ by a time-dependent source $J$. Given retarded boundary conditions, the equation of motion $\del^2 \phi = J$ is solved by 
\begin{equation}
\phi(x) = \int d^4y \, G_r(x-y) J(y)
\end{equation}
where $G_r$ is the retarded Green's function. If the source $J$ corresponds to matter particles that eventually stop accelerating, then at late times the field contains outgoing radiation and the static fields of these particles. We can isolate the radiation in a covariant way by defining
\begin{equation} \label{eq:phi_decomp}
\phi(x) = \phi_{\text{rad}}(x) + \phi_a(x)
\end{equation}
where $\phi_a(x)$ is the solution with advanced boundary conditions, corresponding to Green's function $G_a$. The difference $\phi_{\text{rad}}(x)$ contains only radiation, because $\del_x^2 \phi_{\text{rad}} = \del_x^2 (\phi - \phi_a) = 0$, so it has the null plane wave expansion 
\begin{equation} \label{eq:phi_rad_exp}
\phi_{\text{rad}}(x) = \int \frac{d^3 \v{k}}{(2\pi)^3 \, 2 |\v{k}|} \, (a(\v{k}) e^{- i k \cdot x} + \text{c.c.}) \bigg|_{k^0 = |\v{k}|}.
\end{equation}
On the other hand, in momentum space the retarded propagator corresponds to integrating along a $k^0$ contour that passes above the poles at $k^0 = \pm |\v{k}|$, while the advanced propagator passes below them. Their difference corresponds to integrating along a contour $C$ that simply encircles both poles clockwise, so 
\begin{align}
\phi_{\text{rad}}(x) &= -\int \frac{d^3\v{k}}{(2\pi)^3} \int_C \frac{dk^0}{2\pi} \frac{J(k) e^{-ik\cdot x}}{k^2} \\
&= \int \frac{d^3\v{k}}{(2\pi)^3 \, 2|\v{k}|}\left( i J(k) e^{- i k \cdot x} + \text{c.c.} \right) \bigg|_{k^0 = |\v{k}|}
\end{align}
from which we read off $a(\v{k}) = i J(|\v{k}|, \v{k})$. 

For example, if the source is a single particle that receives an instantaneous kick at the origin, then the particle's worldline is 
\begin{equation} \label{eq:kicked_worldline}
z^\mu(\tau) = \begin{cases} (\gamma \tau, \gamma \v{v} \tau) & \tau < 0 \\ (\gamma' \tau, \gamma' \v{v}' \tau) & \tau > 0\end{cases}.
\end{equation}
Integrating the worldline current~\eqref{eq:scalar_current} yields 
\begin{equation} \label{eq:scalar_worldline_J}
J(\omega, \v{k}) = \frac{ig}{\gamma (\omega - \v{k} \cdot \v{v})} - \frac{ig}{\gamma' (\omega - \v{k} \cdot \v{v}')}
\end{equation}
which corresponds to a radiation amplitude
\begin{equation} \label{eq:scalar_amplitude}
a(\v{k}) = (2mg) \left( \frac{1}{2 k \cdot p'} - \frac{1}{2 k \cdot p} \right)_{k^0 = |\v{k}|}
\end{equation}
where $m$ is the mass of the particle, and $p$ and $p'$ are its initial and final four-momenta. 

We can derive essentially the same result in quantum field-theoretic language by taking a matter field $\Phi$ of mass $m$ with Yukawa coupling $y \phi^2 \Phi$, where matching to the particle theory sets $y = 2mg$. Then in any process with a $\Phi$ particle of initial momentum $p$ and final momentum $p'$, the amplitude to emit a soft $\phi$ particle of momentum $k$ obeys
\begin{equation} \label{eq:scalar_soft}
\frac{\mathcal{M}(p \to p' + k)}{\mathcal{M}_0(p \to p')} = (2mg) \left( \frac{1}{2 k \cdot p'} - \frac{1}{2 k \cdot p} \right)
\end{equation}
where the ``soft factors'' on the right-hand side can be determined solely using unitarity, locality, and Lorentz invariance \cite{Weinberg:1964ew}. Of course, this matches~\eqref{eq:scalar_amplitude} in form. To make the connection more explicit one can consider amplitudes with an arbitrary number of outgoing $\phi$ particles, yielding a coherent final state for the $\phi$ field with $\langle a(\v{k}) \rangle$ given by~\eqref{eq:scalar_amplitude}. We thus recover the exact same result for the radiation field in the classical limit of many emissions.

It is straightforward to generalize these results to a massless vector field $A^\mu$. Defining the radiation field $A^\mu_{\text{rad}}$ analogously, in Lorenz gauge we have 
\begin{equation}
A^\mu_{\text{rad}}(x) = \int \frac{d^3\v{k}}{(2\pi)^3 \, 2|\v{k}|}\left( i J^\mu(k) e^{- i k \cdot x} + \text{c.c.} \right) \bigg|_{k^0 = |\v{k}|}
\end{equation}
and the corresponding plane wave expansion is 
\begin{equation}
A^\mu_{\text{rad}}(x) = \int \frac{d^3 \v{k}}{(2\pi)^3 \, 2 |\v{k}|} \, \left( \sum_{\lambda = \pm} \hat{\epsilon}_\lambda^\mu a_\lambda(\v{k}) e^{- i k \cdot x} + \text{c.c.} + \text{gauge} \right) \bigg|_{k^0 = |\v{k}|}.
\end{equation}
Here we define $\hat{\epsilon}_\lambda = \epsilon_\lambda / \sqrt{2}$ to reach the usual normalization for helicity $\pm 1$ modes. Now,
\begin{equation}
J^\mu = (k \cdot J) q^\mu + (q \cdot J) k^\mu - (\hat{\epsilon}_- \cdot J) \hat{\epsilon}_+^\mu - (\hat{\epsilon}_+ \cdot J) \hat{\epsilon}_-^\mu
\end{equation}
where the first term is zero by current conservation, and the second term corresponds to the residual gauge freedom in Lorenz gauge. The last two terms yield
\begin{equation}
a_\pm(\v{k}) = (i \hat{\epsilon}_\pm)^* \cdot J(k) \bigg|_{k^0 = |\v{k}|}.
\end{equation}
Returning to the example of a single kicked particle, coupled to the vector field by~\eqref{eq:vector_current}, a derivation analogous to the previous one gives 
\begin{equation}
J^\mu(\omega, \v{k}) = \frac{ie\, (\gamma, \gamma \v{v})}{\gamma (\omega - \v{k} \cdot \v{v})} - \frac{ie \, (\gamma', \gamma' \v{v}')}{\gamma' (\omega - \v{k} \cdot \v{v}')}
\end{equation}
which corresponds to a radiation amplitude
\begin{equation} \label{eq:vector_amplitude}
a_\pm(\v{k}) = e \left( \frac{\hat{\epsilon}_\pm^* \cdot p}{k \cdot p} - \frac{\hat{\epsilon}_\pm^* \cdot p'}{k \cdot p'} \right)_{k^0 = |\v{k}|}.
\end{equation}
Of course, this matches the ratio $\mathcal{M}(p \to p' + k) / \mathcal{M}_0(p \to p')$ from field-theoretic soft factors. 

\subsubsection*{Radiated Power}

\begin{figure}[t]
\includegraphics[width=0.6\columnwidth]{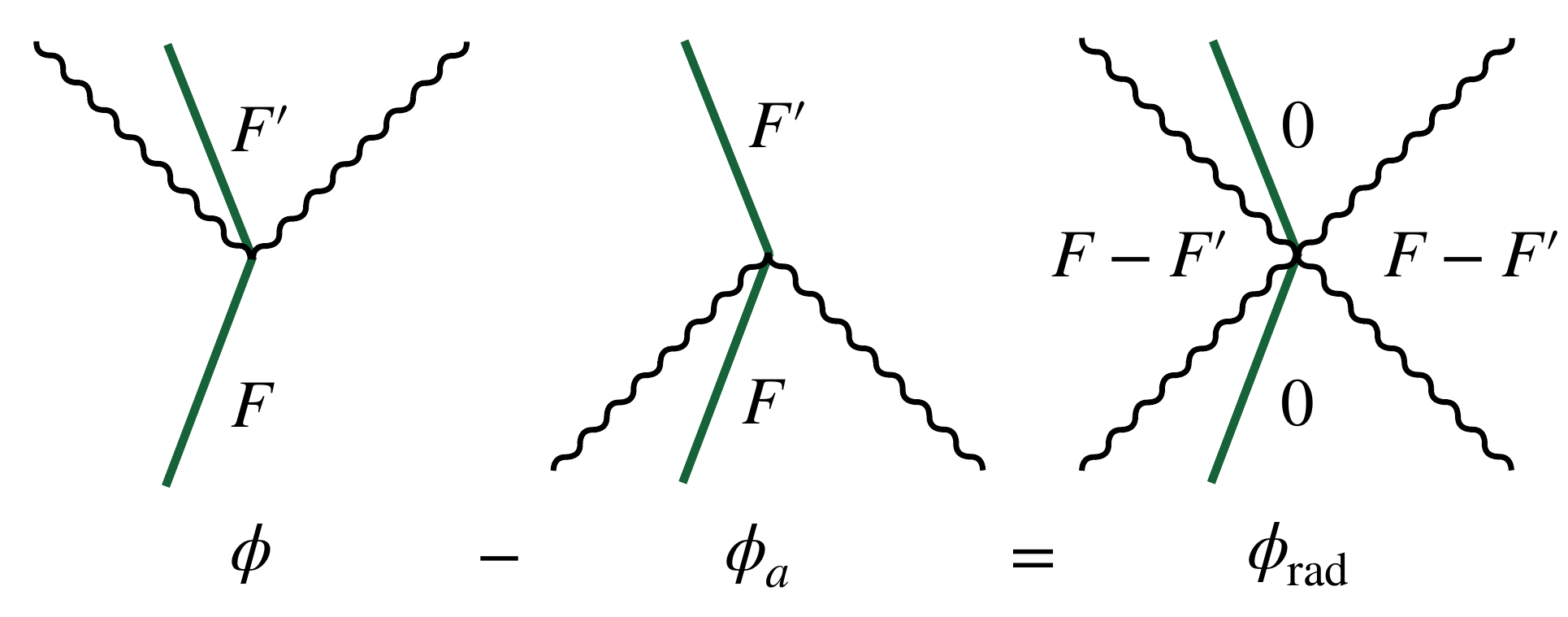}
\caption{The field $\phi$ of a kicked particle has outgoing radiation on the forward light cone, the final static field $F'$ inside, and the initial static field $F$ outside. Subtracting off the advanced field $\phi_a$ gives a pure radiation field, which has incoming radiation on the backward light cone, outgoing radiation on the forward light cone, and the difference $F - F'$ outside both light cones.
}
\label{fig:phi_rad}
\end{figure}

Defining $x^\mu = (t, r \hat{\v{r}})$, we can simplify~\eqref{eq:phi_rad_exp} in the far field limit $r \to \infty$ using
\begin{equation}
e^{i \v{k} \cdot \v{r}} = \frac{2 \pi e^{i|\v{k}|r}}{i|\v{k}|r} \, \delta^{(2)}(\hat{\v{k}}, \hat{\v{r}}) - \frac{2 \pi e^{-i|\v{k}|r}}{i|\v{k}|r} \delta^{(2)}(\hat{\v{k}}, - \hat{\v{r}})
\end{equation}
which holds since both sides are integrated against smooth functions of $\hat{\v{k}}$ and $\hat{\v{r}}$. This yields
\begin{align}
\lim_{r \to \infty} \phi_{\text{rad}}(x) &= \frac{1}{4 \pi r} \int_0^\infty \frac{d|\v{k}|}{2\pi} \left( - i a(|\v{k}| \hat{\v{r}}) \, e^{- i |\v{k}| (t-r)} + i a(- |\v{k}| \hat{\v{r}}) \, e^{- i |\v{k}| (t+r)} + \mathrm{c.c.} \right) \\
&= \frac{1}{4 \pi r} \int \frac{d\omega}{2\pi} \left( J(\omega, \omega \hat{\v{r}}) \, e^{- i \omega (t-r)} - J(\omega, - \omega \hat{\v{r}}) \, e^{- i \omega (t+r)}\right). \label{eq:phi_rad_two}
\end{align}
Much of the information in this expression is irrelevant to computing the radiated power. To see this, consider a particle kicked at time $t = 0$, and let $F$ ($F'$) be the static field that would exist if the particle had its initial (final) velocity for all time. The spacetime profile of $\phi_{\text{rad}}$ is shown in Fig.~\ref{fig:phi_rad}, and contains outgoing radiation (from the first term of~\eqref{eq:phi_rad_two}), incoming radiation (from the second term), and a combination of static fields (from both terms). 

Static fields never contribute to radiated power in the far field, since their derivatives are suppressed by $1/r$. The incoming radiation is unphysical and vanishes for $t > 0$, and more generally vanishes at late times for any situation where the particles eventually stop accelerating. Thus, at late times we can compute the radiated power from the physical, outgoing radiation in $\phi$ by working with only the first term of~\eqref{eq:phi_rad_two}. This prescription even works for periodic motion as long as we consider its amplitude to eventually damp as $t \to \infty$, which we will leave implicit below.

To compute the angular distribution of radiated power, we note that in the far field limit, the derivatives in the Poynting vector $T^{0i} = \del^0 \phi \, \del^i \phi$ effectively only act upon the exponential $e^{- i \omega (t-r)}$, as other contributions are suppressed by $1/r$. Then at late times,
\begin{align}
\frac{dP(t)}{d\hat{\v{r}}} &= \lim_{r \to \infty} r^2 \, (\del^0 \phi(x)) \, (\del^r \phi(x)) \\
&= \left| \int \frac{d\omega}{2 \pi} \frac{e^{-i \omega (t-r)}}{4 \pi} \, \omega J(\omega, \omega \hat{\v{r}}) \right|^2. \label{eq:angular_power_dist}
\end{align}
Similarly, for a vector field in Lorenz gauge, the radiation field in the far field limit is
\begin{equation}
\lim_{r \to \infty} A^\mu_{\text{rad}}(x) = \frac{1}{4 \pi r} \int \frac{d\omega}{2\pi} \left(-\sum_{\lambda = \pm} \hat{\epsilon}_\lambda^\mu \, (\hat{\epsilon}_\lambda^* \cdot J(\omega, \omega \hat{\v{r}})) \, e^{- i \omega (t-r)} + \text{incoming} + \text{gauge} \right).
\end{equation}
We may again discard the incoming term at late times, and since the Poynting vector $T^{0r} = -F^{0\mu} F^{r}_{\ \, \mu}$ is gauge invariant, we may also discard the pure gauge term proportional to~$k^\mu$. Now, in the far field limit, a derivative $i \del^\mu$ yields a factor of $k^\mu = (\omega, \omega \hat{\v{r}})$, so
\begin{align}
\frac{dP(t)}{d\hat{\v{r}}} &= \lim_{r \to \infty} r^2 \, (k^0 A^\mu - k^\mu A^0) (k^r A_\mu - k_\mu A^r)\\
&= \lim_{r \to \infty} \omega^2 r^2 \, A^\mu A_\mu \label{eq:angular_power_second}\\
&= \sum_{\lambda = \pm} \left| \int \frac{d\omega}{2 \pi} \frac{e^{-i \omega (t-r)}}{4 \pi} \, \omega \, \hat{\epsilon}_\lambda^* \cdot J(\omega, \omega \hat{\v{r}}) \right|^2 \label{eq:angular_power_dist_vec}
\end{align}
where we arrived at~\eqref{eq:angular_power_second} using $k^2 = 0$ and $k \cdot \hat{\epsilon}_\pm = 0$. We may also perform the sum over polarizations using $k \cdot J = 0$ to get a compact expression for the total power, 
\begin{equation} \label{eq:angular_power_spectrum_vec}
\frac{dP(t)}{d\hat{\v{r}}} = -\left| \int \frac{d\omega}{2 \pi} \frac{e^{-i \omega (t-r)}}{4 \pi} \, \omega J^\mu(\omega, \omega \hat{\v{r}}) \right|^2.
\end{equation}

As a concrete example, we can again consider a kicked particle. Since the radiation is emitted instantaneously, it is natural to integrate~\eqref{eq:angular_power_dist} over time. We then strip off the remaining frequency integration, pairing positive and negative $\omega$, to find the energy spectrum
\begin{equation} \label{eq:scalar_kick_total}
\frac{dE}{d \omega \, d \hat{\v{r}}} = \frac{\omega^2}{16 \pi^3} |J(\omega, \omega \hat{\v{r}})|^2 = \frac{g^2}{16 \pi^3} \left( \frac{1}{\gamma (1 - \hat{\v{r}} \cdot \v{v})} - \frac{1}{\gamma' (1 - \hat{\v{r}} \cdot \v{v}')} \right)^2
\end{equation}
for $\omega \geq 0$. Similarly, integrating the vector result~\eqref{eq:angular_power_spectrum_vec} yields
\begin{equation} \label{eq:vector_kick_total}
\frac{dE}{d\omega \, d \hat{\v{r}}} = - \frac{\omega^2}{16 \pi^3} |J^\mu(\omega, \omega \hat{\v{r}})|^2 = - \frac{e^2}{16 \pi^3} \left( \frac{(1, \v{v})}{1 - \hat{\v{r}} \cdot \v{v}} - \frac{(1, \v{v}')}{1 - \hat{\v{r}} \cdot \v{v}'} \right)^2
\end{equation}
which is a well-known result (e.g.~see Eq.~(15.2) of Ref.~\cite{jackson1999classical}).

\subsubsection*{Particles in Periodic Motion}

Next, we consider the radiation emitted by a particle undergoing periodic motion with period $T = 2 \pi/\omega_0$. In the scalar case, the current and field amplitude have the form
\begin{equation} \label{eq:fourier_component_defs}
J(\omega, \v{k}) = 2 \pi \sum_n j_n(\v{k}) \, \delta(\omega - n \omega_0), \qquad a(\v{k}) = 2 \pi \sum_n a_n(\hat{\v{k}}) \, \delta(|\v{k}| - n \omega_0)
\end{equation}
where the Fourier components can be extracted by 
\begin{equation}
a_n(\hat{\v{r}}) = i j_n(n \omega_0 \hat{\v{r}}) = i\int_0^T \frac{dt}{T} \, e^{i n \omega_0 t} J(t, n \omega_0 \hat{\v{r}}).
\end{equation}
The time-averaged power distribution is
\begin{equation}
\frac{d\bar{P}}{d \hat{\v{r}}} = \int_0^T \frac{dt}{T} \frac{dP(t)}{d \hat{\v{r}}} = \frac{1}{8 \pi^2} \sum_{n > 0} n^2 \omega_0^2 |a_n(\hat{\v{r}})|^2
\end{equation}
where we paired terms with positive and negative $n$, and dropped the static $n = 0$ term, which does not contribute to the radiated power. In general, we have 
\begin{equation} \label{eq:scalar_osc_amps}
a_n(\hat{\v{r}}) = ig \int_0^T \frac{dt}{T} \, e^{i n \omega_0 (t - \hat{\v{r}} \cdot \v{z}(t))} \, \sqrt{1 - |\v{v}(t)|^2}
\end{equation}
where the final factor comes from changing variables from $\tau$ to $t$. This shows that the leading contribution to $a_n$ is of order $v_0^n$, where $v_0$ is the typical speed, so that in the nonrelativistic limit radiation is predominantly emitted at the fundamental frequency. 

For nonrelativistic sinuosidal motion $\v{z}(t) = \ell \cos(\omega_0 t) \, \hat{\v{x}}$ with $v_0 = \omega_0 \ell \ll 1$, we have 
\begin{align} \label{eq:initial_scalar_amp}
a_1(\hat{\v{r}}) = ig \int_0^T \frac{dt}{T} \, e^{i \omega_0 t} e^{- i v_0 (\hat{\v{r}} \cdot \hat{\v{x}}) \cos(\omega_0 t)} \, \sqrt{1 - v_0^2 \sin^2(\omega_0 t)} = \frac{g v_0}{2} (\hat{\v{r}} \cdot \hat{\v{x}}) + O(v_0^2).
\end{align}
This radiation is emitted primarily longitudinally, and the total averaged power is 
\begin{equation} \label{eq:scalar_larmor}
\bar{P} = \frac{g^2 \omega_0^2 v_0^2}{32 \pi^2} \int d \hat{\v{r}} \, (\hat{\v{r}} \cdot \hat{\v{x}})^2 = \frac{g^2 a_0^2}{24 \pi}
\end{equation}
where $a_0 = \omega_0 v_0$. This is simply the scalar Larmor formula.

For the vector case, we can define the components $j_n^\mu(\v{k})$ and $a_{\pm, n}(\hat{\v{k}})$ analogously to~\eqref{eq:fourier_component_defs}, and they are now related by 
\begin{equation} \label{eq:rad_amp_j}
a_{\pm, n}(\hat{\v{r}}) = (i \hat{\epsilon}_\pm)^* \cdot j_n(n \omega_0 \hat{\v{r}}).
\end{equation}
To quickly recover familiar results, we time average the power summed over helicities~\eqref{eq:angular_power_spectrum_vec},
\begin{equation}
\frac{d\bar{P}}{d \hat{\v{r}}} = -\frac{1}{8 \pi^2} \sum_{n > 0} n^2 \omega_0^2 |j^\mu_n(n \omega_0 \hat{\v{r}})|^2.
\end{equation}
The leading contribution to radiation at the $n^{\text{th}}$ harmonic is still of order $v_0^n$, since
\begin{equation}
j_n^\mu(n \omega_0 \hat{\v{r}}) = e \int_0^T \frac{dt}{T}  \, e^{i n \omega_0 (t - \hat{\v{r}} \cdot \v{z}(t))} \, (1, \v{v}(t)).
\end{equation}
Again specializing to nonrelativistic sinusoidal motion, we have
\begin{equation}
j_1^\mu(n \omega_0 \hat{\v{r}}) = e \int_0^T \frac{dt}{T} \, e^{i \omega_0 t - i v_0 (\hat{\v{r}} \cdot \hat{\v{x}}) \cos(\omega_0 t)} \, (1, - v_0 \sin(\omega_0 t) \, \hat{\v{x}}) = - \frac{i e v_0}{2} (\hat{\v{r}} \cdot \hat{\v{x}}, \hat{\v{x}}) +  O(v_0^2).
\end{equation}
Now the radiation is emitted primarily along the transverse direction, and the total is 
\begin{equation}
\bar{P} = \frac{e^2 \omega_0^2 v_0^2}{32 \pi^2} \int d \hat{\v{r}} \, |\hat{\v{r}} \times \hat{\v{x}}|^2 = \frac{e^2 \, a_0^2}{12 \pi}.
\end{equation}
which is the usual Larmor formula, since $\langle a^2 \rangle = a_0^2/2$. We can also recover the polarization content of the radiation by applying~\eqref{eq:rad_amp_j}, which shows that the radiation is linearly polarized, along the projection of $\hat{\v{x}}$ in the plane transverse to $\hat{\v{r}}$. Of course, these are all familiar results, but we are now prepared to see how they are modified at nonzero $\rho$. 

\subsection{Defining Continuous Spin Radiation}
\label{sec:radiation_CSP}

\subsubsection*{The Radiation Field and Soft Factors}

Our analysis of continuous spin radiation will closely mirror the previous subsection. We work in strong harmonic gauge, where the equation of motion~\eqref{eq:strong_harmonic_def_2} is solved by 
\begin{equation}
\Psi(\eta, x) = -\int \frac{d^3\v{k}}{(2\pi)^3} \int_{C_r} \frac{dk^0}{2\pi} \frac{J(\eta, k) e^{-ik\cdot x}}{k^2}
\end{equation}
where the contour $C_r$ passes above the poles at $k^0 = \pm |\v{k}|$. We define $\Psi_a$ with a contour $C_a$ that passes below these poles but is otherwise identical, so that both contours enclose the same essential singularities. Then the difference,
\begin{equation} 
\Psi_{\text{rad}}(\eta, x) = \Psi(\eta, x) - \Psi_a(\eta, x) = \int \frac{d^3\v{k}}{(2\pi)^3 \, 2|\v{k}|} \left( i J(\eta, k) e^{- i k \cdot x} + \text{c.c.} \right) \bigg|_{k^0 = |\v{k}|}
\end{equation}
is a pure radiation field, with support on only null momenta. We have argued in subsection~\ref{ssec:CSP_currents} that the $\Psi_{\text{rad}}$ produced by a current depends only on the function $\hat{g}$, and is universal for scalar-like or vector-like currents. This can also be seen directly by plugging the general decomposition~\eqref{eq:general_universality_decomposition} of the current into the equation above. The $k^2 X$ term vanishes since $k$ is null, and the final term is of the form $D \xi$, which corresponds to a pure gauge field. 

The mode expansion of $\Psi_{\text{rad}}$ is given by~\eqref{eq:modeexpansion}, and the gauge invariant coefficients can be found by projecting against a helicity wavefunction by~\eqref{eq:mode_projection}, 
\begin{equation}
a_h(\v{k}) = \int [d^4\eta] \, \delta'(\eta^2 + 1) \, i \psi^*_{h, k}(\eta) J(\eta, k) \bigg|_{k^0 = |\v{k}|}.
\end{equation}
We can now readily derive soft factors by considering a kicked particle with worldline~\eqref{eq:kicked_worldline}.

First, for any scalar-like current we have
\begin{equation}
J(\eta, k) = \frac{ig \, e^{- i \rho \eta \cdot p / k \cdot p}}{\gamma (\omega - \v{k} \cdot \v{v})} - \frac{ig \, e^{- i \rho \eta \cdot p' / k \cdot p'}}{\gamma' (\omega - \v{k} \cdot \v{v}')}.
\end{equation}
This corresponds to a radiation amplitude
\begin{equation}
a_h^S(\v{k}) = (2mg) \left( \frac{s_h^S(k, p')}{2 k \cdot p'} - \frac{s_h^S(k, p)}{2 k \cdot p} \right)_{k^0 = |\v{k}|}
\end{equation}
where the scalar-like soft factor is 
\begin{align}
s_h^S(k, p) &= \int [d^4\eta] \, \delta'(\eta^2 + 1) \, \psi^*_{h, k}(\eta) e^{- i \rho \eta \cdot p / k \cdot p} \\
&= \left( \frac{\epsilon_- \cdot p}{|\epsilon_- \cdot p|} \right)^h J_h\left(\rho |\epsilon_- \cdot p / k \cdot p| \right). \label{eq:scalar_soft_step}
\end{align}
Here we evaluated the integral using~\eqref{eq:helicity_identity} with $V = \rho (q - p / k \cdot p)$. Up to differences in phase and normalization conventions, this matches the scalar-like CSP soft factors from Ref.~\cite{Schuster:2013vpr}. Explicitly, defining $z = \epsilon_- \cdot p / k \cdot p$ for brevity, the leading terms are
\begin{equation}
s_h^S(k, p) = \begin{cases} \frac18 \, \rho^2 z^2 & h = 2 \\ \frac12 \rho z & h = 1 \\ 1 - \frac14 \rho^2 z^* z & h = 0 \\ -\frac12 \rho z^* & h = -1 \\ \frac18 \rho^2 (z^*)^2 & h = -2 \end{cases} \ + O(\rho^3)
\end{equation}
In general, the leading contribution to $s_h^S$ is of order $\rho^{|h|}$, so in the $\rho \to 0$ limit all the soft factors vanish except for $s_0^S = 1$, recovering the result~\eqref{eq:scalar_soft} for ordinary scalars. 

For vector-like currents, the results are identical up to the substitution $g \to (\sqrt{2} \, e / \rho) (i k \cdot \dot{z})$. Then the radiation amplitude is
\begin{equation} \label{eq:vector_csp_rad}
a_h^V(\v{k}) = e \left( \frac{s_h^V(k,p)}{k \cdot p} - \frac{s_h^V(k,p')}{k \cdot p'} \right)_{k^0 = |\v{k}|}
\end{equation}
where the vector-like soft factors are
\begin{equation}
s_h^V(k, p) = \frac{\sqrt{2} \, k \cdot p}{i \rho} s_h(k, p).
\end{equation}
This agrees with the vector-like CSP soft factors of Ref.~\cite{Schuster:2013vpr}, up to differences in conventions. As already noted there, the soft factor for $h = 0$ diverges as $1/\rho$, but this contribution cancels between the two terms of~\eqref{eq:vector_csp_rad}, and more generally it will cancel from the soft emission amplitude for any process by charge conservation. Thus discarding this term, the leading contributions are 
\begin{equation}
s_h^V(k, p) = -i k \cdot p \times \begin{cases} \frac{1}{4 \sqrt{2}} \, \rho z^2 & h = 2 \\ \frac{1}{\sqrt{2}} z - \frac{1}{8 \sqrt{2}} \rho^2 z^2 z^* & h = 1 \\ - \frac{1}{2 \sqrt{2}} \rho z z^* & h = 0 \\ -\frac{1}{\sqrt{2}} z^* + \frac{1}{8 \sqrt{2}} \rho^2 (z^*)^2 z & h = -1 \\ \frac{1}{4\sqrt{2}} \rho (z^*)^2 & h = -2 \end{cases} \ + O(\rho^3)
\end{equation}
In general, the leading physical contribution to $s_h^V$ is of order $\rho^{||h|-1|}$, so in the $\rho \to 0$ limit all the soft factors vanish except for $s^V_{\pm 1} = \mp i \hat{\epsilon}_\pm^* \cdot p$, recovering the result~\eqref{eq:vector_amplitude} for ordinary vectors (up to the phases used to define the helicity basis~\eqref{eq:helicity_wavefunctions}). 

We have therefore recovered all the physical results of Ref.~\cite{Schuster:2013vpr}, up to the tensor-like CSP soft factors, which we defer to future work. The agreement is heartening but unsurprising, since soft factors are highly constrained by symmetries. We now turn to deriving new results.

\subsubsection*{Extracting the Radiated Power}

We can compute the radiated power using the canonical stress-energy tensor~\cite{Schuster:2014hca}
\begin{align}
T^{\mu\nu} &= - g^{\mu\nu} \mathcal{L} + \int [d^4 \eta] \, \frac{\del \mathcal{L}}{\del (\del_\mu \Psi)} \, \del^\nu \Psi \\
&= - g^{\mu\nu} \mathcal{L} + \int [d^4 \eta] \, \delta'(\eta^2 + 1) \, \del^\mu \Psi \del^\nu \Psi - \frac12 \del_{\eta}^{\mu} (\delta(\eta^2 + 1) \Delta \Psi) \, \del^\nu \Psi. \label{eq:stress_energy_terms}
\end{align}
Compared to the examples in section~\ref{sec:radiation_scalar_vector}, there is a new subtlety: the stress-energy tensor is not gauge invariant, even after Belinfante improvement terms are added to yield a symmetric tensor $\Theta^{\mu\nu}$. This is not surprising since the same phenomenon already occurs in linearized gravity; it reflects the fact that both general relativity and continuous spin theories do not have local gauge invariant observables. The resolution here is exactly the same as in linearized gravity~\cite{maggiore2007gravitational}: for radiation and gauge transformations with typical wavelength $\lambda$, the average $\langle \Theta^{\mu\nu} \rangle$ of the stress-energy tensor over a spacetime region of typical size $L \gg \lambda$ is approximately gauge invariant, up to terms suppressed by powers of $\lambda/L$. Thus, one can meaningfully describe energy-momentum flow on scales longer than the wavelength. 

To see how this works, note that averaging and discarding terms suppressed by $\lambda/L$ allows us to perform integration by parts, and more generally to discard total derivative terms such as the Belinfante improvement terms, which yields $\langle \Theta^{\mu\nu} \rangle \approx \langle T^{\mu\nu} \rangle$. Furthermore, the variation of $\langle \mathcal{L} \rangle$ approximately vanishes for the same reason that the action~\eqref{eq:CSP_action} is gauge invariant. The variations of the remaining terms cancel after using the equation of motion~\eqref{eq:free_eom} and integration by parts, as shown by a tedious but straightforward calculation in appendix~\ref{app:gauge_inv}.

There is another subtlety which is unique to continuous spin fields. In section~\ref{sec:radiation_scalar_vector}, we neglected the initial and final static fields $F$ and $F'$ when computing the radiated power because their derivatives fall off as $1/r^2$. But continuous spin fields have currents which are not localized to the worldline, as discussed in subsection~\ref{sec:simple_spacetime}, leading to different behavior at large $r$. For example, the derivatives of the static field of the spatial current~\eqref{eq:Psi_S_pos} do not simply scale as $1/r^2$, but they do fall off faster than $1/r$, and thus do not contribute to radiated power in the far field. We do not know how to prove this in general, partly because our analysis is framed in terms of fields with a high degree of gauge redundancy. For example, the temporal current depends explicitly on a particle's past history, so that even the ``static'' field~\eqref{eq:Psi_T_result} depends explicitly on time and an early time cutoff. These features obscure the asymptotic behavior of the field and its derivatives. In lieu of a proof, we make the physically well-motivated conjecture that $\Psi_{\text{rad}}(\eta, x)$, which encodes ``on-shell'' null modes, provides the sole contribution to the far-field power, for all physically sensible currents satisfying the continuity condition.

We can now derive results. Our discussion of gauge invariance implies that we may work in any gauge, as long as we only compute spacetime averaged quantities. We use harmonic gauge, where only the second term of~\eqref{eq:stress_energy_terms} contributes to the Poynting vector,
\begin{equation}
T^{0r} = \int d^4 \eta \, \delta'(\eta^2 + 1) \, \del^0 \Psi \del^r \Psi.
\end{equation}
In analogy with~\eqref{eq:phi_rad_two}, the part of $\Psi_{\text{rad}}$ containing outgoing radiation in the far field limit is 
\begin{equation}
\lim_{r \to \infty} \Psi_{\text{rad}}(\eta, x) \supset \frac{1}{4 \pi r} \int \frac{d\omega}{2 \pi} \sum_h \psi_{h, k}(\eta) \int [d^4 \eta'] \, \delta'(\eta'^2 + 1) \, \psi^*_{h, k}(\eta') J(\eta', k)
\end{equation}
where we defined $k = (\omega, \omega \hat{\v{r}})$ for brevity. As discussed above, we use this piece alone to compute the far field radiated power flux, omitting contributions from ``static'' fields. Following the same steps that led to~\eqref{eq:angular_power_dist} and applying the orthogonality relation~\eqref{eq:overlap_def} yields the radiated power in each helicity mode,
\begin{equation} \label{eq:csp_scalar_power_h}
\frac{dP_h(t)}{d \hat{\v{r}}} = \left|\int \frac{d\omega}{2 \pi} \frac{\omega \, e^{i \omega (r-t)}}{4 \pi} \int d^4 \eta \, \delta'(\eta^2 + 1) \, \psi_{h,k}^*(\eta) J(\eta, k) \right|^2.
\end{equation}
We can sum over $h$ using the completeness relation~\eqref{eq:completeness}, giving the total 
\begin{equation} \label{eq:csp_scalar_power}
\frac{dP(t)}{d \hat{\v{r}}} = \int d^4 \eta \, \delta'(\eta^2 + 1) \left| \int \frac{d\omega}{2 \pi} \frac{e^{i \omega (r-t)}}{4 \pi} \, \omega J(\eta, k) \right|^2.
\end{equation}
The total radiated power distribution~\eqref{eq:csp_scalar_power} is Lorentz invariant, but its distribution in helicity~\eqref{eq:csp_scalar_power_h} is not. This reflects the fact that the helicity of a continuous spin particle is not Lorentz invariant, as mentioned in section~\ref{sec:primer}.

To render~\eqref{eq:csp_scalar_power_h} and~\eqref{eq:csp_scalar_power} gauge invariant and physically meaningful, it suffices to average in time. This fact follows from the same argument as in the gravitational wave literature~\cite{maggiore2007gravitational}: in the limit $r \to \infty$, evaluating the power emitted through any nonzero solid angle automatically involves a large spatial average over the transverse directions. Furthermore, since the power is written in terms of fields that are purely outgoing spherical waves, an average in $t$ is equivalent to an average in $r$, so no further averaging is needed. 

\subsection{Computing Continuous Spin Radiation}
\label{sec:radiation_computation}

As a first example, we can integrate over all times to find the radiated energy spectrum for a kicked particle. In terms of the radiation amplitudes defined above, we have
\begin{equation}
\frac{dE}{d \omega \, d \hat{\v{r}}} = \sum_h \frac{\omega^2}{16 \pi^3} \times \begin{cases} |a_h(\omega \hat{\v{r}})|^2 & \text{scalar-like} \\ |a_h^V(\omega \hat{\v{r}})|^2/2 & \text{vector-like} \end{cases}
\end{equation}
where the factor of $1/2$ is due to the normalization of the frame vectors $\epsilon_\pm^\mu$. As expected, these results reduce to~\eqref{eq:scalar_kick_total} and~\eqref{eq:vector_kick_total} as $\rho \to 0$.

Our main interest in this section is on particles in periodic motion, for which we can compute a gauge invariant radiated power by averaging over a period. As in section~\ref{sec:radiation_scalar_vector}, the currents and amplitudes satisfy
\begin{equation} \label{eq:csp_fourier_component_defs}
J(\eta, \omega, \v{k}) = 2 \pi \sum_n j_n(\eta, \v{k}) \delta(\omega - n \omega_0), \qquad a_h(\v{k}) = 2 \pi \sum_n a_{h,n}(\hat{\v{k}}) \delta(|\v{k}| - n \omega_0)
\end{equation}
and the Fourier components can be extracted by 
\begin{equation}
a_{h,n}(\hat{\v{r}}) = i \int [d^4\eta] \, \delta'(\eta^2 + 1) \, \psi^*_{h,k_n}(\eta) j_n(\eta, k_n)
\end{equation}
where we define $k_n = n \omega_0 (1, \hat{\v{r}})$. The time-averaged power distribution is
\begin{equation} \label{eq:helicity_power}
\frac{d\bar{P}_{h,n}}{d \hat{\v{r}}} = \frac{n^2 \omega_0^2}{8 \pi^2} \, |a_{h,n}(\hat{\v{r}})|^2.
\end{equation}
With this single expression we can compute the universal radiated power, differential in direction $\hat{\v{r}}$, helicity $h$, and harmonic $n > 0$, for both scalar-like and vector-like currents. 

First, for any scalar-like current,
\begin{align}
a_{h,n}(\hat{\v{r}}) &= ig \int_0^T \frac{dt}{T} \, e^{i n \omega_0 (t - \hat{\v{r}} \cdot \v{z}(t))} \, \sqrt{1 - |\v{v}(t)|^2} \int [d^4 \eta] \, \delta'(\eta^2 + 1) \, \psi_{h,k_n}^*(\eta) e^{- i \rho \eta \cdot \dot{z}(t) / k_n \cdot \dot{z}(t)} \label{eq:a_hn_original} \\
&= ig \int_0^T \frac{dt}{T} \, e^{i n \omega_0 (t - \hat{\v{r}} \cdot \v{z}(t))} \, \sqrt{1 - |\v{v}(t)|^2} \left( \frac{\epsilon_- \cdot \dot{z}}{|\epsilon_- \cdot \dot{z}|} \right)^h J_h\left(\rho |\epsilon_- \cdot \dot{z} / k_n \cdot \dot{z}| \right)
\end{align}
where we performed the integral using the same logic as in~\eqref{eq:scalar_soft_step}. This can be written more transparently by writing $\dot{z} = (1, \v{v}) = (1, v_r \hat{\v{r}} + \v{v}_\perp)$, and choosing frame vectors with $\epsilon_\pm^\mu = (0, \bm{\epsilon}_\pm)$, so that $\bm{\epsilon}_\pm \cdot \hat{\v{r}} = 0$ and $|\bm{\epsilon}_\pm \cdot \v{v}| = v_\perp$. This yields 
\begin{equation} \label{eq:a_hn_scalar_final}
a_{h,n}(\hat{\v{r}}) = ig \int_0^T \frac{dt}{T} \, e^{i n \omega_0 (t - \hat{\v{r}} \cdot \v{z})} \, \sqrt{1 - v^2} \left(- \bm{\epsilon}_- \cdot \hat{\v{v}}_\perp \right)^h J_h\left( \frac{\rho v_\perp}{n \omega_0 (1 - v_r)} \right).
\end{equation}
Notice that while the original expression~\eqref{eq:a_hn_original} had direct dependence on $\rho/\omega_0$, our final result only depends on the combination $\rho v_0 / \omega_0$, which is physically reasonable since this parameter controls the mixing of helicity states under boosts, as discussed in section~\ref{sec:primer_particles}. Thus, the amplitude can be viewed as a series in the dimensionless variables $v_0$ and $\rho \ell$, where $\ell = v_0 / \omega_0$ is the typical length scale of the particle's path. 

We have seen that for nonrelativistic sinusoidal motion, the amplitude~\eqref{eq:initial_scalar_amp} for ordinary $h = 0$, $n = 1$ scalar radiation begins at order $v_0$. By contrast, the leading continuous spin corrections begin at \textit{zeroth} order in $v_0$, and are
\begin{align}
a_{h,n}(\hat{\v{r}}) &= ig \int_0^T \frac{dt}{T} \, e^{i n \omega_0 t} \left(- \bm{\epsilon}_- \cdot \hat{\v{v}}_\perp \right)^h J_h\left( \frac{\rho v_\perp}{n \omega_0} \right) + O(v_0) \\
&= ig \int_0^{2 \pi} \frac{d\phi}{2\pi} \, e^{i n \phi} \left(- e^{i \alpha} \, \text{sign}(\sin \phi) \right)^h J_h\left( \frac{\rho \ell \sin \theta}{n} \left|\sin \phi\right| \right) + O(v_0)
\end{align}
where we changed variables to $\phi = \omega_0 t$, $\theta$ is the angle of $\hat{\v{r}}$ to the direction of motion $\hat{\v{x}}$, and the constant phase $e^{i \alpha}$ depends on the phase convention for the polarization vector $\bm{\epsilon}_-$. Discarding the irrelevant phases by taking the magnitude yields 
\begin{equation} \label{eq:nr_scalar_final}
|a_{h,n}(\hat{\v{r}})| = g \left| \int_0^{2 \pi} \frac{d\phi}{2\pi} \, e^{i n \phi} J_h\left( \frac{\rho \ell \sin \theta}{n} \sin \phi \right) \right| + O(v_0).
\end{equation}
Thus, for nonrelativistic sinuosidal motion the leading contribution to radiation at helicity $h > 0$ has order $(\rho \ell)^{h}$ and appears at harmonics $n = h, h-2, \ldots$, while higher order terms in $\rho \ell$ can also contribute to higher harmonics. 

\begin{figure}[t]
\includegraphics[width=0.49\columnwidth]{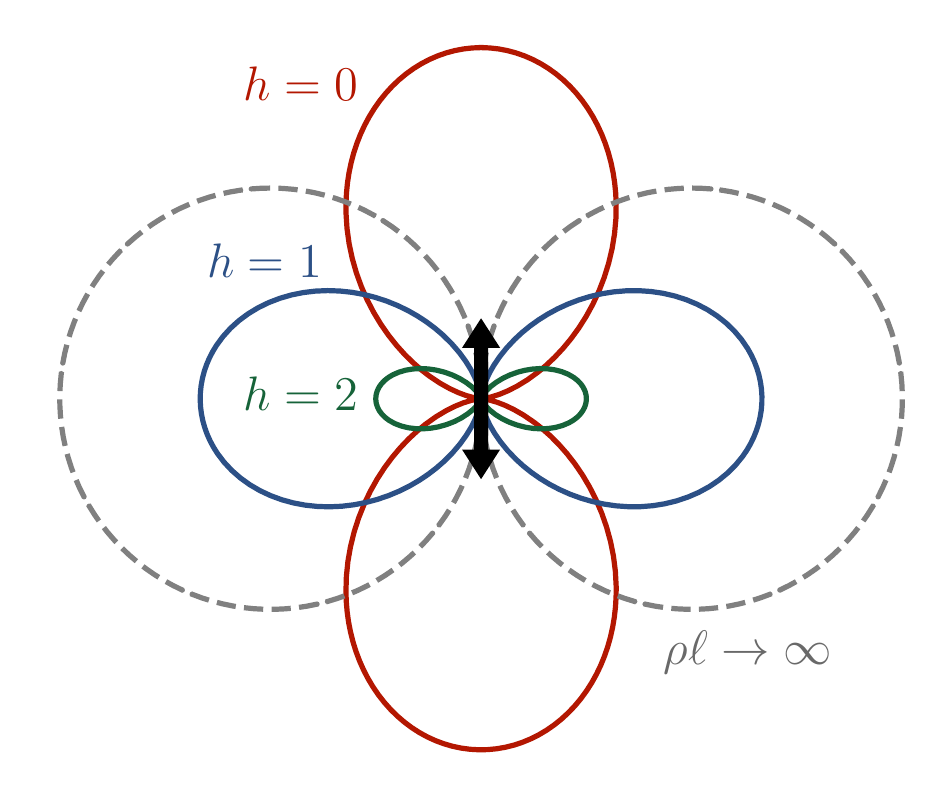} \includegraphics[width=0.49\columnwidth]{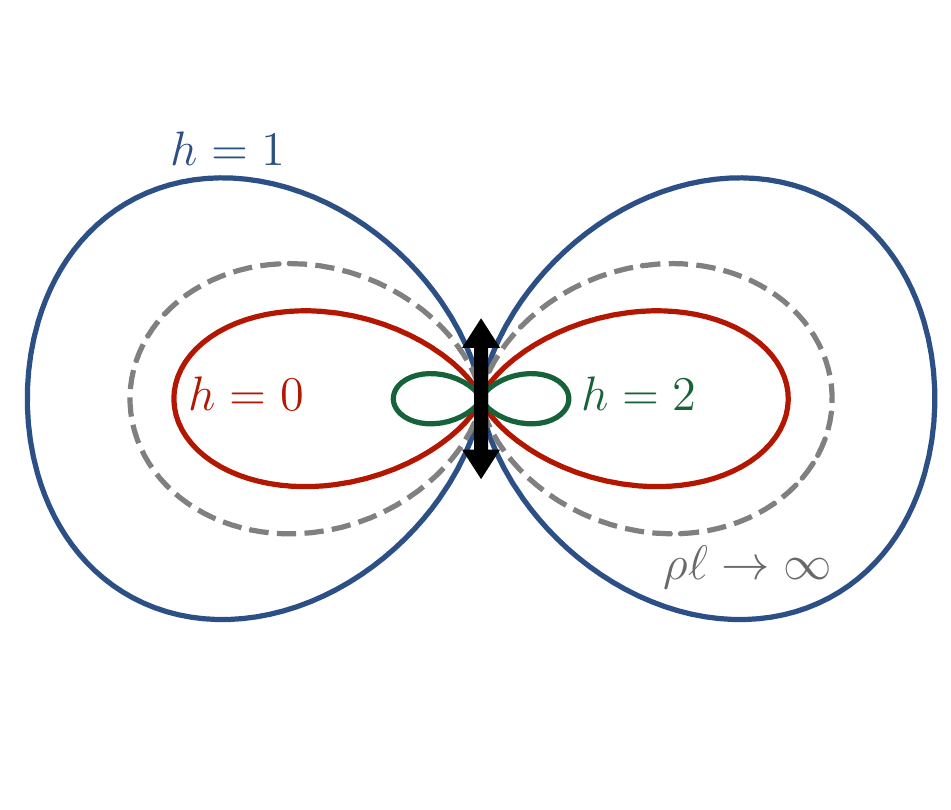}
\caption{Angular distribution of helicity $h$ radiation (not to scale) from a particle in nonrelativistic sinusoidal motion with amplitude $\ell$ and $\rho \ell \ll 1$, for a scalar-like (left) or vector-like (right) current. The dashed curve shows the total power emission in all helicities in the deep infrared limit $\rho \ell \to \infty$. 
}
\label{fig:scalar_rad_angle}
\end{figure}

The resulting radiated power distribution, illustrated at left in Fig.~\ref{fig:scalar_rad_angle}, is 
\begin{equation}
\frac{d\bar{P}_{h,n}}{d \hat{\v{r}}} = \frac{g^2 \omega_0^2}{32 \pi^2} \times \begin{cases} v_0^2 \cos^2 \theta \, (1 - (\rho \ell \sin \theta)^2/8) & h = 0, n = 1 \\
(\rho \ell \sin \theta)^2/4 & |h| = 1, n = 1 \\ (\rho \ell \sin \theta)^4/1024 & |h| = 2, n = 2 \end{cases} + \ldots
\end{equation}
where we show the leading $\rho$-dependent corrections for a few $h$ and $n$. Note that the familiar scalar radiation is primarily emitted longitudinally, while the $h \neq 0$ radiation is primarily emitted in the transverse directions, due to the dependence of the amplitude on $\v{v}_\perp$. As expected, the familiar scalar radiation dominates for small $\rho \ll \omega_0$. 

\begin{figure}[t]
\includegraphics[width=0.49\columnwidth]{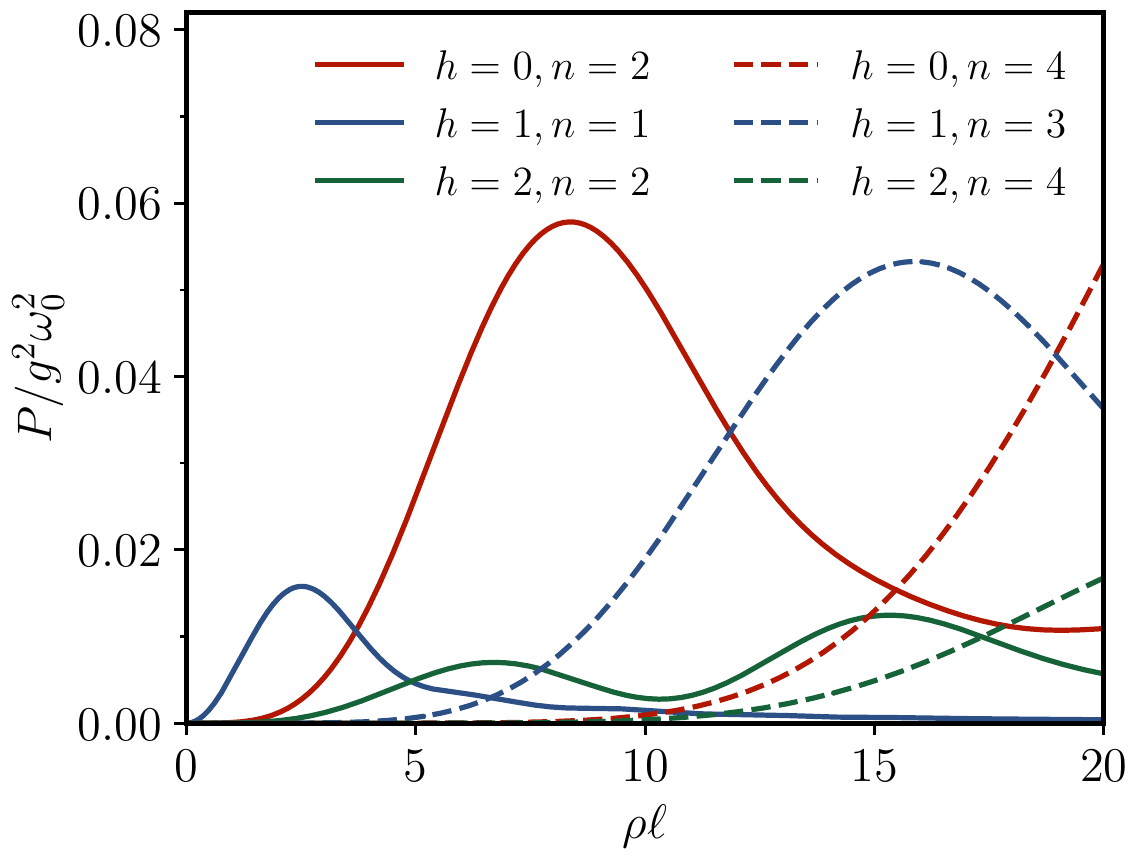}
\includegraphics[width=0.49\columnwidth]{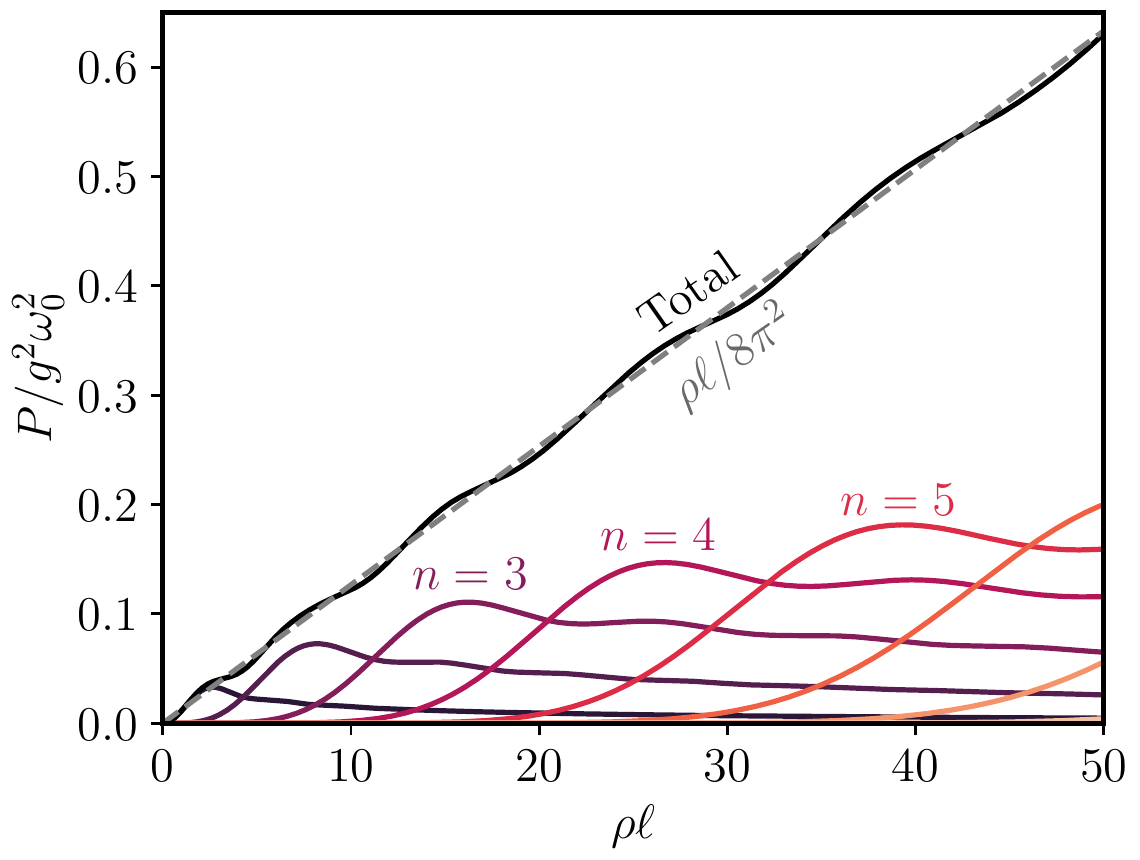}
\caption{Radiation from a particle with a scalar-like current in nonrelativistic sinusoidal motion of amplitude $\ell$. The left panel shows the power emitted in each given harmonic and helicity, evaluated with~\eqref{eq:nr_scalar_final}. It does not show the ordinary, $\rho$-independent scalar radiation at $h = 0$, $n = 1$, as this is suppressed by $v_0^2$. The right panel shows the power summed over helicities, evaluated using~\eqref{eq:summed_power_final}. As discussed in the text, it implies the total power is well-behaved in the limit of low accelerations.}
\label{fig:scalar_rad_power}
\end{figure}

It is also interesting to consider how the radiated power behaves in the deep infrared limit $\rho \ell \to \infty$. As shown at left in Fig.~\ref{fig:scalar_rad_power}, the radiated power in each helicity and harmonic initially increases monotonically with $\rho$, then oscillates and decreases. This is a consequence of the rapid oscillation of the Bessel function in~\eqref{eq:nr_scalar_final}; roughly evaluating the integral with the method of stationary phase, we find that at large $\rho \ell$ the radiated power is distributed democratically among helicities $h \lesssim n$, and dominated by harmonics $n \lesssim \sqrt{\rho \ell}$.

To confirm this behavior numerically, it is easiest to sum the power~\eqref{eq:helicity_power} over all helicities using the completeness relation~\eqref{eq:completeness}, which yields
\begin{align}
\frac{d \bar{P}}{d \hat{\v{r}}} &= \frac{g^2 \omega_0^2}{8 \pi^2} \sum_{n > 0} n^2 \! \int [d^4 \eta] \, \delta'(\eta^2 + 1) \left| \int_0^T \frac{dt}{T} \, e^{i n \omega_0 (t - \hat{\v{r}} \cdot \v{z}(t))} \sqrt{1 - |\v{v}(t)|^2} e^{- i \rho \eta \cdot \dot{z}(t) / k_n \cdot \dot{z}(t)} \right|^2 \\
&= \frac{g^2 \omega_0^2}{8 \pi^2} \sum_{n > 0} n^2 \! \int [d^4 \eta] \, \delta'(\eta^2 + 1) \left| \int_0^T \frac{dt}{T} \, e^{i n \omega_0 t} e^{- i \rho (\eta^0 v_r(t) - \etav \cdot \v{v}(t)) / n \omega_0} \right|^2 \\
&= \frac{g^2 \omega_0^2}{8 \pi^2} \sum_{n > 0} n^2 \! \int_0^T \frac{dt}{T} \int_0^T \frac{dt'}{T} \, e^{i n \omega_0 (t - t')} \, J_0\left( \frac{\rho |\v{v}_\perp(t) - \v{v}_\perp(t')|}{n \omega_0} \right) \label{eq:summed_power_final}
\end{align}
where we took the nonrelativistic limit, cancelled a phase, and evaluated the $\eta$ integral using~\eqref{eq:integral_proof}. In the special case of sinusoidal motion, one can evaluate the integrals in terms of hypergeometric functions, yielding
\begin{equation} \label{eq:scalar_csp_limit}
\lim_{\rho \ell \to \infty} \frac{d\bar{P}}{d\hat{\v{r}}} = \frac{g^2 \omega_0^2}{8 \pi^4} \, \rho \ell \sin \theta.
\end{equation}

This scaling, shown at right in Fig.~\ref{fig:scalar_rad_power}, has an important physical consequence. Working in terms of the independent measurable quantities $v_0$ and $a_0$, where $a_0 = \omega_0 v_0$ is the typical acceleration, the infrared limit $\rho \ell \gg 1$ is precisely the low acceleration regime $a_0 \ll \rho v_0^2$. Combining~\eqref{eq:scalar_larmor} and~\eqref{eq:scalar_csp_limit} yields
\begin{equation}
\bar{P} = \begin{cases} g^2 a_0^2/24 \pi & a_0 \gg \rho v_0^2 \\ g^2 \rho a_0 / 8 \pi^2 & a_0 \ll \rho v_0^2 \end{cases}
\end{equation}
so that the radiated power smoothly goes to zero as the acceleration vanishes, even though radiation is emitted into an unbounded number of helicities. This is a general phenomenon: as shown in appendix~\ref{app:gauge_inv}, one can extract the scaling of~\eqref{eq:scalar_csp_limit} from applying a stationary phase approximation to~\eqref{eq:summed_power_final} for any smooth trajectory. This is another striking example of how continuous spin theories are well-behaved compared to naive expectations.

As in section~\ref{sec:radiation_CSP}, we can easily convert our results to vector-like currents by replacing $g$ with $(\sqrt{2} \, e / \rho) (i k \cdot dz/d\tau)$, giving similar results. The analogue of~\eqref{eq:a_hn_scalar_final} is
\begin{equation} 
a_{h,n}(\hat{\v{r}}) = -\frac{\sqrt{2} \, e n v_0}{\rho \ell} \int_0^T \frac{dt}{T} \, e^{i n \omega_0 (t - \hat{\v{r}} \cdot \v{z})} \, (1 - v_r) \left(- \bm{\epsilon}_- \cdot \hat{\v{v}}_\perp \right)^h J_h\left( \frac{\rho v_\perp}{n \omega_0 (1 - v_r)} \right).
\end{equation}
The $\rho \to 0$ limit appears to diverge for $h = 0$, due to the constant term in the Bessel function. However, that contribution vanishes since the integrand is a total derivative. 

\begin{figure}[t]
\includegraphics[width=0.49\columnwidth]{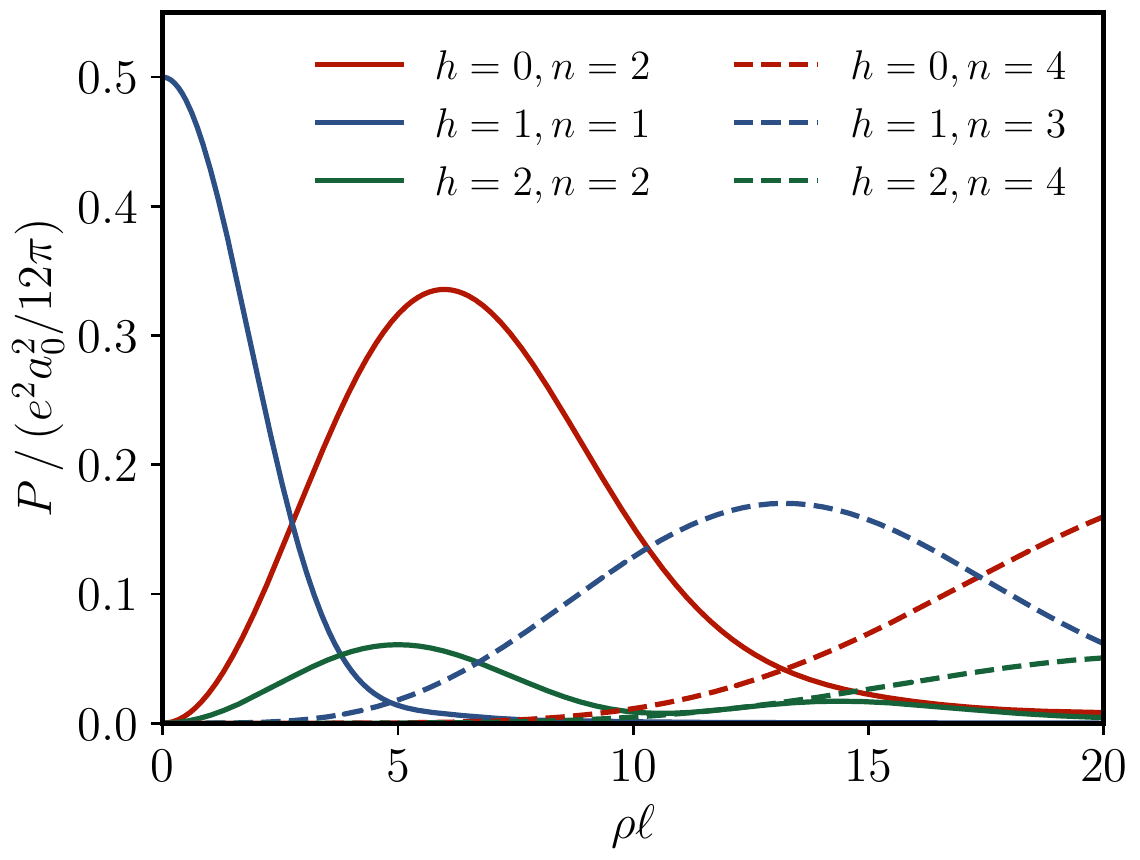}
\includegraphics[width=0.49\columnwidth]{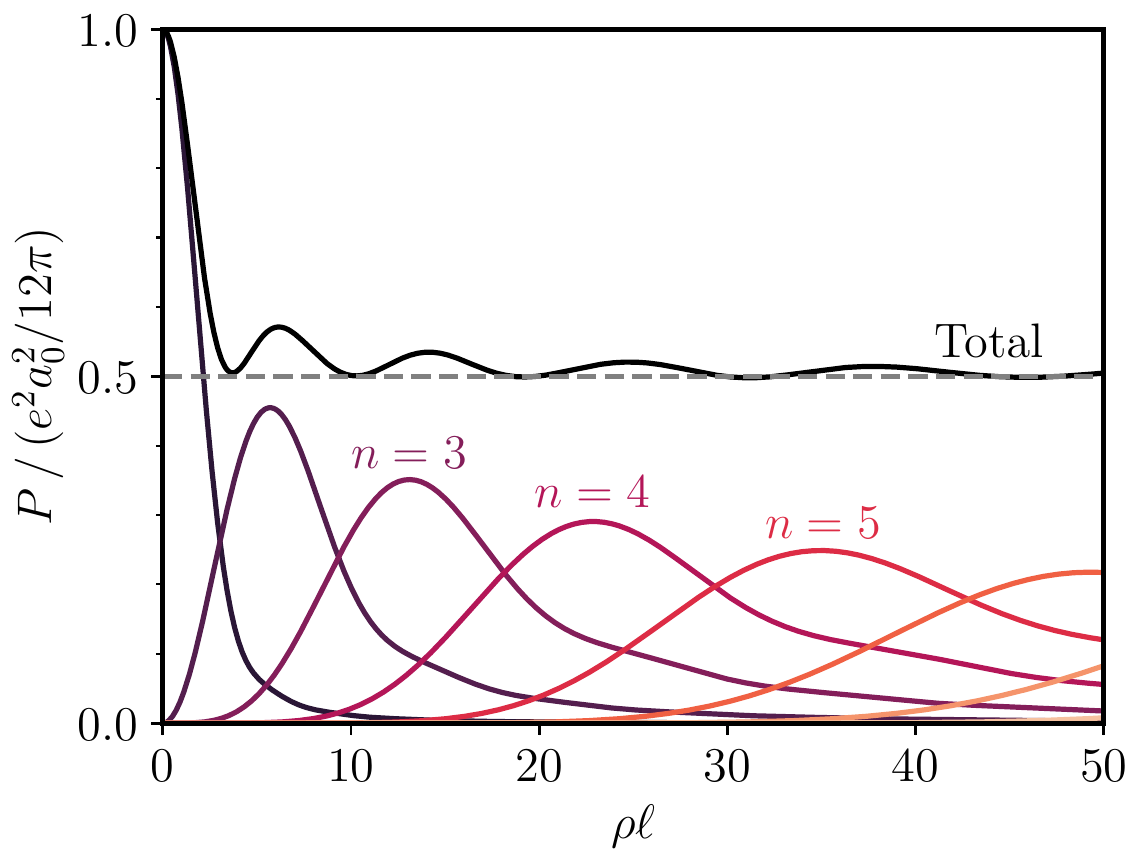}
\caption{Radiation from a particle with a vector-like current in nonrelativistic sinusoidal motion of amplitude $\ell$, normalized to the power at $\rho = 0$. We evaluate~\eqref{eq:nr_vector_final} to find the power emitted in each harmonic and helicity (left), and~\eqref{eq:summed_vector_power_final} to find the power summed over helicities (right).}
\label{fig:vector_rad_power}
\end{figure}

The integral may again be expanded in $\rho \ell$ and $v_0$. Taking the nonrelativistic limit for sinusoidal motion by removing terms subleading in $v_0$, as in~\eqref{eq:nr_scalar_final}, yields 
\begin{equation} \label{eq:nr_vector_final}
|a_{h,n}(\hat{\v{r}})| = \sqrt{2} \, e n v_0 \left| \frac{1}{\rho \ell} \int_0^{2 \pi} \frac{d\phi}{2\pi} \, e^{i n \phi} J_h\left( \frac{\rho \ell \sin \theta}{n} \sin \phi \right) + O(v_0) \right|.
\end{equation}
Unlike the scalar-like case, this expression also contains the usual $\rho$-independent amplitude, since it appears at the same order in $v_0$ as the leading corrections. The leading contribution to the radiation amplitude at helicity $h$ is of order $v_0 (\rho \ell)^{||h| - 1|}$, and the radiated power is 
\begin{equation}
\frac{d\bar{P}_{h,n}}{d \hat{\v{r}}} = \frac{e^2 \omega_0^2 v_0^2}{64 \pi^2} \times \begin{cases} 
\frac{1}{16} (\rho \ell)^2 \sin^4 \theta & h = 0, n = 2 \\ \sin^2 \theta - \frac{3}{16} (\rho \ell)^2 \sin^4 \theta & |h| = 1, n = 1 \\ \frac{1}{64} (\rho \ell)^2 \sin^4 \theta & |h| = 2, n = 2 \end{cases} + \ldots.
\end{equation}
All the radiation is primarily emitted in the transverse directions, as shown at right in Fig.~\ref{fig:scalar_rad_angle}. Integrating this equation over angles yields our earlier stated result~\eqref{eq:introshowcase:rad}.

Numeric results extending to large $\rho \ell$, where high harmonics and many helicities contribute, are shown in Fig.~\ref{fig:vector_rad_power} and earlier in Fig.~\ref{fig:intro_rad_power}. Again, to find the total power emitted it is easiest to sum over helicities. The analogue of~\eqref{eq:summed_power_final} is
\begin{equation} \label{eq:summed_vector_power_final}
\frac{d\bar{P}}{d\hat{\v{r}}} = \frac{e^2 \omega_0^2 v_0^2}{4 \pi^2} \sum_{n > 0} \frac{n^4}{(\rho \ell)^2} \int_0^T \frac{dt}{T} \int_0^T \frac{dt'}{T} \, e^{i n \omega_0 (t - t')} \, J_0\left( \frac{\rho |\v{v}_\perp(t) - \v{v}_\perp(t')|}{n \omega_0} \right) 
\end{equation}
which can again be evaluated in terms of hypergeometric functions, yielding the simple result
\begin{equation}
\lim_{\rho \ell \to \infty} \frac{d\bar{P}}{d\hat{\v{r}}} = \frac{e^2 \omega_0^2 v_0^2}{64 \pi^2} \, \sin^2 \theta.
\end{equation}
This is precisely half the radiated power distribution for an ordinary vector field, though the radiation is now distributed over many helicities. Again, for a general smooth motion one can derive the same result, up to a different numeric constant, by applying a stationary phase approximation to~\eqref{eq:summed_vector_power_final}. This again illustrates how continuous spin physics is surprisingly well-behaved, yielding sensible results for arbitrary accelerations. 

\addtocontents{toc}{\vspace{-0.2\baselineskip}}
\section{Discussion and Future Directions}
\label{sec:conclusion}
\subsubsection*{Is Continuous Spin a Fundamental Feature of Nature?}

All observables in the continuous spin theories computed in this work reduce to those of familiar theories when $\omega \gg \rho$. This motivates a bold conjecture: all massless degrees of freedom in nature could have continuous spin. One could begin exploring this possibility along two independent directions. First, the Abelian gauge theory of this paper must be generalized to theories with nonzero $\rho$ that reduce as $\rho\to 0$ to non-Abelian gauge theories and general relativity. Second, one must embed the Higgs mechanism within a continuous spin theory, which could initially be pursued in the context of Abelian Higgs models.

Intriguingly, a ``continuous spin Standard Model'' may provide insight on the gauge hierarchy problem. In familiar theories, a massless scalar field that mediates a $1/r$ potential is unstable against radiative corrections from interactions with other massive degrees of freedom. Continuous spin fields with scalar correspondence behave just like such fields as $\rho \to 0$, but have a gauge symmetry which forbids a mass term, and thus could render a light scalar technically natural. This simple observation strongly motivates developing interacting continuous spin theories to the point where one can compute radiative corrections. The result is not clear a priori; for instance, the gauge symmetry might be anomalous at the quantum level, or supersymmetry might be required. Or perhaps an entirely new mechanism, related to analytic structure in $\eta$-space, will play a role. Even if $\rho$ is zero in our universe, such investigations may inspire new resolutions to the hierarchy problem.

Investigating the possibility that the graviton is a CSP requires developing a non-Abelian continuous spin theory with tensor correspondence, i.e.~one which reduces to general relativity as $\rho \to 0$. This challenging task is motivated by the cosmological constant problem. In general relativity, a cosmological constant term is allowed by general covariance, and quantum effects are generically expected to produce large corrections to such a term. However, continuous spin fields have a larger gauge symmetry which forbids this term when $\rho = 0$, which could render a small cosmological constant technically natural. A first step toward developing this idea would be to construct currents which reduce to the contribution of vacuum energy to $T_{\mu\nu}$ in the limit $\rho \to 0$. An even more radical possibility is that the cosmological constant is completely forbidden, and the accelerated expansion of the universe instead results from $\rho$-dependent corrections to the static force law. 

\subsubsection*{Future Theoretical Developments}

We highlighted some long-term goals of a continuous spin research programme above; we now list some theoretical developments which can serve as intermediate steps, and may also open new avenues for experimental measurements and detection.

First, in this paper we have focused on spinless matter particles, since these give rise to the leading contributions to long-distance forces and radiation. It would be interesting to see how our universal decomposition of currents is modified for matter particles with spin. Understanding the role of spin would also enable precision tests using nuclear or electronic spin precession. 

Second, we have only computed classical observables in this work, but nonzero $\rho$ could also produce distinctive signatures at microscopic scales, such as forbidden atomic transitions through emission of partner polarizations, or modifications to atomic energy levels. As already discussed in Ref.~\cite{Schuster:2014hca}, the free continuous spin field is straightforward to quantize, and the mode amplitudes $a^*_h(\v{k})$ and $a_h(\v{k})$ simply become raising and lowering operators for field modes with momentum $\v{k}$ and helicity $h$. There are well-known techniques for computing quantum amplitudes for matter particles and gauge fields (e.g.~see Refs.~\cite{Strassler:1992zr,Schubert:2001he,Edwards:2019eby}) which in fact predate~\cite{PhysRev.80.440} the traditional approach using matter fields. These formalisms could be a starting point for computing matter-CSP transition amplitudes from our action~\eqref{eq:introaction}. It is not obvious if such quantized theories will be fully consistent, but the completeness relation \eqref{eq:completeness} strongly suggests that correct unitarity cuts will be obtained at least for tree-level CSP-mediated matter scattering. 

Third, as already discussed in subsection~\ref{sec:eom_causality}, it would be interesting to explore causality more thoroughly, as well as local completions of the action involving intermediate fields. This may shed light on the physical interpretation of our delocalized effective currents and might be necessary to properly define and compute radiative corrections.

Fourth, it would be interesting to tie our work to the broader field theory literature. For instance, one could generalize our interacting formalism to supersymmetric continuous spin fields, or investigate its relation to interactions in higher spin theory, which also often require towers of arbitrarily high helicity modes. In addition, while massless CSPs do not appear in the perturbative treatment of the familiar bosonic string~\cite{Font:2013hia}, it would be interesting to see if they can emerge from other types of string theory. 

Fifth, we have focused on the coupling of continuous spin fields to matter particles, and it would also be interesting to consider their coupling to matter fields. As discussed in appendix~\ref{sec:fields}, existing currents built from matter fields do not obey gauge invariance, do not contain the most relevant minimal couplings, and appear to have vanishing coupling to radiation. We suspect that suitable currents can be built from matter fields, but they must be either nonlocal or involve additional intermediate fields. Understanding this structure is likely a prerequisite to developing non-Abelian continuous spin theories, which require matter-like couplings for the continuous spin field itself. 

Finally, since CSPs are known to exist in (A)dS spaces, it would be interesting to study continuous spin fields in curved spacetimes, such as black holes. For a black hole with radius $r_s \ll 1/\rho$, a photon CSP should be equivalent to one particle of each integer helicity, yielding an unambiguous deviation from the ordinary Hawking radiation rate. More spectacularly, since $\rho$ provides a length scale, a continuous spin field might be exponentially amplified around a black hole through the superradiant instability. Alternatively, our formalism could simply be inconsistent in curved spacetime. It might even be that the only fully consistent gravitational theories of interacting CSPs require the graviton to be a CSP as well. 

\subsubsection*{Towards Experimental Tests of the Photon and Graviton Spin Scale}

In the near-term, the formalism explored in this work can already be used to devise probes of the photon or graviton spin scale. These phenomenological tests should be largely independent of the ultimate origin of the spin scale.

If the photon has a nonzero spin scale $\rho_\gamma$, then the familiar charged particles in nature must couple to it through vector-like currents, leading to distinctive deviations from ordinary electromagnetism. One promising avenue is to try to detect the ``nearest neighbor'' partner polarizations with $h = 0$ or $|h| = 2$, which can be both produced and detected by antennas and electromagnetic resonators. For example, if an electromagnetic cavity is excited, then it will emit in the partner polarizations, which pass out through its walls and can be detected by a second cavity. The partner polarizations could also be produced deep within the Sun. Anomalous cooling caused by such emission should set a bound on $\rho$. Even below this bound the partner modes can freely propagate to a ``helioscope'' detector on Earth, where they would appear as X-rays which penetrate conventional shielding. While we have not yet performed detailed calculations, we can certainly say that $\rho_\gamma \lesssim 10^{-6} \, \mathrm{eV}$ due to the nonobservation of strong deviations from ordinary electromagnetism in microwave/radiofrequency technology and we roughly estimate stellar cooling constraints imply $\rho_\gamma \lesssim 10^{-9} \, \mathrm{eV}$. 

Our results strongly suggest that continuous spin fields have well-behaved thermodynamics, and furthermore make it possible to probe $\rho_\gamma$-dependent corrections to ordinary thermodynamics. For example, nonzero $\rho_\gamma$ changes the effective number of light degrees of freedom in the early universe, which might be constrained through precision cosmological measurements, or detected through the cosmic background radiation in the partner polarizations. 

In these respects, searches for the photon's partner polarizations bear a strong resemblance to ongoing searches for weakly coupled, ultralight fields, such as axions, dilatons, and dark photons. One crucial difference is that deviations from radiation-induced forces are purely velocity-dependent. Another difference is that all $\rho$-dependent effects are enhanced in the infrared, with forces from nearest neighbor polarizations suppressed by $\rho_\gamma/\omega$, which might enhance signals for telescopes sensitive to lower frequency radiation. Very low momentum scales could be probed via long distance modifications of force laws, such as from astrophysical or cosmological magnetic fields, though here the deviations are not universal. 

We suspect it is possible to probe the spin scale $\rho_g$ of the graviton within the context of linearized general relativity, by suitably adapting the tensor-like current discussed at the end of subsection~\ref{ssec:CSP_currents}. Using the formalism developed in this work, it would then be relatively straightforward to compute $\rho_g$-dependent corrections to gravitational radiation, recover the tensor-like soft factors derived in Ref.~\cite{Schuster:2013vpr}, and investigate deviations from ordinary gravitational dynamics on galactic and cosmological scales.

We have discussed a broad set of future directions, but it is also likely that the most exciting new ideas are still waiting to be found. This subject is still in its infancy, and we hope this work will serve as a solid foundation for many developments to come.

\subsection*{ACKNOWLEDGMENTS}

We thank David Cyncynates, Shayarneel Kundu, Huiting Liu, Giacomo Marocco, and Gowri Sundaresan for discussions and comments on the draft. We thank Lenny Evans for early collaboration on related work. The authors were supported by the U.S. Department of Energy under contract number DE-AC02-76SF00515 while at SLAC, and Kevin Zhou was supported by the NSF GRFP under grant DGE-1656518. 

\newpage
\appendix

\addtocontents{toc}{\vspace{-0.2\baselineskip}}
\section{Vector Superspace Integration}
\label{sec:integrals}
In this section, we define vector superspace integrals of the form
\begin{equation} \label{eq:initial_eta_integrals}
\int [d^4\eta] \, \delta(\eta^2 + 1) \, F(\eta), \qquad \int [d^4 \eta] \, \delta'(\eta^2 + 1) \, F(\eta)
\end{equation}
for arbitrary analytic functions $F(\eta)$, and present two equivalent procedures for evaluating them. These approaches were introduced in Ref.~\cite{Schuster:2014hca}, but we include a streamlined summary to make this paper self-contained and provide complementary derivations. We also derive several powerful new identities \eqref{eq:master_identity},~\eqref{eq:helicity_identity}, and \eqref{eq:completeness} used throughout the paper.

Integrals over an ordinary measure $d^4\eta$ diverge, as the hyperboloid $\eta^2+1=0$ has infinite volume. Heuristically, the regulated measure $[d^4\eta]$ is defined by factoring the volume out,
\begin{equation} \label{eq:defd4eta}
d^4 \eta = [d^4 \eta] \int d^4\eta \, \delta(\eta^2+1).
\end{equation}
While this is merely a formal definition, requiring that the integration results obey sensible constraints, such as Lorentz covariance, fully determines all integrals of the form~\eqref{eq:initial_eta_integrals}. This approach is pursued in App.~\ref{app:genFn} and used to write the integrals in terms of generating functions, and derive new identities in App.~\ref{app:usefulIdentities}. Alternatively, the regulated measure can be concretely defined by analytic continuation to Euclidean signature, as shown in App.~\ref{app:anacont}.

These two formulations are completely equivalent, but complementary. The generating function approach can always be straightforwardly applied, but often produces tedious combinatorics. By contrast, the analytic continuation approach requires more geometric insight to use, but can yield elegant shortcuts. We will show how both approaches can be used to derive the ``master identity'' \eqref{eq:master_identity} from which our other new identities follow. Finally, although both approaches use analyticity of the integrand $F(\eta)$, they only require analyticity on an appropriate region, as we show in an example at the end of App.~\ref{app:anacont}. 

\subsection{Generating Functions}\label{app:genFn}

We begin by deriving generating functions for the integrals \eqref{eq:initial_eta_integrals} from six basic properties they must satisfy. First, the heuristic definition \eqref{eq:defd4eta} implies the normalization
\begin{equation}
\int [d^4 \eta] \, \delta(\eta^2 + 1) = 1. \label{eq:normalization}
\end{equation}
We further require integrals over $\delta(\eta^2 + 1)$ to respect the delta function and integration by parts identities,
\begin{alignat}{2} 
&\int [d^4 \eta] \, \delta(\eta^2 + 1) (\eta^2+1) F(\eta) &&= 0, \label{eq:delta_func} \\
&\int [d^4 \eta] \, \ \ \del_\eta^\mu \left( \delta(\eta^2 + 1) F(\eta) \right) &&= 0. \label{eq:ibp_int}
\end{alignat}
To relate these to integrals over $\delta'(\eta^2 + 1)$, we impose the distributional identity
\begin{equation} \label{eq:distro_id}
\delta'(\eta^2 + 1) (\eta^2 + 1) = -\delta(\eta^2 + 1)
\end{equation}
underneath an $\eta$ integral. Finally, we assume that integration is linear and the results are Lorentz covariant, in a way we will describe more concretely shortly.

The above properties determine the integrals~\eqref{eq:initial_eta_integrals} for any analytic function $F$. To show this, note that by linearity, it suffices to define the integral for each term in the Taylor expansion of $F$, which in turn follow from integrals of the form
\begin{equation} 
\int [d^4 \eta] \, \delta(\eta^2 + 1) \, \eta^{\mu_1} \eta^{\mu_2} \cdots \eta^{\mu_m}.
\end{equation}
By Lorentz symmetry, the result must be a fully symmetric rank-$m$ tensor built from $g^{\mu\nu}$. Thus, it must be zero for odd $m$, while for even $m = 2n$, we must have 
\begin{equation}
\int [d^4 \eta] \, \delta(\eta^2 + 1) \, \eta^{\mu_1} \eta^{\mu_2} \cdots \eta^{\mu_{2n}} = c_n \, (g^{\mu_1 \mu_2} \cdots g^{\mu_{2n-1} \mu_{2n}} + \text{perms.}) \label{eq:cn_def}
\end{equation}
where the right-hand side has $(2n)!$ terms. Contracting this with $g_{\mu_1 \mu_2}$, the left-hand side is
\begin{equation}
g_{\mu_1 \mu_2} \int [d^4 \eta] \, \delta(\eta^2 + 1) \, \eta^{\mu_1} \eta^{\mu_2} \cdots \eta^{\mu_{2n}} = - c_{n-1} (g^{\mu_3 \mu_4} \cdots g^{\mu_{2n-1} \mu_{2n}} + \text{perms.})
\end{equation}
where we used~\eqref{eq:delta_func}. The right-hand side has the same tensor structure; to find the numeric coefficient, note there are $2n (2n-2)!$ terms containing $g^{\mu_1 \mu_2}$ or $g^{\mu_2\mu_1}$, which each produce a factor of $D = 4$ in the contraction, and the other $(2n)(2n-2)(2n-2)!$ terms yield a product of metric tensors with unit coefficient. Comparing the two sides yields the recursion relation
\begin{equation}
- c_{n-1} = (2n)(4 + (2n-2)) \, c_n.
\end{equation}
Since $c_0 = 1$, the general solution is
\begin{equation}
c_n = \frac{(-1/4)^n}{n! \, (n+1)!} = \begin{cases} -1/8 & n = 1 \\ 1/192 & n = 2 \end{cases}.
\end{equation}

Both types of $\eta$ integral can be written economically in terms of generating functions, 
\begin{align}
\int [d^4\eta] \, \delta(\eta^2 + 1) \, F(\eta) &= G\left(\sqrt{\del_\eta^2} \right) F(\eta) \bigg|_{\eta = 0} \\
\int [d^4\eta] \, \delta'(\eta^2 + 1) \, F(\eta) &= G'\!\left(\sqrt{\del_\eta^2} \right) F(\eta) \bigg|_{\eta = 0}
\end{align}
where we have just shown that
\begin{equation} \label{eq:G_fn_def}
G(x) = \sum_{n=0}^\infty c_n x^{2n} = \frac{2 J_1(x)}{x}.
\end{equation}
To address the other integral, note that the above properties imply 
\begin{equation}
\int [d^4 \eta] \, \delta'(\eta^2 + 1) F(\eta) = \int [d^4 \eta] \, \delta(\eta^2 + 1) \left( \frac12 \, \del_\eta \cdot \eta -1 \right) F(\eta), \label{deltaPrime}
\end{equation}
which is readily verified by integrating the right-hand side by parts and applying \eqref{eq:distro_id}. This result holds for general $D$, and for $D = 4$, the term in parentheses produces a factor of $n+1$ when acting on $\eta^{2n}$. Thus, the Taylor series of $G'$ has coefficients 
\begin{equation} \label{eq:d_n_coeffs}
d_n = \frac{(-1/4)^n}{(n!)^2} = \begin{cases} -1/4 & n = 1\\ 1/64 & n = 2 \end{cases}
\end{equation}
which sums to 
\begin{equation} \label{eq:primed_generating_function}
G'(x) = \sum_{n=0}^\infty d_n x^{2n} = J_0(x).
\end{equation}

The appearance of Bessel functions is not surprising from a geometric standpoint, and we can see more directly why they must appear with an alternate derivation. Note that showing \eqref{eq:primed_generating_function} is equivalent to showing
\begin{equation} \label{eq:integral_proof}
I(V) = \int [d^4 \eta] \, \delta'(\eta^2 + 1) \, e^{i \eta \cdot V} = J_0(\sqrt{-V^2})
\end{equation}
for any four-vector $V$, since each factor of $\del_\eta^2$ yields a factor of $-V^2$. Now, by~\eqref{eq:distro_id} we have
\begin{align}
\del_V^2 I(V) & = \int [d^4\eta] \, \delta'(\eta^2+1) \, (-\eta^2) e^{i \eta \cdot V} \\
& = \int [d^4\eta] \, \delta'(\eta^2+1) \, e^{i \eta \cdot V} + \int [d^4\eta] \, \delta(\eta^2+1) \, e^{i \eta \cdot V}.
\end{align}
On the other hand, we can use integration by parts~\eqref{eq:ibp_int} to show that 
\begin{align}
\del_V^\mu I(V) & = \int [d^4\eta] \, \delta'(\eta^2+1) \, (i\eta^\mu) e^{i \eta \cdot V} \\
& = \frac{i}{2} \int [d^4\eta] \, (\del_\eta^\mu \delta(\eta^2+1)) \, e^{i \eta \cdot V} \\
& = \frac{i}{2} \int [d^4\eta] \, \delta(\eta^2+1) \, (-\del_\eta^\mu) e^{i \eta \cdot V} \\
& = \frac12 \int [d^4\eta] \, \delta(\eta^2+1) \, V^\mu e^{i \eta \cdot V}.
\end{align}
Combining these results yields the differential equation 
\begin{equation} \label{eq:I_diffeq}
V^2 \, \del_V^2 I(V) = 2 \, V\cdot \del_V I(V) + V^2 I(V).
\end{equation}
Since $I(V)$ is a scalar function of $V^\mu$, it must be solely a function of $x = \sqrt{-V^2}$. In $D$ spacetime dimensions,~\eqref{eq:I_diffeq} implies $x^2 I''(x)+(D-3)x I'(x)+x^2 I(x) = 0$, which for $D=4$ is the Bessel differential equation of order zero. The normalization $I(0)=1$ fixes $I(x)=J_0(x)$. 

\subsection{Useful Identities}
\label{app:usefulIdentities}

With the generating functions in hand, we can prove the concrete identities used in the main text. Most of these identities involve the null frame vectors defined in the conventions. They can be efficiently handled in the generating function approach because~\eqref{eq:frame_metric} implies
\begin{equation} \label{eq:del_squared_expr}
\del_\eta^2 = 2 (q \cdot \del_\eta) (k \cdot \del_\eta) - (\epsilon_+ \cdot \del_\eta) (\epsilon_- \cdot \del_\eta).
\end{equation}
In other words, applying $\del_\eta^2$ produces contractions between the $\eta \cdot q$ and $\eta \cdot k$ components of the integrand, as well as between the $\eta \cdot \epsilon_+$ and $\eta \cdot \epsilon_-$ components. 

As a simple example of this, consider the overlaps of the helicity modes,
\begin{align} \label{eq:overlap_def}
\braket{\psi_{h,k}}{\psi_{h',k}} &\equiv \int [d^4 \eta] \, \delta'(\eta^2+1) \, \psi_{h,k}^*(\eta) \psi_{h',k}(\eta) \\
&= \int [d^4 \eta] \, \delta'(\eta^2+1) \, (\mp i \eta \cdot \epsilon_{\mp})^{|h|} (\pm i \eta \cdot \epsilon_{\pm})^{|h'|}
\end{align}
where the top and bottom signs apply for positive and negative $h$ and $h'$. When $h \neq h'$, the result must be zero because there are an unequal number of copies of $\epsilon_+$ and $\epsilon_-$. (Similarly, the result would automatically be zero if the conjugation in~\eqref{eq:overlap_def} was not present, since the factors of $e^{- i \rho \eta \cdot q}$ would not cancel.) If $h = h'$, then for positive $h$ the norm is 
\begin{align}
\braket{\psi_{h,k}}{\psi_{h,k}} &= d_h \, (\del_\eta^2)^h \, (\eta \cdot \epsilon_-)^h (\eta \cdot \epsilon_+)^h \bigg|_{h=0} \\ 
&= d_h \, (-1)^h \, (\epsilon_+ \cdot \del_\eta)^h (\eta \cdot \epsilon_-)^h \, (\epsilon_- \cdot \del_\eta)^h (\eta \cdot \epsilon_+)^h \bigg|_{h=0} \\
&= d_h \, (-1)^h \, ((\epsilon_+ \cdot \epsilon_-)^h \, h!)^2 = 1. \label{eq:ortho_final}
\end{align}
with an identical result for negative $h$. This shows the helicity basis is orthonormal. 

Using this result, we can derive a powerful ``master'' identity, which reduces certain vector superspace integrals to integrals over the unit circle in the plane spanned by $\epsilon_+$ and $\epsilon_-$,
\begin{equation} \label{eq:master_identity}
\int [d^4 \eta] \, \delta'(\eta^2 + 1) \, F(\eta) = \int \frac{d\theta}{2\pi} \, F(\eta) \bigg|_{\eta = \mathrm{Re}(e^{i \theta} \epsilon_-)} \ \text{ when } \delta(\eta^2+1) \, k \cdot \del_\eta \, F = 0
\end{equation}
for some null $k$. The condition above is equivalent to letting
\begin{equation} \label{eq:weaker_condition}
k \cdot \del_\eta \, F = (\eta^2 + 1) \, \beta(\eta)
\end{equation}
for an arbitrary analytic function $\beta$. 

To prove~\eqref{eq:master_identity} in the special case $\beta = 0$, consider Taylor expanding $F$ in the variables $\eta \cdot k$, $\eta \cdot q$, and $\eta \cdot \epsilon_\pm$. Terms involving $\eta \cdot q$ would contribute to the Taylor expansion of $k \cdot \del_\eta \, F$, and so must vanish by assumption. But then~\eqref{eq:del_squared_expr} implies that $\eta \cdot k$ terms do not contribute to the left-hand side, since they have to be contracted with $\eta \cdot q$ terms. Only terms of the form $(\eta \cdot \epsilon_+ \, \eta \cdot \epsilon_-)^h$ contribute, with unit coefficients as we have just shown, and this is reproduced by the integral on the right-hand side.

It is possible, but tedious, to verify~\eqref{eq:master_identity} for nonzero $\beta$ by Taylor expanding both $F$ and $\beta$ and evaluating the left-hand side using the generating function. Alternatively, consider decomposing $F = F_0 + F_\beta$, where the ``homogeneous'' solution $F_0$ satisfies $k \cdot \del_\eta \, F_0 = 0$. The previous paragraph shows that $F_0$ obeys the master identity, so it suffices to construct a ``particular'' solution $F_\beta$ which does not contribute to either side of~\eqref{eq:master_identity}. To do this, let
\begin{equation} \label{eq:inhomog_soln}
F_\beta(\eta) = \int_0^{\eta \cdot q} dx \, (\eta^2 + 1 + 2 (\eta \cdot k)(x - \eta \cdot q)) \, \beta(\eta + (x - \eta \cdot q) k).
\end{equation}
By construction, $k \cdot \del_\eta$ annihilates the integrand, so that $k \cdot \del_\eta \, F_\beta = (\eta^2 + 1) \, \beta(\eta)$ from differentiating the integral's upper bound. Now we show that $F_\beta$ does not contribute to either side of~\eqref{eq:master_identity}. On the right-hand side, $\eta \cdot q = 0$ on the unit circle, so the range of the $x$ integral vanishes. The left-hand side vanishes because its integrand can be rewritten as a total derivative,
\begin{equation}
\int [d^4 \eta] \, k \cdot \del_\eta \left( \delta(\eta^2 + 1) \int_0^{\eta \cdot q} dx\, (x - \eta \cdot q) \, \beta(\eta + (x-\eta \cdot q) k) \right) = 0,
\end{equation}
as can be checked by carrying out the derivative and using~\eqref{eq:distro_id}.

As a first application of the master identity, we prove a result which is often useful when evaluating projections against helicity modes,
\begin{equation} \label{eq:helicity_identity}
\int [d^4 \eta] \, \delta'(\eta^2 + 1) (i\eta \cdot \epsilon_{\pm})^{h} e^{i \eta \cdot V} = e^{i h \arg(\epsilon_\pm \cdot V)} \, J_{h}(\sqrt{-V^2}) \text{ when } V \cdot k = 0
\end{equation}
for any nonnegative integer $h$. Note that the $h = 0$ case is simply our earlier result~\eqref{eq:integral_proof}, and that conjugating the result for the upper sign gives the result for the lower sign. Also note that when applying this identity to helicity modes with negative $h$, it is often convenient to simplify the final result using $J_{-n}(x) = (-1)^n J_n(x)$.

To prove the result, note that the condition $V \cdot k = 0$ implies $V$ can be decomposed as 
\begin{equation} \label{eq:v_decomp}
V = (q \cdot V) \, k - \frac{v^*}{2} \, \epsilon_+ - \frac{v}{2} \, \epsilon_-
\end{equation}
where $v = \epsilon_+ \cdot V$, which in turn implies $|v|^2 = - V^2$. Now, applying the master identity to the left-hand side of~\eqref{eq:helicity_identity} gives
\begin{align}
\int [d^4 \eta] \, \delta'(\eta^2 + 1) (i\eta \cdot \epsilon_+)^h e^{i \eta \cdot V} &= \int \frac{d\theta}{2 \pi} \, (i\eta \cdot \epsilon_+)^h e^{i \eta \cdot V} \bigg|_{\eta = \mathrm{Re}(e^{i \theta} \epsilon_-)} \\
&= \int \frac{d\theta}{2\pi} \, e^{i h (\theta - \pi/2)} e^{i |v| \cos(\theta - \arg v)} \\
&= e^{i h \arg v} \int \frac{d\theta}{2 \pi} \, e^{i h (\theta - \pi/2)} e^{i |v| \cos \theta} \\
&= e^{i h \arg v} J_h(|v|) \label{eq:helicityIdentity_vForm}
\end{align}
where we plugged in $\eta = \mathrm{Re}(e^{i \theta} \epsilon_-) = (e^{i \theta} \epsilon_- + e^{- i \theta} \epsilon_+)/2$, shifted $\theta$, and used the integral representation of the Bessel functions. This is the desired result~\eqref{eq:helicity_identity} for the upper sign.

Of course, we can also show the result by using generating functions to directly evaluate the left-hand side. The first term in~\eqref{eq:v_decomp} produces factors of $\eta \cdot k$ which have no factors of $\eta \cdot q$ to contract with, so it can be dropped. The left-hand side is then
\begin{align}
\sum_{n=0}^\infty \frac{(-1/4)^{n+h}}{(n+h)!^2} &(\del_\eta^2)^{n+h} \, (i \eta \cdot \epsilon_+)^h e^{- (i/2) (\eta \cdot \epsilon_+) v^*} e^{-(i/2) (\eta \cdot \epsilon_-) v} \bigg|_{\eta=0} \\
= &\sum_{n=0}^\infty \frac{(-1/4)^{n+h}}{(n+h)!^2} \binom{n+h}{h} \left( (-2)^h v^h \, h! \right) |v^2|^n \\
= & \left( \frac{v}{|v|} \right)^h \sum_{n=0}^\infty \frac{(-1)^n}{n! \, (n+h)!} \left( \frac{|v|}{2} \right)^{2n+h}
\end{align}
which in the first step, $h$ of the copies of $\del_\eta^2$ acted on an $i \eta \cdot \epsilon_+$ and the first exponential, and the other $n$ copies acted on both exponentials. This is the desired right-hand side.

As another example, we prove the completeness relation for the helicity modes,
\begin{equation} \label{eq:completeness}
\sum_h \psi_{h,k}(\eta) \psi_{h,k}^*(\eta') \, \delta'(\eta^2+1) \, \delta'(\eta'^2+1) \simeq \delta^{(4)}(\eta-\eta') \, \delta'(\eta^2+1),
\end{equation}
which holds when both sides are integrated over $\eta$ and $\eta'$ against $F(\eta, \eta', k)$ satisfying 
\begin{align}
(i k \cdot \del_\eta + \rho) F &= (\eta^2 + 1) \, \beta, \\
(-i k \cdot \del_\eta' + \rho) F &= (\eta'^2 + 1) \, \beta'.
\end{align}
A key physical example is $F(\eta, \eta', k) = J(\eta, -k) J'(\eta', k)$ where the currents $J$ and $J'$ satisfy the continuity condition~\eqref{eq:continuity_deltaForm}. In that case,~\eqref{eq:completeness} implies that the continuous spin propagator between two currents is equivalent to a sum over physical polarizations, as expected by unitarity. 

To show the result, first note that the integrated left-hand side of~\eqref{eq:completeness} is
\begin{equation}
\int [d^4 \eta] [d^4 \eta'] \, \delta'(\eta^2 + 1) \delta'(\eta'^2 + 1) \, \sum_h (e^{i \rho (\eta' - \eta) \cdot q} F(\eta, \eta', k)) (\eta \cdot \epsilon_\pm)^{|h|} (\eta' \cdot \epsilon_\mp)^{|h|}
\end{equation}
where the upper and lower signs apply for positive and negative $h$. The function in brackets satisfies the condition~\eqref{eq:weaker_condition} to apply the master identity in both $\eta$ and $\eta'$, giving
\begin{equation}
\int \frac{d\theta}{2 \pi} \frac{d\theta'}{2\pi} F(\eta, \eta', k) \sum_h (\eta \cdot \epsilon_\pm)^{|h|} (\eta' \cdot \epsilon_\mp)^{|h|} \bigg|_{\substack{\eta \hspace{0.5mm} = \hspace{0.5mm} \mathrm{Re}(e^{i \theta} \epsilon_-) \\ \eta' = \mathrm{Re}(e^{i \theta'} \epsilon_-)}} = \int \frac{d\theta}{2 \pi} F(\eta, \eta, k) \bigg|_{\eta = \mathrm{Re}(e^{i \theta} \epsilon_-)}
\end{equation}
as the summation yields a delta function, $\sum_h e^{i h (\theta - \theta')} = 2 \pi \, \delta(\theta - \theta')$. On the right-hand side, we have a single integral over $F(\eta, \eta, k)$, and using the master identity gives the same result. 

\subsection{Regulated Measure From Analytic Continuation}
\label{app:anacont}

In this section, we concretely define the regulated measure $[d^D\eta]$ by analytic continuation. While we have already shown that vector superspace integrals can be determined without specifying a regulator, this new perspective readily generalizes to any spacetime dimension $D$ and provides geometric intuition for the identities proven above.

Specifically, we analytically continue to complex $\eta^0$ and rotate the $\eta^0$ integration contourclockwise, going up the imaginary axis. The resulting integral is naturally expressed in terms of Wick rotated coordinates $\bar\eta^M = (\eta^1, \ldots, \eta^{D-1}, - i\eta^0)$, where $M = 1, \ldots, D$. Euclidean inner products are always implied for barred quantities, e.g.~the quantity $\eta^2$ analytically continues to $-\bar{\eta}^2$. Integrals over the Minkowskian unit hyperboloid $\eta^2+1 = 0$ thus continue to integrals over a Euclidean unit sphere $\bar\eta^2 -1 = 0$. This will yield finite results since the unit sphere has finite area $S_{D-1}$.

For any analytic function $F(\eta)$, we therefore define
\begin{equation} \label{eq:defWick}
\int [d^D \eta] \, \delta(\eta^2+ 1) F(\eta) \equiv \frac{2}{S_{D-1}} \int d^D \bar\eta \, \delta(\bar\eta^2 - 1) \bar F(\bar\eta),
\end{equation}
where the prefactor ensures the normalization condition~\eqref{eq:normalization} is satisfied. Similarly,
\begin{equation} \label{eq:def_prime}
\int [d^D \eta] \, \delta'(\eta^2+ 1) F(\eta) = - \frac{2}{S_{D-1}} \int d^D \bar\eta \, \delta'(\bar\eta^2 - 1) \bar{F}(\bar\eta),
\end{equation}
where the minus sign appears because $\delta'(-x) = -\delta'(x)$, and ensures consistency with~\eqref{eq:distro_id}. This is the same definition as used in Ref.~\cite{Schuster:2014hca}, though the analogue of~\eqref{eq:def_prime} in that work does not show the factor of $2$. Note that $\bar{F}$ is also defined by analytic continuation, i.e.~$\bar F(\bar\eta) = F(i\bar\eta^D, \bar{\eta}^1, \ldots, \bar{\eta}^{D-1}) = F(\eta)$.

\subsubsection*{Recovering the Generating Functions}

As a simple example, we will use analytic continuation to evaluate the integral
\begin{equation}
I(V) = \int [d^D \eta] \, \delta'(\eta^2 + 1) e^{i \eta \cdot V}.
\end{equation}
As explained around~\eqref{eq:integral_proof}, this is equivalent to finding the generating function $G'$, which together with the property~\eqref{eq:distro_id} determines $G$ and thus all vector superspace integrals. It therefore suffices to evaluate the above integral for $D = 4$ to show the equivalence of analytic continuation to the symmetry-based arguments in appendix~\ref{app:genFn}.

We define $\bar V = (V^1, \ldots, V^{D-1},-i V^0)$ so the integrand can be written in terms of Euclidean inner products, $\eta \cdot V = - \bar{\eta} \cdot \bar{V}$. Applying~\eqref{eq:defWick} and the Euclidean analogue of~\eqref{deltaPrime} gives
\begin{align}
% I(V) &= \int [d^D\eta] \, \delta(\eta^2 + 1) \, \frac{1}{2}(\del_\eta\cdot\eta-2) e^{i\eta \cdot V} \\
I(V) &= \frac{2}{S_{D-1}} \int d^D \bar\eta \, \delta(1-\bar\eta^2) \, \frac{1}{2} (\del_{\bar\eta}\cdot\bar\eta-2) e^{-i\bar\eta\cdot\bar V} \\
&= \frac{1}{S_{D-1}} \int d^D \bar\eta \, \delta(1-\bar\eta^2) \, (D - 2 + \bar{\eta} \cdot \del_{\bar{\eta}}) e^{-i\bar\eta\cdot\bar V}.\label{euclideanPhaseIntegral}
\end{align}
Note that $\bar{V}$ is complex in general, but it suffices to evaluate~\eqref{euclideanPhaseIntegral} at real $\bar{V}$ and analytically continue the result. Specializing to $D = 4$, the measure in spherical coordinates is
\begin{equation}
d^4 \bar{\eta} = |\bar{\eta}|^3 d|\bar{\eta}| \, \sin^2 \theta \, d \theta \, d^2 \Omega_2
\end{equation}
where $\theta$ is the angle between $\bar{\eta}$ and $V$ and $\Omega_2$ indicates the remaining angular integral. Then
\begin{align}
I(V) &= \frac{1}{S_3} \int_0^\infty d|\bar{\eta}| \, \delta(1 - |\bar{\eta}|^2) \, |\bar{\eta}|^3 \int_0^\pi d \theta \, \sin^2 \theta \, (2 - i |\bar{\eta}| |\bar{V}| \cos \theta) e^{- i |\bar{\eta}| |\bar{V}| \cos \theta} \int d^2 \Omega_2 \\
&= \frac{1}{\pi} \int_0^\pi d\theta \, \sin^2 \theta \, (2 - i |\bar{V}| \cos \theta) e^{- i |\bar{V}| \cos \theta} \\
&= J_0(|\bar{V}|)
\end{align} 
where we used $S_2 = 4 \pi$ and $S_3 = 2 \pi^2$. This matches our earlier result~\eqref{eq:integral_proof}, as $|\bar{V}| = \sqrt{-V^2}$.

The $D$-dimensional analogue of this result is given in Ref.~\cite{Schuster:2014hca}, and can also be derived using spherical coordinates. However, it turns to be simpler to use polyspherical coordinates, even though they employ weaker symmetry properties of the integrand. To do this, group the $\bar{\eta}$ coordinates into a two-dimensional subspace $\eta_L$ orthogonal to $\bar{V}$ and a $(D-2)$-dimensional subspace $\tilde{\eta}$. Using polar coordinates for $\eta_L$, the measure is 
\begin{equation}
d^D \bar{\eta} = d^{D-2} \tilde{\eta} \, \frac{d|\eta_L|^2 \, d\theta}{2}
\end{equation}
and since $\bar{\eta} \cdot \bar{V} = \tilde{\eta} \cdot \tilde{V}$, the integral~\eqref{euclideanPhaseIntegral} becomes
\begin{align}
I(V) &= \frac{1}{2 S_{D-1}} \int d^{D-2} \tilde{\eta} \, d |\eta_L|^2 \, d\theta \, \delta(1-|\tilde{\eta}|^2 - |\eta_L|^2) \, (D - 2 + \bar{\eta} \cdot \del_{\bar{\eta}}) e^{-i\tilde\eta\cdot\tilde V} \\
&= \frac{\pi}{S_{D-1}} \int_{|\tilde{\eta}| \leq 1} d^{D-2} \tilde{\eta} \, (D - 2 + \tilde{\eta} \cdot \del_{\tilde{\eta}}) e^{-i\tilde\eta\cdot\tilde V} \\
&= \frac{\pi}{S_{D-1}} \int_{|\tilde{\eta}| \leq 1} d^{D-2} \tilde{\eta} \, \del_{\tilde{\eta}} \cdot (\tilde{\eta} \, e^{-i\tilde\eta\cdot\tilde V}) \\
&= \frac{\pi}{S_{D-1}} \int_{|\tilde{\eta}| = 1} d^{D-3} \Omega_{\tilde{\eta}} \, e^{- i \tilde{V} \cdot \tilde{\eta}} \label{eq:I_V_final}
\end{align}
where we performed the integral over $\eta_L$ and used the divergence theorem. The integral thus cleanly reduces to one over a $(D-3)$-dimensional unit sphere. For example, when $D = 4$ the remaining integral is over a unit circle, $d\Omega_{\tilde{\eta}} = d \theta$, giving
\begin{equation}
I(V) = \frac{1}{2 \pi} \int_0^{2\pi} d \theta \, e^{- i |\bar{V}| \cos \theta} = J_0(|\bar{V}|)
\end{equation}
in agreement with our previous result. 

\subsubsection*{Generalizing the Master Identity}

We now use analytic continuation to generalize the master identity~\eqref{eq:master_identity} to $D$ dimensions, and shed light on its geometric meaning. We consider the integral
\begin{equation}
\int [d^D \eta] \, \delta'(\eta^2 + 1) F(\eta) = -\frac{2}{S_{D-1}} \int d^D\bar \eta \, \delta'(\bar\eta^2-1) \bar F(\bar\eta) \label{eq:general_deltaprime_integral}
\end{equation}
where $k \cdot \del_\eta \, F = (\eta^2 + 1) \beta$ for some null $k$. We first take $\beta = 0$, in which case $\bar{k} \cdot \del_{\bar{\eta}} \, \bar{F} = 0$.

The first step is to parallel our earlier derivation, but in Euclidean signature. We decompose $\bar{\eta}$ into a $(D-2)$-dimensional subspace $\tilde{\eta}$ which is orthogonal to $\bar{k}$, and a two-dimensional subspace $\eta_L$ spanned by $\bar{k}$ and $\bar{q}$, which satisfy $\bar{k}^2 = \bar{q}^2 = 0$ and $\bar{k} \cdot \bar{q} = -1$. (This is possible because $\bar{k}$ and $\bar{q}$ are complex Euclidean vectors. And as in Lorentzian signature, there is freedom in the choice of $\bar{q}$.) The requirement $\bar{k} \cdot \del_{\bar{\eta}} \, \bar{F} = 0$ then implies that $\bar{F}$ cannot depend on $\bar{\eta} \cdot \bar{q}$. To handle the $\bar{\eta} \cdot \bar{k}$ dependence, we write
\begin{equation}
\bar F(\tilde \eta, \bar\eta\cdot\bar k) = \tilde F(\tilde\eta) + \bar\eta\cdot\bar k \, G(\tilde{\eta}, \bar{\eta} \cdot \bar{k}) 
\end{equation}
which is always possible since $\bar{F}$ is analytic. The second term does not contribute to the integral~\eqref{eq:general_deltaprime_integral}. To see this, note that $\bar{k} \cdot \del_{\bar{\eta}} \, G = 0$, so we may write
\begin{equation}
\int d^D\bar\eta \, \delta'(\bar\eta^2-1) \, \bar\eta\cdot\bar k \, G = \frac{1}{2} \int d^D\bar\eta \, \bar k\cdot\del_{\bar\eta} \left(\delta(\bar\eta^2-1) G \right). 
\end{equation}
The integrand is a total derivative, which vanishes at infinity due to the delta function.

Therefore, only $\tilde{F}(\tilde{\eta})$ contributes to the right-hand side of~\eqref{eq:general_deltaprime_integral}. Now, note that in the derivation of~\eqref{eq:I_V_final}, the only property of the function $e^{- i \bar{\eta} \cdot \bar{V}}$ we used was that it did not depend on $\eta_L$. Since this is also true of $\tilde{F}(\tilde{\eta})$, we may use the exact same reasoning to show
\begin{equation}
\int [d^D \eta] \, \delta'(\eta^2 + 1) F(\eta) = \frac{\pi}{S_{D-1}} \int_{|\tilde{\eta}| = 1} d^{D-3} \Omega_{\tilde{\eta}} \, \tilde{F}(\tilde{\eta}).
\end{equation}
Using the recurrence relation $S_{D-1} = 2 \pi S_{D-3}/(D-2)$ and analytically continuing the right-hand side back to Lorentzian signature (which is trivial when the $\tilde{\eta}$ subspace does not contain $\eta^0$) yields
\begin{equation} \label{eq:general_master_id}
\int [d^D\eta] \, \delta' (\eta^2+1) \, F(\eta) = \frac{D-2}{2 S_{D-3}} \int_{S_{\tilde{\eta}}} d^{D-3}\Omega_{\tilde\eta} \, F(\eta),
\end{equation}
where $\eta^2 = -1$ and $\eta \cdot k = 0$ on the sphere $S_{\tilde{\eta}}$. This is the generalized master identity. For $D = 4$, $S_{\tilde{\eta}}$ is a unit circle in the plane spanned by $\epsilon_\pm \cdot \eta$, which recovers~\eqref{eq:master_identity}. 

The above proof only applies to $\beta = 0$, but our earlier method to generalize the proof to nonzero $\beta$, based upon constructing an explicit particular solution, carries over unchanged. Also note that for Lorentz scalar $F$, the right-hand side of~\eqref{eq:general_master_id} is not manifestly Lorentz invariant. However, any Lorentz transformation that fixes $k$ simply maps the sphere $S_{\tilde{\eta}}$ to another sphere which satisfies the same condition $\eta \cdot k = 0$, leaving the result unchanged. 

Our derivation above sheds light on why a useful identity should exist specifically for a $\delta'(\eta^2 + 1) F$ integral and a null $k$. If we had a timelike $k$, such as $k^\mu = (1, 0, \ldots, 0)$, then we would only have been able to conclude that $\bar{F}$ does not depend on $\bar{\eta}^0$, and performing the $\bar{\eta}^0$ integral leaves an integral over a $(D-1)$-dimensional ball. For null $k$, we have seen that we can drop dependence on both $\bar{\eta} \cdot \bar{k}$ and $\bar{\eta} \cdot \bar{q}$, leaving an integral over a $(D-2)$-dimensional ball. Finally, $\delta'(\eta^2 + 1) F$ turns out to be a total derivative, which would not have been true for $\delta(\eta^2 + 1) F$, reducing the integral to one over a unit $(D-3)$-sphere. 

This is an enormous simplification, especially when $D = 4$. It is connected to the observation in Ref.~\cite{Schuster:2014hca} that, to encode spin in a scalar function of $\eta$ and $x$, the function must be defined on the first neighborhood of a hyperboloid in $\eta$-space. Indeed, the final integral over $S_{\tilde{\eta}}$ is an integral over the continuous basis for CSP representations of the massless little group $ISO(D-2)$ in general dimension, e.g.~for $D = 4$ it is an integral over the angle basis $\ket{\theta}$ mentioned below~\eqref{eq:Tpm_action}. This is the fundamental reason the action~\eqref{eq:introaction} may be written solely in terms of integrals over $\delta(\eta^2 + 1)$ and $\delta'(\eta^2 + 1)$, without requiring $\delta''(\eta^2 + 1)$ or higher derivative terms.

\subsubsection*{A Non-Entire Example}

While most expressions we encounter are analytic everywhere in $\eta$, we can also encounter expressions with branch cuts, such as the static field~\eqref{eq:Psi_S_pos} which depends directly on $|\etav|=\sqrt{\etav^2}$. Integrals of such functions are still well-defined by either the generating functional or analytic continuation approach because they are analytic in an appropriate region. As a simple example, consider the integral
\begin{equation}
I = \int [d^4 \eta] \, \delta(\eta^2 + 1) \, |\etav|.
\end{equation}
The integrand does not have a Taylor expansion about $\eta = 0$ because of the branch cut starting at the surface $\sum_i\eta_i^2 = 0$, which passes thorugh the origin. However, on the support of the delta function the integrand is equivalent to $\sqrt{1 + (\eta^0)^2}$, which does have a Taylor expansion. We can thus use the generating function~\eqref{eq:G_fn_def} to conclude
\begin{equation}
I = \int [d^4 \eta] \, \delta(\eta^2 + 1) \, \sqrt{1 + (\eta^0)^2} = -\sum_{n=0}^\infty \frac{(2n)! (2n-3)!!}{2^{3n} \, (n!)^2 \, (n+1)!} = \frac{8}{3\pi}.
\end{equation}
Alternatively, $|\etav|$ is analytic in $\eta^0$, so that by the definition~\eqref{eq:defWick} we have
\begin{equation}
I = \frac{2}{S_3} \int d^4 \bar{\eta} \, \delta(\bar{\eta}^2 - 1) \, |\bar{\etav}| = \frac{S_2}{S_3} \int_0^\pi \sin^3 \theta = \frac{8}{3 \pi}
\end{equation}
by analytic continuation. By similar reasoning we can make sense of any integral over $\delta(\eta^2 + 1)$ of a function of $|\etav|$, and by applying~\eqref{deltaPrime}, any integral over $\delta'(\eta^2 + 1)$ as well.

We do not have a precise definition of the space of functions $F(\eta)$ for which the integrals \eqref{eq:initial_eta_integrals} are well-defined, but this toy example illustrates that the space includes non-entire functions. It may be valuable to define this space more carefully, and in particular to understand whether allowing fields $\Psi(\eta,x)$ non-entire in $\eta$ would admit physically inequivalent solutions to the free equation of motion \eqref{eq:free_eom}.

\addtocontents{toc}{\vspace{-0.2\baselineskip}}
\section{Vector Superspace Computations}
\label{sec:computations}
This appendix collects derivations of more technical results. In App.~\ref{app:tensors} we prove the statements made in subsections~\ref{sec:primer_fields} and~\ref{ssec:CSP_currents}, which relate the continuous spin theory to familiar theories when $\rho = 0$. In App.~\ref{app:classification} we parametrize the set of all worldline currents obeying the continuity condition, and in App.~\ref{sec:fields} we compare the results to currents built from matter fields. In App.~\ref{app:es_ints} we compute the spacetime profiles of some simple currents by evaluating integrals with essential singularities, deriving results quoted in subsection~\ref{sec:simple_spacetime}. Finally, in App.~\ref{app:gauge_inv} we establish some results for radiation emission quoted in section~\ref{sec:radiation}.

\subsection{Tensor Decompositions}
\label{app:tensors}

\subsubsection*{Absence of Cross-Couplings of Tensor Fields}

In subsection~\ref{sec:primer_fields}, we claimed that for $\rho = 0$, the action and equation of motion do not contain cross-couplings between distinct tensor fields. To show this, we first recall that gauge symmetry allows us to add terms proportional to $(\eta^2 + 1)^2$ to the field $\Psi$. This freedom is sufficient to eliminate the double trace of every tensor component, i.e.~${\phi^{(n)}}'' = 0$ for $n \geq 4$. 

For fully symmetric, double-traceless tensors $\phi^{(n)}$ and $\psi^{(m)}$ of rank $n$ and $m$, we have
\begin{align}
\int [d^4\eta] \, \delta'(\eta^2+1) \, P_{(n)} \cdot \phi^{(n)} \, P_{(m)}\cdot \psi^{(m)} &\propto \delta_{nm} \label{eq:first_ortho} \\
\int [d^4\eta] \, \delta(\eta^2+1) \, \Delta (P_{(n)} \cdot \phi^{(n)}) \, \Delta(P_{(m)}\cdot \psi^{(m)}) &\propto \delta_{nm}. \label{eq:second_ortho}
\end{align}
These show that there are no cross-couplings in the first and second term, respectively, of the action and equation of motion. 

Let us first motivate~\eqref{eq:first_ortho}, taking $n \geq m$ without loss of generality. As shown in appendix~\ref{app:genFn}, the $\eta$ integrations above produce full contractions of the tensors in the integrand. If $n - m$ is odd, the total number of indices is odd, so no full contractions exist. If $n - m \geq 4$, at least two pairs of indices on $\phi^{(n)}$ must be contracted, which produces a double trace which vanishes by assumption. Similar logic applies to~\eqref{eq:second_ortho}, so in both cases we need only check there is no contribution from $n - m = 2$. 

\subsubsection*{Proof of the Orthogonality Theorems}

To derive~\eqref{eq:first_ortho}, note that since the tensors are double-traceless, the only possible contribution when $n = m + 2$ is of the form ${\phi^{(m+2)}}'\cdot \psi^{(m)}$, so we must show its coefficient vanishes. The coefficient is simplest to calculate by rewriting the main results of appendix~\ref{app:genFn} as
\begin{align}
\int [d^4 \eta] \, \delta(\eta^2 + 1) \, \eta^{\mu_1} \eta^{\mu_2} \cdots \eta^{\mu_{2n}} &= \tilde{c}_n \, (g^{\mu_1 \mu_2} \cdots g^{\mu_{2n-1} \mu_{2n}} + \text{other pairings}) \label{eq:cn_def_2} \\
\int [d^4 \eta] \, \delta'(\eta^2 + 1) \, \eta^{\mu_1} \eta^{\mu_2} \cdots \eta^{\mu_{2n}} &= \tilde{d}_n \, (g^{\mu_1 \mu_2} \cdots g^{\mu_{2n-1} \mu_{2n}} + \text{other pairings}) \label{eq:dn_def_2}
\end{align}
where $\tilde{c}_n = (-1/2)^n / (n+1)!$ and $\tilde{d}_n = (-1/2)^n / n!$, and we sum over the $(2n-1)!!$ distinct ways to pair up indices.

Now, we write the integral in~\eqref{eq:first_ortho} as a sum of three integrals,
\begin{align}
2^{(n+m)/2} \phi^{\nu_1\dots \nu_n} \psi^{\mu_1\dots\mu_m} \times \bigg[ &\int [d^4\eta] \, \delta^\prime(\eta^2+1) \, \eta_{\nu_1} \cdots \eta_{\nu_n}  \eta_{\mu_1} \cdots \eta_{\mu_m} \label{eq:PnPnTerm1} \\
&+ \ \frac{n(n-1)}{4} \ g_{\nu_1 \nu_2} \int [d^4\eta] \, \delta(\eta^2+1) \, \eta_{\nu_3} \cdots \eta_{\nu_n}  \eta_{\mu_1} \cdots \eta_{\mu_m}  \label{eq:PnPnTerm2}\\
&+  \frac{m(m-1)}{4} g_{\mu_1 \mu_2} \int [d^4\eta] \, \delta(\eta^2+1) \, \eta_{\nu_1} \cdots \eta_{\nu_n} \eta_{\mu_3} \cdots \eta_{\mu_m} \bigg] \label{eq:PnPnTerm3}
\end{align}
where we simplified using~\eqref{eq:delta_func} and~\eqref{eq:distro_id} and removed a redundant symmetrization. The final integral~\eqref{eq:PnPnTerm3} cannot produce terms of the form ${\phi^{(m+2)}}' \cdot \psi^{(m)}$, since it already contains a trace of $\psi^{(m)}$, so we need only consider the first two integrals.

In the first integral~\eqref{eq:PnPnTerm1}, we count pairings which contract two $\nu$ indices, and pair all remaining $\nu$ indices with $\mu$ indices; this yields a combinatorial factor of $\binom{m+2}{2} \, m!$. The second integral~\eqref{eq:PnPnTerm2} already has a trace on $\phi^{(m+2)}$, so we only count pairings where all remaining $\nu$ indices go with $\mu$ indices; this yields a combinatorial factor of $m!$. The overall coefficient is 
\begin{equation}
2^{m+1} \left[ \tilde{d}_{m+1} \binom{m+2}{2} \, m! + \tilde{c}_m \, \frac{(m+2)(m+1)}{4} \, m! \right] = 0
\end{equation}
which establishes the result. 

It will also shortly prove useful to derive the value of the integral when $m = n$, 
\begin{equation}
\int [d^4\eta] \, \delta'(\eta^2+1) \, P_{(n)} \cdot \phi^{(n)} \, P_{(n)}\cdot \psi^{(n)} = (-1)^n \left( \phi^{(n)}\cdot \psi^{(n)} -\frac{n(n-1)}{4} {\phi^{(n)}}'\cdot {\psi^{(n)}}' \right) \label{eq:P_orthonormality}
\end{equation}
The two terms on the right-hand side are the only ones that could have appeared, so we need only compute their coefficients. Only~\eqref{eq:PnPnTerm1} can produce $\phi^{(n)} \cdot \psi^{(n)}$ terms, with coefficient $2^n \tilde{d}_n \, n! = (-1)^n$ as desired. By contrast, all three integrals can produce ${\phi^{(n)}}'\cdot {\psi^{(n)}}'$ terms. The relevant combinatorial factor for~\eqref{eq:PnPnTerm1} is $\binom{n}{2}^2 (n-2)!$, and the combinatorial factor for each of the other two integrals is $\binom{n}{2} (n-2)!$, giving a total coefficient 
\begin{equation}
2^n \left[ \tilde{d}_n \binom{n}{2}^2 (n-2)! + 2 \, \tilde{c}_{n-1} \, \frac{n(n-1)}{4} \binom{n}{2} (n-2)! \right] = (-1)^{n-1} \, \frac{n(n-1)}{4}
\end{equation}
as desired. One immediate consequence of~\eqref{eq:P_orthonormality} is that the action~\eqref{eq:CSP_action} has canonically normalized kinetic terms for the tensor fields $\phi^{(n)}$. The alternating sign $(-1)^n$ is due to the mostly negative metric, as the physical components of $\phi^{(n)}$ have $n$ spatial indices.

The other orthogonality theorem~\eqref{eq:second_ortho} can also be derived with straightforward combinatorics, but there is a faster route. The integral contains terms of the form 
\begin{equation}
\Delta(P_{(n)} \cdot \phi^{(n)}) \propto \del_\eta^\nu \left( \eta^{\mu_1} \cdots \eta^{\mu_n} - \frac{n(n-1)}{4} \eta^{\mu_1} \cdots \eta^{\mu_{n-2}} g^{\mu_{n-1} \mu_n} (\eta^2 + 1) \right) \del_{\nu} \phi^{(n)}_{\mu_1 \ldots \mu_n}
\end{equation}
where the $\eta$ derivative can hit any of the $n$ factors of $\eta$ in the first term, but must hit the factor of $\eta^2 + 1$ in the second term due to the delta function. The result is 
\begin{equation}
n \, \eta^{\mu_1} \cdots \eta^{\mu_{n-1}} (\del \cdot \phi^{(n)})_{\mu_1 \ldots \mu_{n-1}} - \frac{n(n-1)}{2} \eta^{\mu_1} \cdots \eta^{\mu_{n-2}} \eta^\nu \del_\nu {\phi^{(n)}}'_{\mu_1 \cdots \mu_{n-2}}
\end{equation}
which, by symmetry, is equal to 
\begin{equation}
n \, \eta^{\mu_1} \cdots \eta^{\mu_{n-1}} \left( (\del \cdot \phi^{(n)})_{\mu_1 \ldots \mu_{n-1}} - \frac12 \sum_{k=1}^{n-1} \del_{\mu_k} {\phi^{(n)}}'_{\mu_1 \ldots \mu_{k-1} \mu_{k+1} \ldots \mu_{n-1}} \right).
\end{equation}
The quantity in parentheses is traceless. Thus, the integral~\eqref{eq:second_ortho} involves contractions of traceless tensors, which can only be nonzero if $m = n$. 

\subsubsection*{Dual Polynomials for the Current Expansion}

Now we motivate the ``dual'' polynomials~\eqref{eq:Pn_dual_def} used to define the tensor components $J^{(n)}$ of the current. If we expand $J(\eta, x)$ in the same polynomials as the continuous spin field, 
\begin{equation}
J(\eta,x) = \sum_{n \geq 0} P_{(n)}(\eta)\cdot j^{(n)}(x)
\end{equation}
and use the freedom to add terms proportional to $(\eta^2 + 1)^2$ to the current $J$ to eliminate the double trace of every $j^{(n)}$, then~\eqref{eq:P_orthonormality} immediately implies 
\begin{equation}
S_{\text{int}} = \sum_{n \geq 0} (-1)^n \left(\phi^{(n)}\cdot j^{(n)} - \frac{n(n-1)}{4} {\phi^{(n)}}'\cdot {{{j}^{(n)}}}' \right)
\end{equation}
The unwanted trace terms for $n \geq 2$ can be removed if we define tensors $J^{(n)}$ by
\begin{equation}
J^{(n)}_{\mu_1\dots \mu_n} = { j}^{(n)}_{\mu_1\dots \mu_n} - \frac{1}{4 \, (n-2)!} \, g_{(\mu_1 \mu_2}  j^{(n)'}_{\mu_3\dots \mu_n)}
\end{equation}
where $J^{(2)}$ is the trace reversal of $j^{(2)}$, and generally ${J^{(n)}}' = (1 - n) \, {j^{(n)}}'$ with double traces still vanishing. This produces an interaction~\eqref{eq:canonical_sint} in canonical form, and is equivalent to expanding the current in the dual polynomials $\bar{P}_{(n)}$ defined in~\eqref{eq:Pn_dual_def}. For completeness, we note that the $\bar{P}_{(n)}$ obey an orthogonality relation like~\eqref{eq:first_ortho}, and the analogue of~\eqref{eq:P_orthonormality} is
\begin{equation}
\int [d^4\eta] \, \delta'(\eta^2+1) \, \bar{P}_{(n)} \cdot J^{(n)} \, \bar{P}_{(n)} \cdot K^{(n)} = (-1)^n \left(J^{(n)} \cdot K^{(n)} - \frac{n(n+1)}{4} {J^{(n)}}' \cdot {K^{(n)}}' \right)
\end{equation}
for double-traceless $J^{(n)}$ and $K^{(n)}$. 

% The $\bar P_{(n)}$'s satisfy a similar orthonormality condition to the $P_{(n)}$'s, but with different coefficients
% \begin{equation}
% \int [d^4\eta] \, \delta'(\eta^2+1) \, \bar P_{(n)} \cdot J^{(n)}  \bar P_{(m)}\cdot J^{(m)} = \delta_{mn} \left( J^{(n)}\cdot J^{(m)} -\frac{m}{4} {J^{(n)}}' \cdot {J^{(m)}}'\right)
% \end{equation}
% under the assumption that the $J^{(n)}$'s are double-traceless.
% \kz{I think this result was incorrect}

\subsubsection*{Conservation and Continuity Conditions}

Finally, we derive the tensor form~\eqref{eq:continuity_tensors} of the continuity condition. Note that for a symmetric tensor $A$ of rank $n$, the trace subtraction is defined for $n \geq 2$ by
\begin{equation}
\langle A^{\mu_1 \ldots \mu_n} \rangle = A^{\mu_1 \ldots \mu_n} - \frac{1}{4n \, (n-2)!} \, g^{(\mu_1 \mu_2} {A'}^{\mu_3 \ldots \mu_n)}
\end{equation}
and has no effect for $n < 2$. 

We warm up by considering the case $\rho = 0$, where the result reduces to $\langle \del \cdot J_{(n)} \rangle = 0$. To derive this result, note that the continuity condition~\eqref{eq:continuity_deltaForm} expands to 
\begin{align}
0 &= \delta(\eta^2+1) \sum_{n \geq 1} 2^{n/2} n \, \eta_{\mu_3} \cdots \eta_{\mu_{n-1}} \! \left[\eta_{\mu_1} \eta_{\mu_2} (\del\cdot J_{(n)})^{\mu_1 \ldots \mu_{n-1}} + \frac{n-2}{4} (\del\cdot J'_{(n)})^{\mu_3 \ldots \mu_{n-1}} \right]\\
&= \delta(\eta^2 + 1) \sum_{n \geq 1} 2^{n/2} n \, \eta_{\mu_1} \cdots \eta_{\mu_{n-1}} \langle \del \cdot J_{(n)} \rangle^{\mu_1 \ldots \mu_{n-1}}. \label{eq:weak_second_line}
\end{align}
where the second term in the first line is only nonzero for $n \geq 3$, and we reached the second line by multiplying that term by $-\eta^2$, which is equal to one on the support of the delta function. We can show that each term in the sum individually vanishes by induction. Integrating~\eqref{eq:weak_second_line} over $\eta$ immediately produces $\langle \del \cdot J_{(1)} \rangle = 0$, because all other terms would be proportional to vanishing traces. Next, suppose we have shown $\langle \del \cdot J_{(n)} \rangle = 0$ for all $n \leq m$. Then if one multiplies~\eqref{eq:weak_second_line} by $\eta_{\nu_1} \cdots \eta_{\nu_m}$ before integrating over $\eta$, the only term that can contribute is $n = m + 1$, and performing the integration yields $\langle \del \cdot J_{(m+1)} \rangle = 0$. 

For $\rho \neq 0$, a similar calculation to the one above shows the continuity condition is
\begin{equation}
0 = \delta(\eta^2 + 1) \sum_{n \geq 1} 2^{n/2} n \, \eta_{\mu_1} \cdots \eta_{\mu_{n-1}} \left\langle \del \cdot J_{(n)} + \frac{\rho}{n \sqrt{2}} (J_{(n-1)} + \frac12 J'_{(n+1)}) \right\rangle^{\mu_1 \ldots \mu_{n-1}}
\end{equation}
and an identical inductive argument shows that each term in the sum vanishes. 

\subsection{General Worldline Currents}
\label{app:classification}

In this section we parametrize the general $f(\eta, k, \dot{z})$ obeying the continuity condition~\eqref{eq:continuity_fourier}, and derive the general decomposition~\eqref{eq:general_universality_decomposition}. We also give a simpler parametrization specialized to null $k$, and another suited for static particles.

\subsubsection*{General Solution for Null Momenta}

We warm up by writing the general solution to the continuity condition for only null $k$. First, note that any solution can be decomposed as
\begin{equation}
f(\eta, k, \dot{z}) = e^{- i \rho \eta \cdot \dot{z} / k \cdot \dot{z}} \left(\tilde{f}_0(\eta, k, \dot{z}) + \tilde{f}_\alpha(\eta, k, \dot{z})\right)  \label{eq:phasePrefactorForm}
\end{equation}
where the ``homogeneous'' solution $\tilde{f}_0$ and the ``particular'' solution $\tilde{f}_\alpha$ obey
\begin{align}
k \cdot \del_\eta \, \tilde{f}_0 &= 0, \\ 
k \cdot \del_\eta \, \tilde{f}_\alpha &= (\eta^2 + 1) \, \alpha(\eta, k, \dot{z}).
\end{align}
We have factored the phase of the scalar temporal current~\eqref{eq:J_T_intro} out of tilded variables, which is not necessary, but yields the most convenient parametrization for radiation problems. 

To find the general homogeneous solution for null $k$, note that its dependence on $\eta$ can be written solely in terms of the variables $\eta \cdot \dot{z}$, $\eta \cdot k$, and $\eta^2 + 1$. We drop terms with more than one power of $\eta^2 + 1$, as they can never couple to the continuous spin field due to the delta function in~\eqref{eq:interaction}. Since $k \cdot \del_\eta$ annihilates $\eta \cdot k$, the general solution is parametrized by two analytic functions of $\eta \cdot k$ and the single kinematic variable $k \cdot \dot{z}$,
\begin{equation} \label{eq:f_0_null}
\tilde{f}_0 = g_0(\eta \cdot k, k \cdot \dot{z}) + \left( \eta^2 + 1 - 2 \, \eta \cdot k \, \frac{\eta \cdot \dot{z}}{k \cdot \dot{z}} \right) g_1(\eta \cdot k, k \cdot \dot{z}).
\end{equation}
For null $k$, we can construct a particular solution using the same trick as in appendix~\ref{app:usefulIdentities},
\begin{equation} \label{eq:f_a_null}
\tilde{f}_\alpha = \int_0^{\eta \cdot q} dx \, (\eta^2 + 1 - 2 \, \eta \cdot k \, (\eta \cdot q - x)) \, \alpha(\eta - (\eta \cdot q - x) k, k \cdot \dot{z}).
\end{equation}
While many other particular solutions are possible, this one has the special feature that it vanishes when $\eta \cdot q = 0$, i.e.~all terms in its Taylor expansion have at least one power of $\eta \cdot q$. 

\subsubsection*{General Solution for Arbitrary Momenta}

It is straightforward to extend the above derivation to general $k$. As before, we may drop terms with more than one power of $\eta^2 + 1$, and the remaining $\eta$ dependence is through $\eta \cdot \dot{z}$ and $\eta \cdot k$. The general solution still depends on only three functions $g_0$, $g_1$, and $\alpha$, but they can now depend on two independent kinematic variables $k^2$ and $k \cdot \dot{z}$.

In general, $k \cdot \del_\eta$ does not annihilate $\eta \cdot k$. However, if we define
\begin{equation}
u = \eta \cdot k - \frac{\eta \cdot \dot{z}}{k \cdot \dot{z}} \, k^2, \qquad v = \frac{\eta \cdot \dot{z}}{k \cdot \dot{z}}
\end{equation}
then $k \cdot \del_\eta \, u = 0$. In these variables, we have $k \cdot \del_\eta = \del_v$, so the homogeneous solution is 
\begin{equation} \label{eq:f_0_full}
\tilde{f}_0 = g_0(u, k \cdot \dot{z}, k^2) + \left( \eta^2 + 1 - 2 u v - v^2 k^2 \right) g_1(u, k \cdot \dot{z}, k^2).
\end{equation}
As for the particular solution, we can no longer use the null frame vector $q$, but the same role can be played by $\dot{z} / k \cdot \dot{z}$ as it also has unit inner product with $k$,
\begin{equation} \label{eq:f_a_full}
\tilde{f}_\alpha = \int_0^v dx \, (\eta^2 + 1 - 2 u (v-x) - (v^2-x^2) k^2) \, \alpha(\eta - (v-x)k, k \cdot \dot{z}, k^2).
\end{equation}
These expressions contain the full freedom in the solutions to the continuity condition.

To derive the decomposition~\eqref{eq:general_universality_decomposition} of the general solution, it suffices to show that
\begin{equation} \label{eq:intermediate_universality_decomposition}
f(\eta,k,\dot z) = e^{- i \rho \eta \cdot \dot{z} / k \cdot \dot{z}} \, \hat{g}(k\cdot\dot z) + k^2 X(\eta, k,\dot z) + D \xi(\eta, k,\dot z)
\end{equation}
where $X$ and $\xi$ are both regular as $k^2\to 0$. This is because the continuity condition automatically implies a relation between $X$ and $\xi$,
\begin{align} \label{eq:continuity_relation}
0 = \delta(\eta^2 + 1) \Delta f = \delta(\eta^2 + 1) (k^2 \Delta X - k^2 \xi)
\end{align}
where we used the momentum space form of~\eqref{eq:Delta_D_identity}. Since we have dropped terms in $X$ and $\xi$ with more than one power of $\eta^2 + 1$, this implies $\xi = \Delta X$ and hence~\eqref{eq:general_universality_decomposition}. Equivalently, we can always add terms proportional to $(\eta^2 + 1)^2$ to set $\xi = \Delta X$ for any current.

It is convenient to work entirely in terms of tilded variables, which all have the temporal current's phase factored out. Defining
\begin{equation}
\tilde{D} = \frac{i}{2} (\eta^2 + 1) (k \cdot \del_\eta) - i \eta \cdot k
\end{equation}
we aim to show that $\tilde{f} = \hat{g} + k^2 \tilde{X} + \tilde{D} \tilde{\xi}$. First, using $-2 i \tilde{D} v = \eta^2 + 1 - 2uv - 2v^2k^2$, we can write the contributions of $g_1$ and $\alpha$ as
\begin{align}
\tilde{f}_0 &\supset \tilde{D}(- 2 i g_1 v) + k^2 (g_1 v^2), \\
\tilde{f}_\alpha &= \tilde{D} \int_0^v dx \, (-2i) (v-x) \alpha(\eta - (v-x)k) + k^2 \int_0^v dx \, (v-x)^2 \alpha(\eta - (v-x)k).
\end{align}
From these results we can read off $\tilde{X}$ and $\tilde{\xi}$, and also readily confirm the expected result $\tilde{\xi} = \tilde{\Delta} \tilde{X} = (- i k \cdot \del_\eta) \tilde{X}$, by using $\tilde \Delta v=-i $ and $\tilde \Delta (\eta-(v-x) k) = 0$.

As for $g_0$, we can decompose it by separating out terms with powers of $u$ or $k^2$, 
\begin{equation}
g_0(u, k \cdot z, k^2) = \hat{g}(k \cdot \dot{z}) + u g_u (u, k \cdot \dot{z}) + k^2 g_k(u, k \cdot \dot{z}, k^2)
\end{equation}
which we can write in the desired final form as
\begin{equation}
\tilde{f}_0 \supset \hat{g} + (k^2 + \tilde D \tilde \Delta) (g_k - g_u v).
\end{equation}

\subsubsection*{General Solution for Static Particles}

For computations involving static particles, where $\dot{z}^\mu = (1, \bm{0})$, a time integration often sets the frequency to zero, so that $k^\mu = (0, \v{k})$ and $k \cdot \dot{z} = 0$. In this case the expressions above are not useful because they factor out a divergent phase, which is tied to the problems with defining a static limit for the temporal current. We instead parametrize the general static solution by factoring out the phase of the static spatial current,
\begin{equation}
f(\eta, \v{k}) = e^{- i \rho \etav \cdot \hat{\v{k}} / |\v{k}|} \left(\bar{f}_0(\eta, \v{k}) + \bar{f}_\alpha(\eta, \v{k}) \right)
\end{equation}
so that the homogeneous and particular solutions satisfy
\begin{align}
\v{k} \cdot \nabla_{\etav} \, \bar{f}_0 &= 0, \label{eq:homog_eq_static} \\ 
\v{k} \cdot \nabla_{\etav} \, \bar{f}_\alpha &= (\eta^2 + 1) \, \alpha(\eta, \v{k}).
\end{align}
To compactly write the general solution, note that the dependence on $\eta$ can be written in terms of the variables $\eta^0$, $\etav \cdot \hat{\v{k}}$, and $|\etav \times \hat{\v{k}}|^2$. In these variables,~\eqref{eq:homog_eq_static} states that the homogeneous solution must be independent of $\etav \cdot \hat{\v{k}}$, and the decoupling of higher powers of $\eta^2 + 1$ implies we do not need to consider terms with more than one power of $|\etav \times \hat{\v{k}}|^2$. Then the general homogeneous solution is simply
\begin{equation} \label{eq:general_static_homogeneous}
\bar{f}_0 = \bar{g}_0(\eta^0, |\v{k}|) + |\etav \times \hat{\v{k}}|^2 \, \bar{g}_1(\eta^0, |\v{k}|).
\end{equation}
For example, the scalar-like inhomogeneous current~\eqref{eq:J_I_def} has $\bar{g}_0 = g \cos(\eta^0 \beta \rho^2 / |\v{k}|^2)$ and the vector-like spatial current~\eqref{eq:J_S^V} has $\bar{g}_0 = \sqrt{2} \, e \eta^0$, with both having zero $\bar{g}_1$. Finally, a particular solution is 
\begin{equation} 
\bar{f}_\alpha = \int_0^{\etav \cdot \hat{\v{k}} / |\v{k}|} dx \, (\eta^2 + 1 - (\etav \cdot \hat{\v{k}}) (\etav \cdot \hat{\v{k}} - |\v{k}| x)) \, \alpha(\eta - (\etav \cdot \hat{\v{k}} - x |\v{k}|), \v{k}).
\end{equation}

\subsection{Comparison to Currents From Matter Fields}
\label{sec:fields}

In section~\ref{sec:worldlines} we coupled continuous spin fields to matter particles and identified the most relevant interactions, which reduce to familiar scalar and vector minimal couplings as $\rho \to 0$. It is interesting to compare to previous work which instead described the matter with fields~\cite{Metsaev:2017cuz,Metsaev:2018moa,Bekaert:2017xin,Rivelles:2018tpt}. The most comprehensive analysis for scalar matter fields is Ref.~\cite{Rivelles:2018tpt}, which gives a current quadratic in matter fields $\phi_1$ and $\phi_2$ of mass $M$, parametrized by nonnegative integers $n_0$, $n_1$, and $n_2$, and two real parameters $\lambda_1$ and $\lambda_2$ related by $\rho = (\lambda_1 + \lambda_2) M^2$. For illustration, let us consider the case $n_0 = 2$, $n_1 = n_2 = 0$, which in our conventions is 
\begin{equation} \label{eq:example_J}
J(\eta, x) = g(\eta \cdot \del_x)^2 (\bar{\phi}_1 \bar{\phi}_2) + g (\eta^2 + 1) \left(2 M^2 \, \bar{\phi}_1 \bar{\phi}_2 - \left( 2 + \frac{\lambda_1 + \lambda_2}{2} \, \eta \cdot \del_x \right) \del^\mu \bar{\phi}_1 \, \del_\mu \bar{\phi}_2 \right).
\end{equation}
Above, we have defined $\bar{\phi}_i = \phi_i(x + \lambda_i \eta)$, which should be interpreted as a series in $\lambda_i$. It is straightforward to show this current obeys the continuity condition for on-shell matter fields, but only to first order in the coupling $g$, i.e.~only when one uses the free equations of motion $(\del^2 + M^2) \bar{\phi}_i = 0$. This feature is shared by all currents previously found, and it is therefore unclear to what degree they can be trusted when computing physical observables.  

Furthermore, these currents have no relation to the phenomenologically interesting scalar-like or vector-like currents we have identified. The leading terms in the currents have $n = n_0 + n_1 + n_2$ derivatives, while a minimal scalar or vector current has $0$ or $1$ derivative, respectively. Thus, the currents could only be scalar-like or vector-like for $n < 2$, but when $n_0 < 2$ the current contains undefined negative powers of $\eta$. As discussed in Refs.~\cite{Bekaert:2017xin,Rivelles:2018tpt}, if one treats these negative powers as equivalent to zero, then the continuity condition cannot be satisfied even at first order in $g$ for equal mass scalar fields. 

One might hope the current~\eqref{eq:example_J} is somehow a tensor-like current, but this is not the case either. In the limit $\rho, \lambda_i \to 0$ it instead reduces to a combination of a nonminimal current $J = \del^2(\phi^2)/2$ for a scalar field, and a trivially conserved current $T_{\mu\nu} = (\del_\mu \del_\nu - g_{\mu\nu} \del^2) \phi^2$ for a tensor field. Furthermore, the current in general can be rewritten in the form~\eqref{eq:intermediate_universality_decomposition},
\begin{equation}
J(\eta, x) = - \del_x^2 \left( \frac{g}{2} (\eta^2 + 1) \, (\bar{\phi}_1 \bar{\phi}_2) \right) + D(g \, \eta \cdot \del_x (\bar{\phi}_1 \bar{\phi}_2))
\end{equation}
which indeed obeys~\eqref{eq:continuity_relation}. As discussed below~\eqref{eq:general_universality_decomposition}, this implies the current does not affect radiation emission or the response of the matter to a free field background; in our classification it corresponds to the trivial case $\hat{g} = 0$.

It would be interesting to see if the currents in our work, which satisfy the continuity condition to all orders in $g$, can be recovered in terms of matter fields by relaxing the assumptions of Ref.~\cite{Metsaev:2017cuz,Metsaev:2018moa,Bekaert:2017xin,Rivelles:2018tpt}. In particular, one could either allow explicitly nonlocal Lagrangians or resolve the nonlocality by introducing appropriate auxiliary fields. 

\subsection{Fields and Currents in Spacetime}
\label{app:es_ints}

\subsubsection*{The Scalar-Like Spatial Current}

To evaluate the scalar-like spatial current in position space, it turns out to be easier to first evaluate the field, which satisfies $\Psi_S(\etav, \v{k}) = J_S(\etav, \v{k}) / |\v{k}|^2$ in strong harmonic gauge. Defining $k = |\v{k}|$, expanding $d\v{k} = (k^2 \, dk) \, d \hat{\v{k}}$ and performing the $\hat{\v{k}}$ integral gives 
\begin{align}
\Psi_S(\etav, \v{r}) &= \frac{g}{2 \pi^2} \int_0^\infty dk \, \frac{\sin\left(|\rho \etav / k - k \v{r}|\right)}{|\rho \etav / k - k \v{r}|} \label{eq:Psi_S_setup} \\ 
&= \frac{g}{2 \pi^2 r} \int_0^\infty d x \, \frac{\sin\big(\sqrt{\alpha \, ((x - 1 / x)^2 + \beta}\big)}  {\sqrt{(x - 1 / x)^2 + \beta}}
\end{align}
where we have switched to the dimensionless variable $x = k \sqrt{r / |\rho \etav|}$, defined $\alpha = |\rho \etav| r$ and $\beta = 2 (1 - \cos \theta)$, and always use the positive branch of the square root. The remaining integral is difficult to evaluate directly because of the essential singularity at $x = 0$, but we can avoid it by exploiting the symmetry of the integrand. Mapping the integration range $x \in [0, 1]$ to $x \in [1, \infty)$ by sending $x \to 1/x$ gives, for any function $f$,
\begin{align}
\int_0^\infty dx \, f((x-1/x)^2) &= \int_1^\infty dx \, \left( 1 + \frac{1}{x^2} \right) f((x - 1/x)^2) = \int_0^\infty du \, f(u^2).
\end{align}
Applying this trick, the static field reduces to 
\begin{align}
\Psi_S(\etav, \v{r}) &= \frac{g}{2 \pi^2 r} \int_0^\infty du \, \frac{\sin\big(\sqrt{\alpha (u^2 + \beta)}\big)}{\sqrt{u^2 + \beta}} \\
&= \frac{g}{2 \pi^2 r} \int_0^\infty dz\, \sin\big( \sqrt{\alpha \beta} \, \cosh z \big) \\
&= \frac{g}{4 \pi r} J_0(\sqrt{\alpha \beta}) 
\end{align}
where we let $u = \sqrt{\beta} \, \cosh z$ and used an integral representation of the Bessel function. This is precisely~\eqref{eq:Psi_S_pos}, applying $-\nabla^2$ yields~\eqref{eq:J_S_pos}, and removing the time integration yields~\eqref{eq:f_S_expr}.

\subsubsection*{Temporal Currents}

The frequency integral~\eqref{eq:f_T_initial} for the scalar-like temporal current encloses an essential singularity at $\omega = 0$. While this situation may be unfamiliar, such integrals are well-known in the literature (e.g. see section 4.6 of Ref.~\cite{titchmarsh1939theory}) and can be handled straightforwardly because the integrand has a convergent Laurent series for all $z \neq 0$. The basic identity we will need is 
\begin{equation} \label{eq:es_identity}
\int_C \frac{dz}{2\pi} \, (iz)^p e^{-ibz} e^{-ia/z} = \left( \frac{a}{b} \right)^{(p+1)/2} \, J_{p+1}(2 \sqrt{ab}) 
\end{equation}
where $C$ is a counterclockwise contour encircling the origin, and $p$ is a nonnegative integer. It is derived by simply expanding the integrand in $z$ and applying the residue theorem, 
\begin{align}
\int_C \frac{dz}{2\pi} \, (iz)^p e^{-ibz} e^{-ia/z} &= \int_C \frac{dz}{2\pi} \, \sum_{n, m \geq 0} (iz)^p \frac{(-ibz)^n}{n!} \frac{(-ia/z)^m}{m!} \\
&= - i a^{p+1} \int_C \frac{dz}{2\pi z} \, \sum_{n = 0}^\infty \frac{(-ab)^{n}}{n! \, (n+p+1)!} \\
&= a^{p+1} \, \sum_{n = 0}^\infty \frac{(-ab)^{n}}{n! \, (n+p+1)!}.
\end{align}
Note that we would find the same result if any of the $p+1$ initial terms in the Laurent expansion of $e^{-ia/z}$ were removed, because such terms cannot yield simple poles. 

Now, to evaluate~\eqref{eq:f_T_initial} we rewrite the integrand as 
\begin{equation}
\int \frac{d\omega}{2 \pi} \, e^{- i \omega t} e^{- i \rho \eta^0 / \omega} = \int \frac{d\omega}{2 \pi} \, e^{- i \omega t} + \int \frac{d\omega}{2 \pi} \, e^{- i \omega t} (e^{- i \rho \eta^0 / \omega} - 1)
\end{equation}
where the first integral yields $\delta(t)$, and the second falls off at large $\omega$, so that the contour integral can be closed at infinity. We may therefore close the contour in the upper-half plane for $t < 0$, giving zero, and in the lower-half plane for $t > 0$. In the latter case, we can shrink the contour to a clockwise one enclosing the essential singularity and apply~\eqref{eq:es_identity}, which yields the result~\eqref{eq:f_T}. Similarly, to derive~\eqref{eq:f_T^V} for the vector-like temporal current, write
\begin{equation}
j_T^V(\eta^0, \v{r}, t) = \frac{\sqrt{2} \, e}{\rho} \left(  \int \frac{d\omega}{2 \pi} \, \rho \eta^0 e^{-i \omega t} + \int \frac{d\omega}{2\pi} \, i \omega \, e^{- i \omega t} \, (e^{- i \rho \eta^0 / \omega} - (1 - i \rho \eta^0/\omega)) \right)
\end{equation}
where the first term produces the delta function and the second can be evaluated using~\eqref{eq:es_identity}.

Incidentally, for real $a$, integrals like~\eqref{eq:es_identity} are well-defined even when the contour passes directly through the essential singularity along the real axis, where the integrand is bounded. When $a$ is positive (negative), the contour can be deformed off the origin in the positive (negative) imaginary direction, and the remaining integral can be evaluated using~\eqref{eq:es_identity}.

\subsubsection*{Currents for Arbitrary Motion}

For completeness, we note that it is straightforward to write expressions for the currents of particles in arbitrary motion, once one has evaluated the current of a worldline element at rest. For example, if we expand $k^\mu = k_\ell \dot{z}^\mu(\tau) + k_p^\mu$, where $\dot{z} \cdot k_p = 0$, and do the same for $x$ and $z$, then the scalar-like temporal current is 
\begin{equation}
J_T(\eta, x) = g \int d \tau \int \frac{d^3k_p}{(2\pi)^3} \, e^{- i k_p \cdot (x_p - z_p)} \int \frac{d k_\ell}{2\pi} \, e^{- i \rho \eta_\ell / (k_\ell + i \epsilon)} e^{- i k_\ell (x_\ell - z_\ell)}.
\end{equation}
By the same reasoning used to derive~\eqref{eq:f_T} above, we have
\begin{equation}
J_T(\eta, x) = g \int d\tau \, \delta^{(4)}(x - z) - \theta(x_\ell - z_\ell) \, \delta^{(3)}(x_p - z_p) \sqrt{\frac{\rho \eta^0}{x_\ell - z_\ell}} J_1\!\left(2 \sqrt{\rho \eta^0 (x_\ell - z_\ell)}\right).
\end{equation}
Performing the $\tau$ integral for a static particle with a cutoff $\tau = -T$ yields~\eqref{eq:J_T_result}. Similarly, the scalar-like spatial current is 
\begin{equation}
J_S(\eta, x) = g \int d\tau \int \frac{d^3 k_p}{(2\pi)^3} \, \delta(x_\ell - z_\ell) \, e^{- i k_p \cdot (x_p - z_p)} e^{-i \rho k_p \cdot \eta_p / k_p^2}
\end{equation}
where the $k_p$ integral can be evaluated analogously to~\eqref{eq:f_S}. However, while expressions like these make the spacetime support of the current manifest, they are seldom useful for calculating physical quantities, where the momentum space expressions are more convenient. 

\subsection{Radiation Emission Results}
\label{app:gauge_inv}

\subsubsection*{Gauge Invariance of the Averaged Stress-Energy Tensor}

As discussed in section~\ref{sec:radiation_CSP}, it suffices to consider the variations of the second and third terms of $T^{\mu\nu}$ in~\eqref{eq:stress_energy_terms} under an infinitesimal gauge transformation $\delta_\epsilon \Psi = D \epsilon$. Since we are implicitly taking a spacetime average, we can freely integrate spacetime derivatives by parts, and we can simplify terms involving the operator $D$ using the identities~\eqref{eq:Delta_D_identity} and~\eqref{eq:deltapr_D_identity}. 

First, defining $\epsilon^\nu = \del^\nu \epsilon$ for brevity, the variation of the second term is 
\begin{align}
\delta T^{\mu\nu}_{(a)} &= - 2 \int [d^4 \eta] \, \delta'(\eta^2 + 1) \, \Psi \, D(\del^\mu \del^\nu \epsilon) \\
&= - \int [d^4 \eta] \, \Psi \, \Delta (\delta(\eta^2 + 1) \, \del^\mu \epsilon^\nu) \\
&= - \int [d^4 \eta] \, \delta(\eta^2 + 1) \, (\Delta \Psi) \, \del^\mu \epsilon^\nu
\end{align}
where we integrated by parts and used~\eqref{eq:deltapr_D_identity}. The variation of the third term includes
\begin{align}
\delta T^{\mu\nu}_{(b)} &= - \frac12 \int [d^4 \eta] \, \del_{\eta}^\mu (\delta(\eta^2 + 1) \, \Delta D \epsilon) \, \del^\nu \Psi \\
&= \frac12 \int [d^4 \eta] \, \del_{\eta}^\mu (\delta(\eta^2 + 1) \, \del_x^2 \epsilon^\nu) \, \Psi \\
&= \int [d^4 \eta] \, (\del_x^2 \Psi) \left( \frac12 \delta(\eta^2 + 1) \, \del_\eta^\mu \epsilon^\nu + \delta'(\eta^2 + 1) \, \eta^\mu \epsilon^\nu \right) 
\end{align}
where we integrated by parts and used~\eqref{eq:Delta_D_identity}. The other part of the third term's variation is 
\begin{align}
\delta T^{\mu\nu}_{(c)} &= \frac12 \int [d^4\eta] \, \delta(\eta^2 + 1) \, (\Delta \Psi) (\del_\eta^\mu \del^\nu D \epsilon) \\
&= \frac12 \int [d^4\eta] \, \delta(\eta^2 + 1) \, (\Delta \Psi) \left((\eta \cdot \del_x) (\del_\eta^\mu \epsilon^\nu) + (\del^\mu - \eta^\mu \Delta) \epsilon^\nu \right) \\
&= \int [d^4\eta] \, \delta(\eta^2 + 1) \, (\Delta \Psi) \left(\frac12 (\eta \cdot \del_x) (\del_\eta^\mu \epsilon^\nu) + \del^\mu \epsilon^\nu - \frac12 \Delta \eta^\mu \epsilon^\nu \right)
\end{align}
where we simply performed the $\eta$ derivative using the definition of $D$. The middle term of this expression cancels with $\delta T^{\mu\nu}_{(a)}$. The last term cancels with the last term of $\delta T^{\mu\nu}_{(b)}$ upon using the equation of motion~\eqref{eq:free_eom}. Finally, note that multiplying the equation of motion by $\eta^2 + 1$ and simplifying gives the identity 
\begin{equation}
\delta(\eta^2 + 1) \, \del_x^2 \Psi = \delta(\eta^2 + 1) \, (\eta \cdot \del_x) \Delta \Psi
\end{equation}
which implies the remaining two terms cancel. 

\subsubsection*{Stationary Phase Approximation for Total Power}

Here we extract the scaling behavior of the integral
\begin{equation}
I = \sum_{n > 0} n^2 \int_0^T \frac{dt}{T} \int_0^T \frac{dt'}{T} \, e^{i n \omega_0 (t - t')} \, J_0\left( \frac{\rho |\v{v}_\perp(t) - \v{v}_\perp(t')|}{n \omega_0} \right).
\end{equation}
Consider an arbitrary generic linear motion, meaning that $v_\perp(t) = v_0 f(\omega_0 t) \sin \theta$ where $v_0$ is the typical speed, and $f(\phi)$ is an order one periodic function with order one derivatives. Defining $x = (\rho v_0/\omega_0) \sin \theta$ and changing variables to the phase sum and difference gives 
\begin{align}
I &\sim \sum_{n>0} n^2 \int d \phi' \int d\phi \, e^{i n \phi} J_0\left( \frac{x}{n} (f(\phi + \phi'/2) - f(\phi - \phi'/2))  \right) \\
&= \int d\phi' \, \sum_{n>0} n^2 \int d\phi \, e^{i n \phi} J_0\left(x g_{\phi'}(\phi) / n\right)
\end{align}
where $g_{\phi'}(\phi)$ is another order one periodic function. Since $x$ and $n$ are large, the integrand is rapidly oscillating and thus dominated by contributions from points of stationary phase. Using the asymptotic expansion of the Bessel function gives, up to overall phases,
\begin{equation}
I \sim \int d\phi' \, \sum_{n>0} n^2 \int d\phi \, \sqrt{\frac{n}{x}} \, e^{i (n \phi + x g_{\phi'}/n)}.
\end{equation}
The integrand has stationary phase when $n + x g'_{\phi'}/n = 0$, which is generically possible only when $n \lesssim \sqrt{x}$, and at such points, the rate of change of phase is $x g''_{\phi'}/n$. Then we have
\begin{equation}
I \sim \int \frac{d\phi'}{\sqrt{g''_{\phi'}}} \, \int_0^{\sqrt{x}} dn \, n^2 \, \frac{n}{x} \sim x
\end{equation}
where the $\phi'$ integral is of order one because $g_{\phi'}''$ is generically of order one. This recovers the scaling exhibited in~\eqref{eq:scalar_csp_limit}, as well as the rough distribution of the power in harmonics. 

\newpage
\bibliography{CSPAmplitudes}

\end{document}